\documentclass[journal,comsoc]{IEEEtran}
\usepackage[hyphens,spaces,obeyspaces]{url}
\usepackage[T1]{fontenc}
\usepackage{tikz}
\usepackage{blindtext, graphicx}
\usepackage{multirow}
\usepackage{float}
\usepackage{color,soul}
\usepackage{graphicx}
\graphicspath{{figs/}}
\usepackage{ctable}
\usepackage{footnote}
\makesavenoteenv{tabular}
\makesavenoteenv{table}
\usepackage{longtable}
\usepackage{threeparttable}
\usepackage{ctable}
\usepackage{cite}
\usepackage{amsfonts}
\usepackage{comment}
\usepackage{colortbl}
\usepackage{diagbox}
\usepackage[super]{nth}
\usepackage{subcaption}
\usepackage{dblfloatfix}
\usepackage{mathtools}
\usepackage{tcolorbox}
\usepackage{amssymb}
\usepackage{xcolor}
\usepackage{soul}
\newcommand{\mathcolorbox}[2]{\colorbox{#1}{$\displaystyle #2$}}

\urldef{\footurl}\url{https://www.marketwatch.com/press-release/97-of-people-globally-unable-to-correctly-identify-phishing-emails-2015-05-12}

\ifCLASSINFOpdf
\else
\fi
\usepackage{amsmath}
\usepackage[cmintegrals]{newtxmath}
\renewcommand\hl[1]{#1} 

\begin{document}
%
\title{SOK: A Comprehensive Reexamination of Phishing Research from the Security Perspective}
%
%
%

\author{Avisha Das, Shahryar Baki, Ayman El Aassal, Rakesh Verma, and Arthur Dunbar
\thanks{All the authors are with the Department
of Computer Science, University of Houston,
TX, 77204 USA. e-mail: \{rverma, adas5, sbaki2, aelaassal\}@uh.edu}
}

\newtcolorbox[list inside=mybox,auto counter,number within=section]{MyBox}{colbacktitle=blue,coltitle=white,title={Example \thetcbcounter: Example of a phishing email header}}

\newtcolorbox[list inside=mybox,auto counter,number within=section]{NotationBox}{colbacktitle=blue,coltitle=white,title={Notations \thetcbcounter: Summary of the notations used in the tables}}

%
%


\markboth{Reexamining Phishing Research}%
{Shell \MakeLowercase{\textit{et al.}}: Bare Demo of IEEEtran.cls for IEEE Communications Society Journals}
%



\maketitle

\IEEEpubid{\begin{minipage}{\textwidth}\ \\ [10pt] \centering \copyright 2019 IEEE. Personal use of this material is permitted.  Permission from IEEE must be obtained for all other uses, in any current or future media, including reprinting/republishing this material for advertising or promotional purposes, creating new collective works, for resale or redistribution to servers or lists, or reuse of any copyrighted component of this work in other works. \end{minipage}} 
\IEEEpubidadjcol

\begin{abstract}
Phishing and spear phishing are typical examples of masquerade attacks since trust is built up through impersonation for the attack to succeed. Given the prevalence of these attacks, considerable research has been conducted on these problems along multiple dimensions.
We reexamine the existing research on phishing and spear phishing from the perspective of the unique needs of the security  domain, which we call \textit{security challenges}: real-time detection, active attacker, dataset quality and base-rate fallacy. 
We explain these challenges and then survey the existing phishing/spear phishing solutions in their light. This viewpoint consolidates the literature and  
illuminates several opportunities for improving  
existing solutions. We organize the existing literature based on  detection techniques for different attack vectors (e.g., URLs, websites, emails) along with studies on user awareness. For detection techniques we examine properties of the dataset, feature extraction, detection algorithms used,  and performance evaluation metrics. This work can help guide
the development of more effective defenses for phishing, spear phishing and email masquerade 
attacks of the future, as well as provide a framework for a thorough evaluation and comparison.
\end{abstract}

\begin{IEEEkeywords}
Phishing, spear phishing, usable security, email, website, URL, dataset properties, unique challenges of security
\end{IEEEkeywords}

%
\IEEEpeerreviewmaketitle

\section{Introduction}
\label{sec-intro}


Internet users continue to be plagued by many attacks, which include: spam, phishing, spear phishing, masquerade, and malware delivery. \textit{Spam} is an advertisement and its most pernicious effect is the loss of time and productivity. \textit{Phishing} and \textit{spear phishing} are more damaging. In these attacks, the attacker impersonates a trusted entity with an intent to steal sensitive information or the digital identity of the target, e.g., account credentials, credit card numbers, etc. The difference between them is that spear phishing is more targeted and phishing is more indiscriminate. 
\hl{Both of them are cases of masquerade attacks which involve impersonation. However, this type of attack (masquerade) tends to have broader objectives.}
Examples include: planting fake news, sowing divisions in communities (e.g., the case of WhatsApp in India), and swaying opinions (e.g., the case of stealing elections). \hl{The phrase} \textit{malware delivery} is self-explanatory.

Email continues to be one of the most convenient and popular vectors of choice for the above attacks. An email is usually embedded with a poisoned link to a fraudulent website set up to trick the victim.
However, with the growing popularity of social networks, a new medium for spreading malicious links is also available in the form of instant messaging applications, chat services, etc.

The persistent popularity of phishing and spear phishing with attackers~\cite{ferreiraM18, vermaKM15}, comes from the fact that they exploit the human element (``weakest link'')~\cite{daswani07} and do not require any actual intrusion into the system or network. Phishing/spear phishing attacks have been successfully used to bring well-protected companies to their knees (e.g., the attack on RSA as described in \cite{diehl16}) and estimates of losses from phishing alone run into several hundred million dollars in the US. \hl{In addition to the monetary loss, there is also a loss of time, productivity, and  damage to reputation.} 
Besides stealing sensitive information, email attachment and web links in the emails are the most common way of spreading malware, for example, 9 out of 10 phishing emails detected in Verizon network on March 2016 carried ransomware \cite{Barkly2016}. \hl{Previous phishing studies have found that phishing is ``far more successful than commonly thought'' and it is the main mechanism for manual account hijacking} \cite{burszteinBM14,ferreiraM18}.

Despite more than a decade of research on phishing, it still continues to be a serious problem. There could be several reasons for this: the problem itself may be intractable, the technical approaches so far may have missed important parameters of the problem, phishing exploits the human as the weakest link so purely technical approaches may not be sufficient, or some combination of these. To expose the reasons for the continued success of phishing, we survey the detection literature and user studies on phishing and spear phishing.

We found that researchers have tackled phishing/spear phishing in many papers, and there are several surveys of these attempts. However, we \hl{discovered} that there are certain methodological issues with phishing detection research. For example, we \hl{notice}: 
\begin{itemize}
    \item The use of balanced datasets and inappropriate metrics
    \item Unreported training and testing times
    \item Lack of generalization studies
    \item Also in user studies, we find the multiple comparison issue
\end{itemize}


Many surveys have missed quite a few of these issues, e.g., the dataset diversity issue is never mentioned. Challenges such as base-rate fallacy, active attacker and generalization studies, which we collectively call ``security challenges'' are rarely mentioned, let alone emphasized. \hl{We identify the required set of challenges in cybersecurity based on those outlined in} \cite{vermaKM15,paxson}.

\newpage


Our goal is to reinvigorate research on these problems and reorient it towards solving the urgent, practical needs of the \hl{security domain}~\cite{vermaKM15}. Hence, we reexamine the previous literature on phishing and spear phishing \textit{from the viewpoint of the unique needs of the security domain}, which we elaborate upon in Section \ref{sec-challenges}, to determine the appropriateness of the proposed solutions. 
To our knowledge, a comprehensive evaluation of the appropriateness of the previous research on phishing and spear phishing from the security perspective has not been done before. A better understanding of the previous studies in this light will foster research on effective and practical defenses for these problems. Such a perspective will also provide a framework for a thorough evaluation of current and future solutions. 
User studies are also important for stopping the phishing attacks since attackers try to elude computer users. No matter how good the detection system works, end users should be prepared for the different types of attacks. So, we also review the phishing user studies \hl{in addition to} the detection techniques.
We observed that there are quite a few existing surveys on phishing emails, significantly fewer on phishing websites and Uniform Resource Locators (URLs), and none on user studies. However, we did not find any survey on phishing or spear phishing that emphasized the unique needs of the security domain and examined the research from this perspective. 
Our contributions are as follows:
\begin{itemize}
    \item We adapt the security challenges in cybersecurity \cite{vermaKM15,paxson} to the field of detecting phishing URLs, websites, emails and user studies. 
    \item We collect and review a comprehensive list of literature on phishing and spear phishing detection systems \textit{as well as user studies}.
    \item We conduct a systematic review of the phishing detection research with a focus on: the features extracted, detection methods used, properties of the evaluation dataset (source, size, class ratio, diversity, recency), evaluation metrics for studying system performance.
    \item We investigate the diversity of several popular URL, website and email datasets.
    \item We select and discuss well-cited literature based on their contributions from the perspective of the security challenges addressed by the authors.
    \item We summarize the best practices observed in the studied phishing detection literature.
\end{itemize}

\noindent\textbf{Paper Organization}:

Section~\ref{sec:keyword} describes our search for relevant literature, which is important for the reader to understand the exact contours of our systematization. Then, we discuss essential background (Section~\ref{sec-bground}) and security challenges (Section~\ref{sec-challenges}). Common notations for a thorough review of detection research and user studies are in Section~\ref{sec:notations}. We investigate phishing detection research in Sections~\ref{sec-techniques-urls},~\ref{sec-techniques-websites}, and~\ref{sec-tech-emails}.   Opportunities for future work are discussed in Section~\ref{sec:gik-uw}. We analyze user studies in Section~\ref{sec-userstudies}, and compare our work with related work in Section~\ref{sec-related}. 
Section~\ref{sec-concl} concludes the paper. 
\section{Systematic Literature Collection}
\label{sec:keyword}
We followed a systematic methodology to search the relevant research papers and surveys, which address phishing and spear phishing \hl{attacks. 
According} to DBLP (\hl{Digital Bibliography \& Library Project}), the first phishing papers appeared in 2004,\footnote{There is a 1997 MIT thesis starting \hl{with PHISH}, but this is an acronym and means something completely different. There is also a book called Phish.net in 1997, but this is a web community regarding the music band called Phish.} and there are 732 papers from 2004-2017.\footnote{Keyword search, phish, on 6 May 2018.} For comparison, a publication search with keyword `spam' on DBLP yields 2,213 papers (the query, `spam\$', exact word match for spam, yielded 1,785 papers), but this is an estimate, since it includes opinion/review spam papers and also more than 100 matches with author last name ``Spampinato.''

To gather the research papers for this study, at first we used the queries \textit{``phish URL, phish link, phish site, phish web, 
phish email/e-mail''} independently 
on four databases: DBLP, ACM Digital Library, IEEE Xplore, and Google Scholar (\textit{allintitle} query). We also found many papers that study phishing as part of malicious and malware centric behavior. Therefore, we added the following additional queries (without quotes) - \textit{``malware URL, malicious URL, malware link, malicious link, malware site, malicious site, malware email/e-mail, malicious email/e-mail.''} However, we only consider research specific to phishing attack vectors -- URL, email and websites. Papers which propose mainly malware detection techniques are beyond the scope of this survey. \hl{To search for relevant literature on spear phishing, we used the queries \textit{``spear phishing''} (this query also covers spear-phishing) and \textit{``spearphishing.''}} The queries \textit{``spear phish''} and \textit{``spearphish''}  did not yield any additional results.
Later, we realized that authors sometimes just use \textit{``phishing detection''} or \textit{``phishing attack,''} or some other variation. So we expanded our search using \textit{``phish''} and \textit{``phishing''} to the above databases.


\begin{figure*}[h]
\centering
 \includegraphics[width=0.9\linewidth]{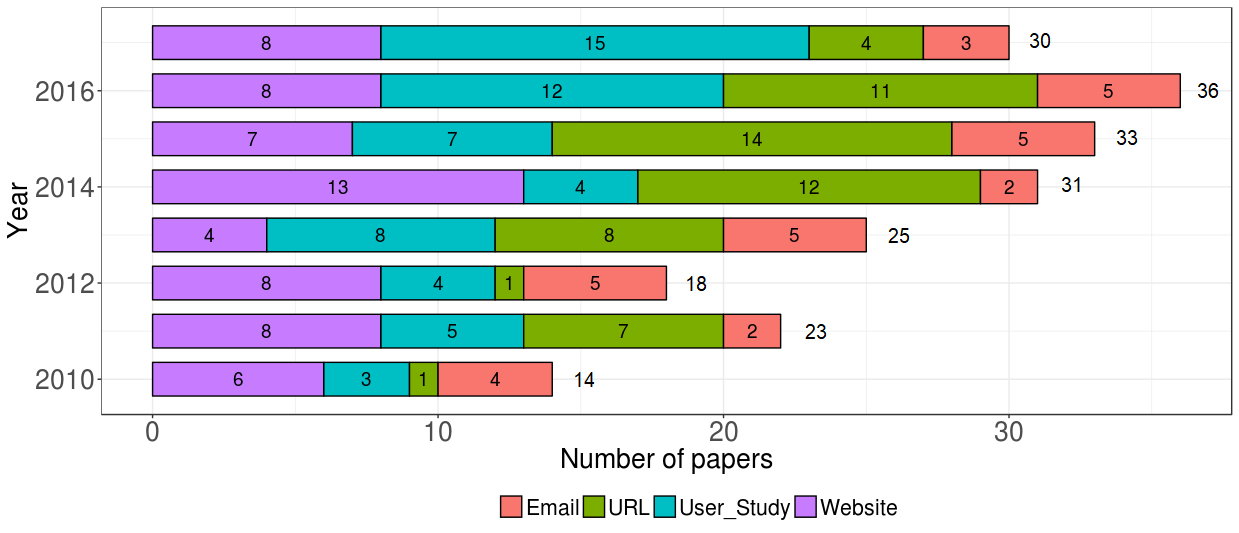}
 \caption{The number of papers that we cover in this survey for each year based on their type (from 2010 to 2017). Numbers inside the bars denote the count of phishing URL, website, email and user study papers in each year. The numbers in front of each bar is the total count of papers in that year.}
 \label{fig:totalpapers}
 \end{figure*}


For this paper, we mainly focus on research published between the years 2010-2017. We cover papers appearing up to March 2018, and any pre-2010 paper that is highly cited or appeared in a major security venue. \hl{We also cover general phishing surveys appearing up to 2018.}\footnote{This list includes ACM's CCS, ASIACCS, CODASPY, TISSEC journal, IEEE's Trans. on DSN, Trans. on Forensics, Symp. on Security and Privacy, Security and Privacy Mag., USENIX Security, NDSS, ACSAC and ESORICS.} 
Our search gave us over 734 papers on {\em phishing detection and user studies}, 
 which we then reduced using the conference/journal CORE rankings of B or higher to approximately 
300 papers.\footnote{If a conference or journal was unranked by CORE, we use Google Scholar H5 index of $\geq 20$ as our guideline.}

Figure~\ref{fig:totalpapers} shows the distribution of papers covered based on the publication year and type: URL detection, website detection, email detection and user studies (papers published between 2010-2017). 
It shows a significant increase in attention to the user studies (a 20\% increase from 2014 to 2016) compared to other areas. This could be due to the increasing use of spear phishing attacks. It is also clear that more papers focused on phishing URL detection during 2014-2016. Surprisingly, phishing email detection seems to be falling out of favor.  
Note that if an email detection paper also used URL features for detection (70\% of email detection papers do this), we classify it as an email paper for this figure, and similarly if a phishing website detection paper used URL features (64\%), then we classify it as a website paper. 


During our search, we also found a number of surveys on various aspects of phishing.
Before we begin, we should clarify that it is very difficult to directly compare the proposed systems in literature, since most of them used different datasets for evaluation. Even if the same sources were used, e.g., Phishtank and Alexa, we still cannot compare them directly because these datasets are updated regularly. The situation is a little different for emails since several research papers use the same phishing email datasets, which are publicly available. However, the legitimate datasets used are different because researchers use their own or a private company's emails. We could  compare the detection rate for the phishing email datasets used, but it would not be an accurate comparison since the training is done using both datasets.

\section{Background}\label{sec-bground}
\subsection{Phishing and Spear Phishing}

As reported by many surveys, the first phishing attack is supposed to have occurred in 1995 on America Online (AOL). 
Phishing has been one of the preferred methods used by attackers to lure unsuspecting victims. There is ample literature on the steps involved in a phishing attack (see \cite{dou2017systematization, gupta17TJ} for example). As suggested in \cite{dou2017systematization}, the phishing process can be divided into five steps:  reconnaissance, weaponization, distribution, exploitation, exfiltration. It starts with the attacker, disguised as a legitimate entity (reconnaissance). Then they host a website similar to the target (weaponization) and send an attack vector (usually an email) to the victim (distribution). The attackers may also spread such links using social networking sites, instant messaging applications, etc. The attackers use innovative methods to exploit the weakness of humans to make them think the fraudulent websites are legitimate (exploitation), e.g., using URLs which are similar to the original one (paypa\textbf{1}.com instead of paypal.com). In the last step, attackers collect the sensitive information exposed by victims (exfiltration). For spear phishing, we refer the reader to \cite{caputo2014going}.  

\begin{figure}[!htb]
\centering
 \includegraphics[width=\linewidth]{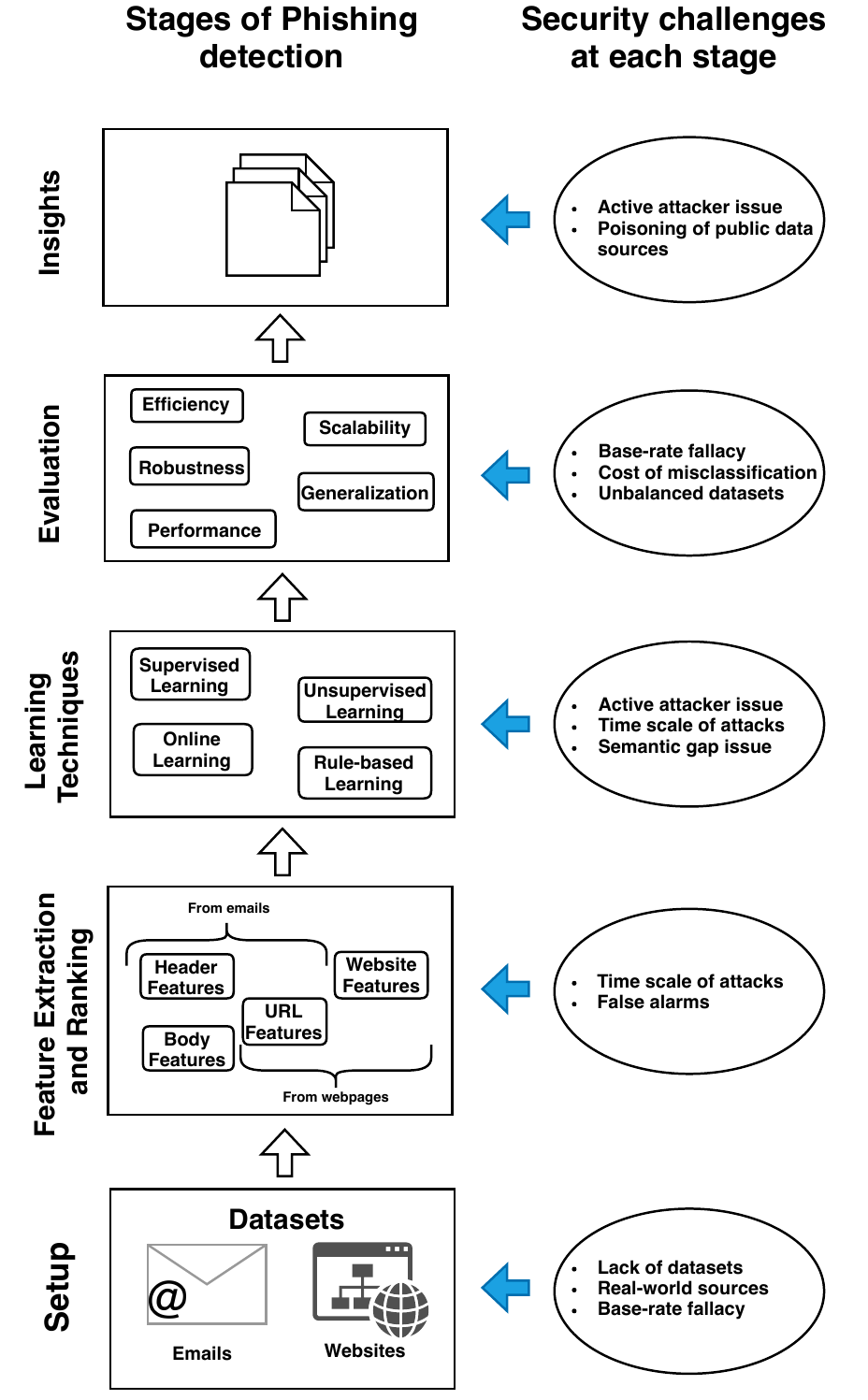}
 \caption{Different stages of a typical phishing detection system with the critical challenges associated at each stage}
 \label{fig:stagechallenge}
 \end{figure}

 Estimates of the economic damage caused by phishing can be found in \cite{guptaAP17}, and the growth or prevalence of phishing can be seen from reports by Anti-Phishing Working Group (APWG) \cite{apwg}. 
 Besides these papers, noteworthy is a study of phishing networks by ``spidering'' chat rooms across major Internet Relay Chat (IRC) servers \cite{abad05} and a study of the economics on phishing from the perspective of all phishers treated as a group \cite{herleyF08}. Authors argue that the average payoff for a phisher is likely to be small under certain assumptions. This result does not tell us about the distribution, which could be highly skewed with a high median for example.  
Also, if the payoff is low for some phishers, many of them may be discouraged and move to greener pastures until the situation ``improves'' for the rest, which will attract more phishers. Thus, the cycle repeats. 

\subsection{Modus Operandi of Phishing} 
Several studies examined aspects of how phishers create their attacks and how they deceive their victims \cite{mcgrathG08,covaKV08,mccalleyWW11,HanKB16,chiewYT18}. In \cite{mcgrathG08}, researchers analyzed a proprietary dataset and found that phishing domain names have a different length distribution from legitimate domains and a different character frequency distribution from that of standard English text. Cova et al. \cite{covaKV08} and McCalley et al. \cite{mccalleyWW11} analyzed phishing kits, which are packages that contain easy-to-deploy, complete phishing websites downloaded from the Internet. They found in these kits examples of obfuscation and also backdoors, which send information collected to a third party. Thus, they were being used to fool gullible users as well as na\"ive phishers. More recently, Han et al. \cite{HanKB16} created a honeypot to attract phishers so that they could analyze live phishing kits. They have several interesting results on the duration and other characteristics of the phishing life cycle from deployment to blacklisting. In \cite{chiewYT18}, authors have made a detailed study of vulnerabilities exploited and technical approaches in phishing attacks. 

\subsection{Variations on Phishing}
\hl{Because it exploits the weakest link in the security chain, i.e. humans}~\cite{daswani07}, phishing is now moving to social networks, and also SMS/text messages (smishing).\hl{\footnote{\url{https://www.wandera.com/mobile-security/phishing/mobile-phishing-attacks/}}} Other variants of phishing include vishing (voice-based phishing) and QRishing (phishing using QR codes). However, there are only a handful of papers on these topics, as shown in \cite{ferreiraM18,chiewYT18}. 

\subsection{Phishing Detection}

Blacklists are a commonly used method to thwart phishing attacks \cite{sahoo2017malicious}. The popular blacklists like PhishTank are created using crowd-sourcing but \hl{have} two shortcomings - i. \hl{manual maintenance}, ii. \hl{newer} phishing attacks launched for short time windows \hl{may remain undetected} \cite{vermaK15, verma2017s}. Thus, to overcome these issues, data science methods have gained popularity with researchers to detect phishing websites and emails. They use dataset of legitimate and phishing URLs/emails to train different classification models \cite{dou2017systematization}. Then, these models can be used to either build and update a blacklist or detect phishing URLs/emails in real-time while internet users open a website or email.

Figure \ref{fig:stagechallenge} illustrates the stages of a machine learning based phishing detection system. It starts with one or more datasets and extracts features from different components of website/email (e.g., URL, HTML, email header, etc.). In the next step, the extracted features are fed to the learning algorithms to extract the patterns/rules for distinguishing legitimate email/URLs from phishing. In the fourth step, the effectiveness of the model will be evaluated using different metrics. 
Finally, the results will be interpreted and the best models and features will be determined, with insights on what worked best or worst. We review the different techniques proposed/used by researchers in more details in Sections \ref{sec-techniques-urls}, \ref{sec-techniques-websites} and \ref{sec-tech-emails}.

\subsection{Human Vulnerability}
Machine learning techniques are not 100\% accurate. Moreover, models trained on historical data may fail to identify newer types of attacks. Therefore, the user also becomes an important element in the detection cycle. Understanding the users' behavior as well as training them to properly detect attack vectors play an important role in the prevention of phishing attacks.

There are some vulnerabilities in human cognition which help attackers in deceiving people. Researchers in \cite{parsons2015design} categorized emails based on their intention into four groups: (1) Risk or Loss (2) Benefit or Gain (3) Account Information and (4) Information Only. The first category uses the sense of urgency and panic to make people click on a link and expose their sensitive information. The second group uses people's excitement about receiving a big prize to lure them. The last two groups are not as pressing as the first two which make them look more legitimate for non-expert users. A study of phishing emails sent to about 6 million people, conducted by KnowBe4 \cite{report-scmagazine}, revealed that people fall for attacks belonging to groups 1 and 2 more frequently than the other types. Their results showed that the click rate is higher for the emails which promise money or threaten loss of money.

In this paper, we also cover the user studies to have a thorough overview of the different aspects of the detection cycle. Before reexamining the phishing detection literature and user studies from the security viewpoint, we examine the specific needs of the security domain, which we call security challenges. 

\section{Security Challenges}
\label{sec-challenges}

The security domain imposes a number of challenges on data science methods that are typically used to detect phishing and spearphishing attacks \cite{vermaKM15,paxson}. In \cite{vermaKM15}, the following challenges were identified. 

\begin{enumerate}
\item \textit{Active attacker}: In the security domain, there is an adversary who is constantly trying to learn the defensive methods being employed and working to defeat them. For this reason, defenders need to design methods that can detect new attacks (zero-day attacks) or variations of existing attacks, not just the ones seen in the past. Supervised machine learning methods perform well if the test data has the same/similar distribution as the training data, but this ``stationarity'' assumption may not hold in practice because of active attackers (adversarial setting), so retraining of the model or online learning methods are required. 
\item \textit{Base-rate fallacy}: The incidence of an attack, i.e.\ the base-rate, has a bearing on the likelihood that something {\em classified} as an attack is actually an attack. In a big company like Chevron, the number of legitimate emails sent and received is in the millions per day and the attacks may only be a few 10s to 100s \cite{irene17}. For example, if the incidence of an attack is 10\% and a classifier is 90\% accurate, then the probability that a scenario classified as an attack is actually an attack is only 50\%. \hl{The ratio between positive and negative samples used for system evaluation and the use of appropriate metrics are a major concern for security researchers.}
\item \textit{Time scale of attacks}: An attack can be over very quickly, so we may need \textit{real-time}  prevention/detection. In the case of phishing detection, the volume of data that has to be processed, e.g., emails, necessitates efficient and fast methods. 
\item \textit{Dataset Issues}: There are four dataset issues: availability, diversity, recency, and quality. Because companies are worried about the damage to their reputation~\cite{irene17}, there is a reluctance to share attacks. Hence researchers have limited data sources available, which also has an adverse effect on diversity and quality of datasets. By quality, we mean how well the dataset represents the real scenario. For example, if the real scenario is extremely unbalanced as for spear phishing emails, then the test dataset should reflect this. Recency of datasets also plays an important role in system evaluation. 
\end{enumerate}
Other challenges mentioned in \cite{vermaKM15} include: asymmetrical and user-dependent costs of misclassification (e.g., misclassifying phishing emails is less costly than misclassifying legitimate emails for an expert, but for a beginner the opposite may be true\hl{\footnote{An expert is an individual who is familiar with such targeted attacks and the attackers' modus operandi; while a beginner is usually an individual with little to no prior exposure to such attacks.}})%
, unbalanced datasets (this is related to the base-rate fallacy), the pernicious effects of false alarms, the semantic gap (the gap between the classifier's decision and the explanation for that decision; many machine learning methods provide no human-readable explanation of their decision), and the opportunity for attackers to poison publicly-available datasets. 

In \cite{paxson}, the researchers identified the following challenges that are specific to applying machine learning for intrusion detection: outlier detection setting, the high cost of misclassification, semantic gap, diversity of network traffic, and difficulties with evaluation. Difficulties with evaluation were further elaborated upon as data difficulties, mind the gap, and adversarial setting. Figure~\ref{fig:stagechallenge} shows the relevant and critical challenges at each step of phishing detection.

In this paper, we focus on the following challenges, which are the most important in the phishing context: adversarial setting, fast and efficient detection, dataset issues, base-rate fallacy, and evaluation metrics. The semantic gap is also an important issue, but there has been hardly any progress in this dimension. The pernicious effect of false alarms, i.e., flagging a legitimate email as phishing, implies that the false alarm rate should be kept low, but the threshold is subjective and hence hard to treat objectively in scientific research. Similar observations hold for user-dependent misclassification cost.


%
    

\section{Common Notations and Organization}\label{sec:notations}
In reviewing the phishing detection schemes, we focus on the  following key attributes, which are essential for robust detection.
\subsection{Key Attributes for Phishing Detection} \label{sec-techniques}
\begin{itemize}
    \item \textit{Nature and Source of data}: 
    \hl{Nature and source of data: Information about the nature and composition of the data i.e. malware, phishing, or spam. In addition to the sources of the dataset and their availability, datasets' sizes, quality, diversity, and recency.}
    This stems directly from the dataset concerns.
    \item \textit{Feature Extraction}: Description of the types of features extracted from the datasets for detection and analysis of URLs, websites, or emails along with any feature selection techniques like information gain. Feature extraction has implications for speed, efficiency, and robustness. To have a better understanding of the feature's effect on detection systems, we discuss two important properties of the features: feature size 
and processing time. \textit{Feature size} is important since it can affect the detection time of systems. For example, using \textit{term frequency} as a feature increases the number of features dramatically, which may slow down some classifiers (and require more resources) if the feature space becomes too large.
  \textit{Processing time} can affect the response time of the detection system. If generating a feature takes a long time it can delay the system's response. 
    \item \textit{Classification methods}: This includes the type of detection technique (e.g., machine learning, rule/pattern based, or heuristic) used.  We also include fine-grained attributes such as supervised or unsupervised,  running times for training and testing the system, specifications of the architecture used and whether system retraining frequency was specified. 
    These attributes contribute to the speed and efficiency of the proposed system.
    \item \textit{Experimental setup/methodology}: This is related to adversarial issues and replicability of the work.
It includes system evaluation details, e.g., testing it in a real-time setting,  whether the system was tested across various domains (e.g., if a model trained on phishing links was also tested on spam, etc.), and how was ground truth obtained. Moreover, if the proposed system was evaluated for robustness, the parameters or metrics were used for system evaluation, and whether the experiment was realistic.
\end{itemize}

\subsection{Common Notations for Detection Sections} 
\hl{We briefly explain the common notations and terminologies that we will be using throughout the paper in this section.}
\textit{Features Tables.} For each feature, we report two criteria: \textit{Feature Size} (FS) 
and \textit{Processing Time} (PT). These criteria were selected keeping in mind the goals and security challenges mentioned in Sections~\ref{sec-intro} and~\ref{sec-challenges}, respectively. 
We consider feature sizes less than 10 as Small, between 10 to 100 as Medium, and more than 100 as Large. 
For \textit{PT}, we group the processing times of less than 100 milliseconds as short, between 100 milliseconds and 1 second as moderate, and more than 1 second as high. 

A number of phishing website and email detection techniques use URL features along with other host/networking features. Therefore, we highlight the papers that use URL features for phishing website detection, e.g., in Table~\ref{tab:urlfeatures}, under the name `uW.' Papers that study URL features extracted from phishing emails are tagged with `uE'. 

\begin{NotationBox} \label{notations}
\scriptsize
\textbf{Features Tables:} Notations used in Tables~\ref{tab:urltypes},~\ref{table:website-features-short},~\ref{table:email-features-short}\\
\textit{FS:} Feature Size, FS \hl{ $\in \left \{ \mathit{Small}, \mathit{Medium}, \mathit{Large} \right \} $}\\
\textit{PT:} Processing Time, PT \hl{ $\in \left \{ \mathit{Short}, \mathit{Moderate}, \mathit{High} \right \}$}\\
\textbf{Dataset Sources Tables:} Notations used in Tables~\ref{tab:urldatasources1},~\ref{tab:urldatasources2},~\ref{tab:urldatasources3},~\ref{table-website-dataset},~\ref{table-website-dataset_2} \\
\textit{Misc.:} Miscellaneous Sources \\
\textit{pvt.:} Private sources\\
\textit{pub.:} Publicly available sources\\
\textbf{Dataset Sizes Tables:} Notations used in Tables~\ref{tab:sizedatasetscomb},~\ref{table-webpage-dataset-size-summarized},~\ref{tab-sizedatasets-emails}\\
\textit{\textbf{$N$}s:} Where $N$ varies from 100 to 100,000\\
\textit{$\geq$ 1M:} Greater than equal to 1 million\\
\textit{N/A:} Not applicable\\
\textit{Cell colors:} Red: highly used ($\geq$6 papers), yellow: moderately used ($\geq$3 and $<$6), green: rarely used ($\leq$2 papers), white: not used.\\
\textbf{Evaluation Metrics Tables:} Notations used in Tables~\ref{tab:urlmetrics1},~\ref{tab:metricsmap},~\ref{tab:metricsmap-email} \\
\textit{CMx:} Confusion Matrix\\
\textit{Acc:} Accuracy\\
\textit{PRF:} \textbf{P}recision, \textbf{R}ecall, and $F_{1}$-Score \\
\textit{ErR:} Error Rate \\
\textit{AUC:} Area Under Curve \\
\textit{Cell colors:} Same conventions as for Dataset Sizes Tables. Red: highly used ($\geq$ 6 papers), etc.\\
\end{NotationBox}

\textit{Dataset Sources Tables.} If researchers used private or `unnamed' sources, these are listed under `Miscellaneous (pvt.)' and for sparsely used public sources, we include them under `Miscellaneous (pub.)'. We list data collected by the authors under `Author (pvt.)'. Some papers do not reveal dataset source(s) - we include these papers under `N/A' (Not Available). 

\textit{Dataset Sizes Tables.} In the dataset sizes tables, the number `100s' denotes that the total dataset size (training and testing combined) used for evaluation is between 100-999. Similarly, 1000s denotes dataset sizes in the range 1000-9999, etc.

\textit{Metrics Tables.} The notations in all metrics tables (e.g.,\ Table~\ref{tab:urlmetrics1}) are as follows:  \textit{CMx} - Confusion Matrix (a matrix with the number of true positives ($TP$), true negatives ($TN$), false positives ($FP$) and false negatives ($FN$)); \textit{Acc} - Accuracy (\hl{Equation (}\ref{eq-accuracy}\hl{))}, \textit{PRF} - \textbf{P}recision, \textbf{R}ecall, and $F_{1}$-Score (Equations  \hl{(}\ref{eq-precision}\hl{)}, \hl{(}\ref{eq-recall}\hl{)}, and \hl{(}\ref{eq-fscore})\hl{)}, \textit{ErR} - Error Rate \hl{($1-accuracy$)}, \textit{AUC} - Area Under an ROC Curve \hl{(the area under the Receiver Operating Characteristic (ROC) Curve is a measure of system accuracy)}.
There are also some other metrics which are more appropriate for unbalanced datasets, e.g., Geometric Mean (G-Mean), Matthews Correlation Coefficient (MCC), and Balanced Accuracy~\cite{bekkar2013evaluation}. \hl{Equations (}\ref{eq-gmeans}\hl{)}, \hl{(}\ref{eq-mcc}\hl{)} and \hl{(}\ref{eq-baccuracy}\hl{)} are the formulas for these metrics. \textit{P} and \textit{N} are the size of positive and negative classes, respectively.

\small
\begin{equation}
\label{eq-accuracy}
Accuracy = \frac{TP+TN}{TP+TN+FP+FN}
\end{equation}

\begin{equation}
\label{eq-precision}
Precision = \frac{TP}{TP+FP}
\end{equation}

\begin{equation}
\label{eq-recall}
Recall = \frac{TP}{TP+FN}
\end{equation}

\begin{equation}
\label{eq-fscore}
F_{1} = \frac{2*Precision*Recall}{Precision+Recall}
\end{equation}

\begin{equation}
\label{eq-gmeans}
\mathcolorbox{yellow}{G-Mean = \sqrt{\frac{TP}{TP+FN} * \frac{TN}{TN+FP}}}
\end{equation}

\begin{equation}
\label{eq-mcc}
MCC=\frac{TP*TN - FP*FN}{\sqrt{(TP+FP)(TN+FN)(P)(N)}} 
\end{equation}

\begin{equation}
\label{eq-baccuracy}
Balanced\;Accuracy = \frac{TP}{P} + \frac{TN}{N}
\end{equation}
\normalsize

\textit{Dataset Diversity.} We analyze the diversity of content found in websites and email bodies by converting them into vector representation using \textbf{T}erm \textbf{F}requency-\textbf{I}nverse \textbf{D}ocument \textbf{F}requency (TFIDF), \hl{a term weighting scheme that uses term frequency in a document and log of the inverse popularity of the term in the collection} \cite{salton1986introduction}. \hl{This is defined by Equation (}\ref{eq-tfidf}\hl{) below, where $n_{t,d}$ is the number of times term $t$ appears in a document $d$, $N_{d}$ is the total number of terms in $d$, $D$ is the total number of documents, and $d_{t}$ is the number of documents with the term $t$ in it.} 
\begin{equation}
\label{eq-tfidf}
TFIDF(t,d) = \frac{n_{t,d}}{N_{d}} * log_{e}(\frac{D}{1+d_{t}})
\end{equation}

\hl{TFIDF} tries to capture the importance of a word in a document. Words that appear in several documents are devalued since they are worse at distinguishing between the documents. Before converting the content to TFIDF vectors, we remove the uninformative and commonly used words such as \textit{the}, \textit{a}, and \textit{in} (called stopwords). We use the stopwords list provided in Natural Language Toolkit (NLTK).
Then we use cosine similarity,\hl{ Equation (}\ref{eq:cosine}\hl{)}, to measure the similarity between the vectors. Cosine similarity is the cosine of the angle between the two vectors.

\small
\begin{equation}
\label{eq:cosine}
similarity(A, B) = cos\:\theta = \frac{A.B}{||A||\:||B||}
\end{equation}
\normalsize

\textit{\underline{Organization of each detection section.}} We start our discussion by showing the structure of a URL/website/email and the features that can be extracted from it. We then describe the methods used for phishing URL/website/email detection, the datasets employed, and the metrics studied in previous research. We close the section with research that has addressed at least a few of the key security challenges and some opportunities for future research.

\hl{We now organize and assess 
the phishing detection techniques and user studies from the standpoint of the above security challenges. Many researchers have focused on URL analysis as one way (sometimes the only way) of detecting phishing websites. We reexamine this line of research first.}

\section{Phishing URL Detection}
\label{sec-techniques-urls}

Based on the above key attributes (Section~\ref{sec-techniques}), we reexamine  61 relevant research papers on phishing URL detection. 

\subsection{URL Structure} 
\begin{figure}[!htb]
\centering
 \includegraphics[width=0.8\linewidth]{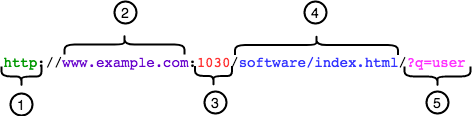}
 \caption{Example of a complete URL with its typical components} 
 \label{fig:urlexample}
 \end{figure}

A URL is defined as a reference or an address to a \textit{resource} on the Internet. \hl{In}  Figure~\ref{fig:urlexample}, we demonstrate the structure of a typical URL\hl{.}\footnote{\hl{"Uniform Resource Identifiers (URI): Generic Syntax", http://www.ietf.org/rfc/rfc2396.txt.}} 

\begin{enumerate}
\item \textit{Scheme}: This identifies the protocol to be used to access the resource on the Internet. \hl{Common} protocols are HyperText Transfer Protocol (HTTP) and \hl{Secure HyperText} Transfer Protocol  (HTTPS).
\item \textit{Host}: The hostname or domain name refers to the human-readable name for the address at which the resource can be accessed using the URL. In the above example, \textcolor{violet}{\textit{www.example.com}} acts as the hostname. 
\item \textit{Port}: The port number follows the hostname or the Internet Protocol (IP) address. The Transfer Control Protocol (TCP) looks at the port to identify the type of communication between the client (asking for resource access) and the server (granting resource access) processes. Here, the port number \textcolor{orange}{\textit{1030}} acts as \hl{the TCP port.} 
\item \textit{Path:} This indicates the path to the exact location of the resource on the host. \hl{In the above} example, the resource resides at the location \textcolor{blue}{`\textit{/software/index.html/}'}. 
\item \textit{Query}: The query string if provided follows the path component and provides a string of information \hl{with a} specific purpose. In the above example, the query string \textcolor{purple}{`\textit{q=user}'} will prompt the server to access the user page. The symbol `?' \hl{separates the} path and the query. 
\end{enumerate}

\begin{table*}[!htb]
\centering
\caption{Features used in phishing URL detection. For uncommon features, we cite the paper which best describes it.}
\label{tab:urltypes}
\begin{threeparttable}
\resizebox{2\columnwidth}{!}{
\begin{tabular}{|l|c|c|l|c|c|lcclcc}
\hline
\multirow{2}{*}{\textbf{Lexical properties}}     & \multicolumn{2}{c|}{\textbf{Criteria}} & \multirow{2}{*}{\begin{tabular}[c]{@{}l@{}} \textbf{Obfuscation and Shortened} \\ \textbf{URL properties}\end{tabular}} & \multicolumn{2}{c|}{\textbf{Criteria}} & \multicolumn{1}{l|}{\multirow{2}{*}{\begin{tabular}[c]{@{}l@{}}\textbf{Hostname and DNS} \\ \textbf{properties}\end{tabular}}} & \multicolumn{2}{c|}{\textbf{Criteria}}                                               & \multicolumn{1}{l|}{\multirow{2}{*}{\begin{tabular}[c]{@{}l@{}}\textbf{IP address and} \\ \textbf{WHOIS properties} \end{tabular}}} & \multicolumn{2}{c|}{\textbf{Criteria\tnote{a}}}                                               \\ \cline{2-3} \cline{5-6} \cline{8-9} \cline{11-12} 
                             & FS        & PT      &                   & FS        & PT    &   \multicolumn{1}{l|}{}                                                                       & \multicolumn{1}{l|}{FS} &\multicolumn{1}{l|}{PT}    &  \multicolumn{1}{l|}{}   & \multicolumn{1}{l|}{FS}  & \multicolumn{1}{l|}{PT} \\ \specialrule{.2em}{.1em}{.1em}
URL Length                   & S               & S       & \begin{tabular}[c]{@{}l@{}} Special Characters\\ Frequency    \end{tabular}                          & M               & M       & \multicolumn{1}{l|}{Token Frequency}                                                           & \multicolumn{1}{l|}{M}    & \multicolumn{1}{l|}{M}  & \multicolumn{1}{l|}{\begin{tabular}[c]{@{}l@{}}Characters and \\ digits in IP\end{tabular}}                    & \multicolumn{1}{l|}{M}  & \multicolumn{1}{l|}{S}  \\ \hline
\begin{tabular}[c]{@{}l@{}} Length of \\ URL parameters\tnote{a}  \end{tabular}   & S                & S       & IP address obfuscation                                    & S               & S      & \multicolumn{1}{l|}{Longest Token}                                                             & \multicolumn{1}{l|}{S}    & \multicolumn{1}{l|}{S}  & \multicolumn{1}{l|}{IP Encoding}                                    & \multicolumn{1}{l|}{S}  & \multicolumn{1}{l|}{M}  \\ \hline
\begin{tabular}[c]{@{}l@{}} Token/Word \\ Frequency  \end{tabular}       & S               & S     & URL encoding                                              & S             & M       & \multicolumn{1}{l|}{Digit frequency}                                                               & \multicolumn{1}{l|}{S}    & \multicolumn{1}{l|}{S}  & \multicolumn{1}{l|}{Blacklisted IP}                                 & \multicolumn{1}{l|}{S}    & \multicolumn{1}{l|}{H}  \\ \hline
\begin{tabular}[c]{@{}l@{}} Blacklisted Word \\ Frequency \end{tabular}    & S              & M       & \begin{tabular}[c]{@{}l@{}}URL shortening\end{tabular} & S              & M       & \multicolumn{1}{l|}{\begin{tabular}[c]{@{}l@{}}Frequency of special \\ characters in domains \end{tabular}}                                            & \multicolumn{1}{l|}{M}   & \multicolumn{1}{l|}{S}  & \multicolumn{1}{l|}{Whitelisted IP}       & \multicolumn{1}{l|}{S}    & \multicolumn{1}{l|}{H}  \\ \hline
\begin{tabular}[c]{@{}l@{}} Freq. or Ratio of \\ Digits and Characters \end{tabular}    & M                & S       & URL path spoofing                                         & S               & M       & \multicolumn{1}{l|}{Port presence}                                                             & \multicolumn{1}{l|}{S}    & \multicolumn{1}{l|}{S}  & \multicolumn{1}{l|}{\begin{tabular}[c]{@{}l@{}}IP records and \\ prefix checking~\cite{ma2011learning}\end{tabular}}                  & \multicolumn{1}{l|}{M}    & \multicolumn{1}{l|}{H}  \\ \hline
Number of Dots (.)           & S              & S       & \begin{tabular}[c]{@{}l@{}} Mismatch in URL \\ source and destination \end{tabular}                              & S               & H      & \multicolumn{1}{l|}{HTTPS/HTTP}                                                                & \multicolumn{1}{l|}{S}    & \multicolumn{1}{l|}{S}  & \multicolumn{1}{l|}{\begin{tabular}[c]{@{}l@{}}WHOIS Registration \\ details \end{tabular}}                     & \multicolumn{1}{l|}{M}   & \multicolumn{1}{l|}{H}  \\ \hline
 \begin{tabular}[c]{@{}l@{}} Character\tnote{b} \\ Frequency   \end{tabular}       & M           & M       & \begin{tabular}[c]{@{}l@{}} IP address instead \\ of domain name \end{tabular}                          & S           & S       & \multicolumn{1}{l|}{Misspelled/Bad domains}                                                    & \multicolumn{1}{l|}{S}   & \multicolumn{1}{l|}{M}  & \multicolumn{1}{l|}{\begin{tabular}[c]{@{}l@{}}Creation, Update, Expiration \\ date of WHOIS info.~\cite{ma2011learning}\end{tabular}}               & \multicolumn{1}{l|}{S}   & \multicolumn{1}{l|}{H}  \\ \hline
\begin{tabular}[c]{@{}l@{}} Kolmogorov \\ Complexity  \end{tabular}      & S              & M       & Hostname obfuscation\tnote{f}                                      & S            & M       & \multicolumn{1}{l|}{TLD features~\cite{ma2011learning}}                                                              & \multicolumn{1}{l|}{M}    & \multicolumn{1}{l|}{H}  & \multicolumn{1}{l|}{AS Number}                                      & \multicolumn{1}{l|}{S}   & \multicolumn{1}{l|}{H}  \\ \hline
Character Ngrams             & L               & H       & \begin{tabular}[c]{@{}l@{}} Entry point URL \\ frequency~\cite{lee2013warningbird} \end{tabular}                                     & S             & M       & \multicolumn{1}{l|}{TTL value\tnote{c}~}                                                                 & \multicolumn{1}{l|}{S}  & \multicolumn{1}{l|}{M}  & \multicolumn{1}{l|}{Status of WHOIS Entry}                          & \multicolumn{1}{l|}{S} & \multicolumn{1}{l|}{H}  \\ \hline
\begin{tabular}[c]{@{}l@{}} Edit Dist.,\\ KL Div., KS-test \end{tabular} & M         & M       & \begin{tabular}[c]{@{}l@{}} Position of entry\\ point URLs~\cite{lee2013warningbird}   \end{tabular}                           & S       & M       & \multicolumn{1}{l|}{Age of domain}                                                             & \multicolumn{1}{l|}{S}  & \multicolumn{1}{l|}{M}  & \multicolumn{1}{l|}{Location of IP origin\tnote{e}~}                          & \multicolumn{1}{l|}{M}    & \multicolumn{1}{l|}{M}  \\ \hline
URL Entropy                  & S     & M       & Number of landing URLs~\cite{lee2013warningbird}                                    & S            & M       & \multicolumn{1}{l|}{\begin{tabular}[c]{@{}l@{}}Ranking based \\ features\tnote{d} \end{tabular}}                                                    & \multicolumn{1}{l|}{S}     & \multicolumn{1}{l|}{M}  & \multicolumn{1}{l|}{\begin{tabular}[c]{@{}l@{}}Nature or Speed \\ of connection~\cite{ma2011learning}\end{tabular}}                          & \multicolumn{1}{l|}{M}  &  \multicolumn{1}{l|}{M}  \\ \hline
Bag-of-words                 & L               & M       & \begin{tabular}[c]{@{}l@{}} Number of domain \\ names and IPs  \end{tabular}                          & S              & M       &    &         &     &     &      &                       \\ \cline{1-6}
\end{tabular}}
\begin{tablenotes}
\item[a] \textit{Parameters are parts of URL like scheme, host, path, query string}
\item[b] \textit{Characters like period, slash, etc.} 
\item[c] \textit{Time to live}
\item[d] \textit{Alexa Ranking, PageRank scores, search engines lookup }
\item[e] \textit{Continent/country/city}
\item[f] \textit{Hexadecimal encoding}
\end{tablenotes}
\end{threeparttable}
\end{table*}

\subsection{Features}\label{url-features}
A large number of features can be extracted from the URL like the presence of certain tokens or words, token length, and frequency, count of special symbols, as well as the presence of any meta-symbols, such as the @ symbol (which is interpreted in a special way by the browser). A URL can also be \hl{considered} as a string of characters, and lexical features\hl{, e.g.,} edit distance, \hl{character n-gram frequencies (i.e., sequences of n characters)}, etc., can be used. 

We group the different features used in phishing URL detection literature into four major classes: \hl{(i)} \textit{Lexical properties}, \hl{(ii)} \textit{Obfuscation} (use of encoding in URL, use of `\&', `\%20' instead of space, etc.) and \textit{Shortened URL properties} (use of goo.gl, bit.ly, tinyurl services \hl{for shortening and obfuscating long malicious links}~\cite{gupta2014bit}), \hl{(iii)} \textit{Hostname and Domain Name System (DNS) based properties}, \hl{and (iv)} \textit{IP address} and \textit{WHOIS properties} (WHOIS server details e.g., the date of registration, update, and expiration). The \hl{aforementioned} taxonomy has been developed after careful  consideration of previous literature including related surveys, e.g.,~\cite{sahooLH17,dou17KK}. 

{Table~\ref{tab:urltypes}} lists  URL feature classes along with feature size and processing time required for each feature. 
A majority of the lexical features -- target word frequency, frequency of alphanumeric characters, character n-grams, bag-of-words, etc., -- \hl{need} higher processing time. 
Similarly, we note that, for obfuscation based properties, the processing time ranges from short to moderate.
IP address based features -- the presence of encoding in IP, is the IP blacklisted or whitelisted, etc., -- have short to moderate feature sizes and processing times. 
Retrieving WHOIS based information (\hl{e.g.,} creation, update and expiration dates of WHOIS information, WHOIS registration details) involves querying WHOIS servers, which in turn increase the feature processing time. 

\begin{figure*}[!htb]
\centering
 \includegraphics[width=1\linewidth]{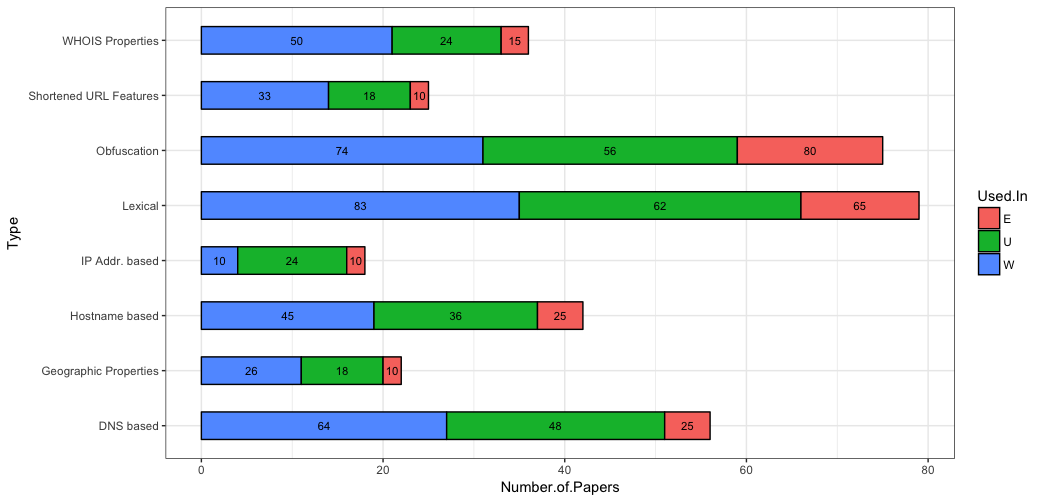}
 \caption{Distribution of URL features used in \hl{the three dimensions of} phishing detection: \textit{URL} (U), \textit{website content} (W) and \textit{email content} (E). Inside each section of the bar, we mention the percentage of papers \hl{that use} a class of URL-based features. For example, 50\% of the phishing website detection papers that use URL features employ WHOIS properties.}  
 \label{fig:urlfeaturedist}
 \end{figure*}

Figure~\ref{fig:urlfeaturedist} shows the distribution of URL feature classes based on their usage across \hl{the} three \hl{dimensions} of phishing detection  -  \textit{URL} (U), \textit{website content} (W) and \textit{email content} (E). According to Figure~\ref{fig:urlfeaturedist}, the most popular feature types used across all \hl{dimensions}  are:  obfuscation-based, lexical-based and DNS-based properties. Table~\ref{tab:urlfeatures} in the Appendix gives a more detailed breakdown of the literature based on types of features extracted. 
We now consider the detection methods that have been used in the phishing URL detection literature. 

\subsection{Detection Methods}\label{sec:url-det}
Researchers have proposed a variety of detection methods for phishing URL detection. These methods vary from blacklist/whitelist  based techniques to machine learning based algorithms.

\begin{figure*}
\centering
 \includegraphics[width=\linewidth]{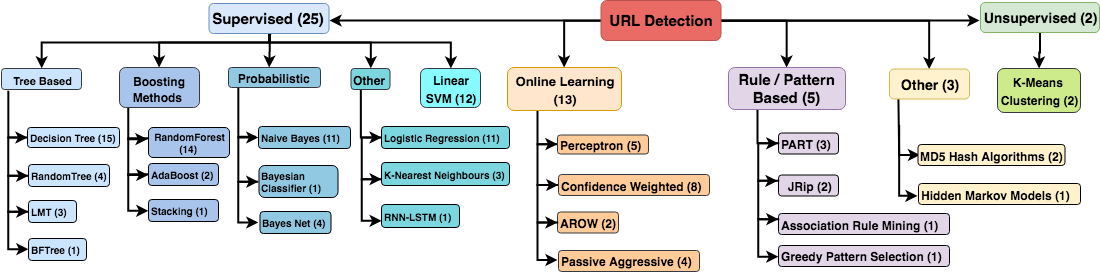}
 \caption{Algorithms used in phishing URL detection. The number of papers using the method is in parentheses.}  
 \label{fig:urlmethodsdet}
 \end{figure*}

We \hl{show} the major classes of machine learning algorithms used in phishing URL detection literature in Figure~\ref{fig:urlmethodsdet}. The papers \hl{are}  categorized \hl{based} on the \hl{type} of learning method -- supervised (popular techniques are Decision Tree, Random Forest, Support Vector Machines, Na\"{i}ve Bayes, and Logistic Regression), Unsupervised (K-Means Clustering), Online learners (Confidence Weighted, \hl{Adaptive Regularization of Weights}, etc.).  A few papers employ rarely-used methods, e.g., association rule mining and Markov models. Clearly, a majority of the proposed systems are supervised. \hl{Other} promising learning approaches, such as rule-based, unsupervised, and semi-supervised approaches, remain largely unexplored. Of the supervised methods, only one paper \cite{bahnsen2017classifying} used deep learning in the form of Recurrent Neural Networks. Table~\ref{tab:urlmethods} (in the Appendix) describes the variety of techniques  used for detection  of malicious URLs along with \hl{citations}.

\hl{The properties of the dataset play an important role in building a robust detection system. We discuss these next.}

\subsection{Dataset Properties}
\hl{Critical properties of a dataset include:} the source(s) of a dataset, its diversity \hl{or representativeness}, its size, its \hl{class} distribution and \hl{its} recency. In this section, we \hl{review} the literature based on these properties.
\subsubsection{\textbf{Dataset sources and availability}}  
\begin{figure}[t]
\centering
 \includegraphics[width=0.8\linewidth]{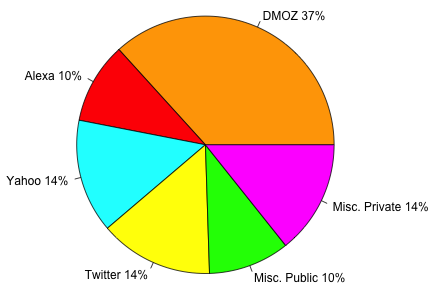}
 \caption{Distribution of papers \hl{based on} legitimate URL source}  
 \label{fig:legiturldist}
 \end{figure}

\begin{figure}
\centering
 \includegraphics[width=0.8\linewidth]{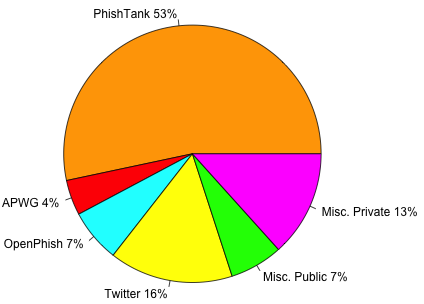}
 \caption{Distribution of papers \hl{based on} phishing URL source} 
 \label{fig:phishurldist}
 \end{figure}

In Tables~\ref{tab:urldatasources1},~\ref{tab:urldatasources2}, and~\ref{tab:urldatasources3}, we categorize papers based on the combination of benign and malicious data sources (phishing, spam, and malware, respectively) along with their availability (public, private, deprecated). Figures~\ref{fig:legiturldist} and~\ref{fig:phishurldist} show data sources and the percentage of \hl{surveyed}  literature using \hl{each source}.


The most popular sources for URL data are: (a) \textbf{Legitimate}: DMOZ (now \textit{deprecated}), Alexa.com, Yahoo URL generator (\textit{deprecated}) and (b) \textbf{Phishing}: PhishTank, APWG and OpenPhish.  
Sources like Alexa.com list \textit{\hl{domain names only}}, which can \hl{artificially inflate} the \hl{results,} since such legitimate domains can be easily distinguished from the complete URLs found in phishing data sources. A more thorough data collection involves crawling the top sites from Alexa \textit{and subsequent links from these sites}.
During our review, we found that many papers that use Alexa's ranking fail to mention any preprocessing that addresses this issue. 
Multiple research papers have also used DMOZ and Yahoo URL generator as legitimate sources, which are currently deprecated. In such a situation, system retraining becomes important and comparing results becomes harder. 
PhishTank is a crowdsourced blacklist where controlling the quality of the URLs\footnote{\hl{If the database contains live websites, absence of repetitive links, only phishing domains, etc., the quality is better for evaluation.}} is critical.
Twitter is also a common platform for sharing malicious URLs.  
We include spam and malware URL data sources both in URL and website sections, primarily because a large number of papers, e.g., \cite{rathod2015comparative,darling2015lexical,egan2011evaluation,zhang2014web,gyawali2011evaluating,xiong2015mird,le2011phishdef,dewan2017facebook}, use a mix of phishing, malware, and spam data, collectively termed as ``\textit{malicious}.'' We observe little variation in the dataset sources used for evaluation in the literature, \hl{with a few exceptions}. Researchers tend to use publicly available sources for building their evaluation datasets.



\begin{table*}[!htb]
\centering
\caption{Phishing and legitimate URL datasets: \hl{\textit{sources and availability}}}
\label{tab:urldatasources1}
\begin{threeparttable}
\resizebox{2\columnwidth}{!}{
\begin{tabular}{|c|c|c|c|c|c|c|c|c|}
\hline
 \multirow{2}{*} {\diagbox{\textbf{Phish.}}{\textbf{Legit.}} }   
         & \multirow{2}{*}{DMOZ (dep.)} & \multirow{2}{*}{Alexa (pub.)} & \multirow{2}{*}{Yahoo\tnote{f} (dep.)} & \multirow{2}{*}{Twitter (pub.)} & \multicolumn{2}{c|} {Miscellaneous} & \multirow{2}{*}{Author\tnote{c}~  (pvt.)} & \multirow{2}{*}{N/A}\\ 
  \cline{6-7}  & \multicolumn{1}{c|}{}   & \multicolumn{1}{c|}{}      & \multicolumn{1}{c|}{}     & \multicolumn{1}{c|}{}    & \multicolumn{1}{c|}{pub.\tnote{a}} & \multicolumn{1}{c|}{pvt.\tnote{b}} &    \multicolumn{1}{c|}{}   & \multicolumn{1}{c|}{}    \\ \specialrule{.2em}{.1em}{.1em}
PhishTank (pub.)       & \begin{tabular}[c]{@{}l@{}} ~\cite{rathod2015comparative,darling2015lexical,xiong2015mird}\\~\cite{le2011phishdef,nguyen2014novel,ma2009beyond,feroz2014examination,sha2015limited,astorino2016malicious,basnet2012mining,patil2016malicious,verma2017s,basnet2015towards,marchal:phishscore,marchal:phishstorm,vermaK15,pradeepthi2014performance,feroz2015phishing} \end{tabular}     & \begin{tabular}[c]{@{}l@{}}  \cite{darling2015lexical,vermaK15}\\~\cite{jeeva2016intelligent,patil2016malicious}\end{tabular}      & \begin{tabular}[c]{@{}l@{}} \cite{egan2011evaluation,ma2009beyond, basnet2012mining}\\\cite{le2011phishdef,basnet2015towards,jeeva2016intelligent} \end{tabular}     &             &   \cite{rathod2015comparative,bahnsen2017classifying}          & ~\cite{huang2014malicious}             &  \begin{tabular}[c]{@{}c@{}} ~\cite{verma2017s,khonji2011novel}\\~\cite{khonjiIJ11,sananse2015phishing} \end{tabular}  &           \\ \hline
APWG (pub.)           & \begin{tabular}[c]{@{}l@{}} \cite{vermaK15,verma2017s}
\end{tabular}     &  \begin{tabular}[c]{@{}l@{}} \cite{vermaK15}\end{tabular}     &       &             &             &              &  ~\cite{verma2017s}   &          \\ \hline
OpenPhish (pub.)      &   \cite{darling2015lexical,patil2016malicious}   & ~\cite{mamun2016detecting}  &       &             &             &              &    &           \\ \hline
Twitter (pub.)     &      &       &       &    \cite{gupta2014bit,sandracoordinator,alsharnouby2015phishing,burnap2015real,lee2013warningbird,nepali2016you}         &             &              &     &          \\ \hline
Misc.\tnote{d}~ (pub.)  &      &       &       &             &   \cite{cao2016detection}           &  ~\cite{blum2010lexical}         &      &   ~\cite{gyawali2011evaluating}      \\ \hline
Misc.\tnote{e}~ (pvt.) &      & ~\cite{jeeva2016intelligent}      & ~\cite{jeeva2016intelligent,chu2013protect}       &             & \begin{tabular}[c]{@{}l@{}}   ~\cite{su2013suspicious,catakoglu2016automatic} \end{tabular}       &   ~\cite{popescu2015study}              &   & ~\cite{lee2014users}            \\ \hline
N/A &      &       &       &             &         &                 &   & \begin{tabular}[c]{@{}l@{}} \cite{rathodanti, jeun2013collecting,gabriel2016detecting,canova2015learn,yuan2013tfd,guan2009anomaly}   \end{tabular}   \\
\specialrule{.2em}{.1em}{.1em}
\end{tabular}}
    \begin{tablenotes}
\item[a] \textit{CommonCrawl~\cite{bahnsen2017classifying}, Gmail directory~\cite{rathod2015comparative}, Weibo Sina API~\cite{cao2016detection}}
\item[b] \textit{URL feeds and logs from Web~\cite{huang2014malicious}, Cyveillance~\cite{blum2010lexical}, Bitdefender Laboratories~\cite{popescu2015study}}
\item[c] \textit{Datasets collected by authors in~\cite{verma2017s,khonjiIJ11,khonji2011novel}}
\item[d] \textit{Phishing messages from Weibo Sina API~\cite{cao2016detection}, URLs from UAB Spam DataMine email messages~\cite{gyawali2011evaluating,blum2010lexical} }
\item[e] \textit{Click-through data from the Trend
Micro research laboratory~\cite{lee2014users}, Private honeypot~\cite{catakoglu2016automatic}, T.Co. Labs~\cite{su2013suspicious}}
\item[f] \textit{Yahoo URL Generator (http://random.yahoo.com/bin/ryl)}
\end{tablenotes}
\end{threeparttable}
\end{table*}

\begin{table}[!htb]
\centering
\caption{Spam and legitimate URL datasets: \hl{\textit{sources and availability }  }}
\label{tab:urldatasources2}
\begin{threeparttable}
\resizebox{\columnwidth}{!}{
\begin{tabular}{|c|c|c|c|c|c|}
\hline
 \diagbox{\textbf{Spam}}{\textbf{Legit.}}  
         & \begin{tabular}[c]{@{}c@{}}DMOZ\\(pub.)\end{tabular} & \begin{tabular}[c]{@{}c@{}}Alexa\\(pub.)\end{tabular} & \begin{tabular}[c]{@{}c@{}}Misc.\tnote{a}\\(pub.) \end{tabular}& \begin{tabular}[c]{@{}c@{}}Author\tnote{b}\\(pvt.)\end{tabular} & N/A\\ \specialrule{.2em}{.1em}{.1em}
         

\begin{tabular}[c]{@{}l@{}}SpamScatter (pub.) \end{tabular}   & \cite{ma2009beyond}     & \cite{mamun2016detecting}      &            &          &                 \\ \hline
\begin{tabular}[c]{@{}l@{}}SpamAssassin (pub.)\end{tabular}  & \cite{rathod2015comparative}     &       &   \cite{rathod2015comparative}        &                 &          \\ \hline
Misc.\tnote{c}~ (pvt.) &      &   \cite{vermaK15}    &       &                  \begin{tabular}[c]{@{}l@{}}\cite{gyawali2011evaluating,jeun2013collecting}\\~\cite{verma2017s,dewan2017facebook} \end{tabular}    & \begin{tabular}[c]{@{}l@{}}~\cite{gyawali2011evaluating,dewan2017facebook}\\~\cite{jeun2013collecting, dewan2015towards} \end{tabular}        \\ 
\specialrule{.2em}{.1em}{.1em}
\end{tabular}
}
\begin{tablenotes}
\item[a] \textit{Gmail directory~\cite{rathod2015comparative} }
\item[b] \textit{Spam URLs from UAB Spam DataMine email messages~\cite{gyawali2011evaluating}}
\item[c] \textit{Private SpamTrap system~\cite{jeun2013collecting}, collected using blacklists and \\ crowd-sourcing~\cite{dewan2015towards,dewan2017facebook} }
\end{tablenotes}
\end{threeparttable}
\end{table}

\begin{table}[!htb]
\centering
\caption{Malware and legitimate URL datasets: \hl{\textit{sources and availability }  }}
\label{tab:urldatasources3}
\begin{threeparttable}
\begin{tabular}{|c|c|c|c|c|}
\hline
  \diagbox{\textbf{Malware}}{\textbf{Legit.}}  & \begin{tabular}[c]{@{}c@{}}DMOZ\\(pub.)\end{tabular} & \begin{tabular}[c]{@{}c@{}}Alexa\\(pub.)\end{tabular} & \begin{tabular}[c]{@{}c@{}}Yahoo\\(pub.)\end{tabular} &  \begin{tabular}[c]{@{}c@{}}Author\\ (pvt.)\end{tabular} \\ 
  \specialrule{.2em}{.1em}{.1em}
MalwarePatrol (pub.)  & \cite{xiong2015mird,le2011phishdef}     &       &  ~\cite{egan2011evaluation}     &                          \\ \hline
DNS-BlackHole (pub.)  &  \cite{darling2015lexical}    & \cite{darling2015lexical,mamun2016detecting}      &       &                       \\\hline
Misc.\tnote{a}~ (pub.)  & \cite{darling2015lexical}     &   \cite{darling2015lexical,patil2016malicious}    &       &    \\\hline
Misc.\tnote{b}~ (pvt.) &      &       &       &   \cite{popescu2016practical}            \\
\specialrule{.2em}{.1em}{.1em}
\end{tabular}
    \begin{tablenotes}
\item[a] \textit{MalwareDomainslist~\cite{darling2015lexical} and malwareurl~\cite{patil2016malicious}}
\item[b] \textit{Malware private sources~\cite{popescu2016practical} }
\end{tablenotes}
\end{threeparttable}
\end{table}

\subsubsection{\textbf{Dataset size and class ratio}}\label{sec:url-size}
We expect that an Internet user will generally encounter a phishing URL significantly less frequently than a benign one. Thus, researchers should evaluate their classifiers with different and unbalanced class ratios. We report the count of \textit{phishing} URL detection papers using various combinations of data sizes in Table~\ref{tab:sizedatasetscomb}. We only compare the sizes of the phishing datasets with the legitimate, since it is the main focus of this paper. 
Table~\ref{tab:sizedatasetscomb} shows that a majority of the papers studied report evaluation results only on balanced datasets\hl{,} since most papers appear on or close to the diagonal in the table. Also, only a small subset (approx. 23\%) of the \hl{surveyed} papers has used \hl{relatively}  large (comprising $\geq$ 100,000s URLs) URL datasets.  

Researchers in \cite{le2011phishdef,lee2013warningbird,dewan2017facebook,eshete2014webwinnow} considered both unbalanced and balanced scenarios in their evaluations. \hl{In}  \cite{verma2017s,le2011phishdef,lee2013warningbird,xiong2015mird}\hl{, researchers} evaluate\hl{d} their \hl{classifiers} on varying ratios (ratios considered were 2:1, 3:2, 8:1, etc.) of benign to malicious URLs, whereas \hl{in} \cite{zhao2013cost,ma2011learning,sha2015limited,su2013suspicious,basnet2015towards,dewan2015towards} \hl{researchers} evaluated their classifiers on \textit{highly unbalanced} datasets (the benign dataset size is almost 50 times or more)  but fixed ratios. Papers  \cite{darling2015lexical,popescu2015study,huang2014malicious,basnet2012mining,basnet2015towards} report moderate benign-to-malicious data ratios (e.g., 2:1, 3:2, 5:1, etc.).
Authors in \cite{garera2007framework,jeeva2016intelligent,astorino2016malicious,vu2016firstfilter,sharmaefficient,nandy2014anti,ma2009beyond,gupta2014bit} used slightly unbalanced\hl{,} or almost balanced\hl{,}\footnote{If one class is $x$ times the size of the other class, where $1 \leq x < 2$\hl{.}} data. 
Systems in \cite{mamun2016detecting,gabriel2016detecting,ma2009beyond} used more malicious URLs than legitimate ones. 

Table~\ref{tab:sizedatasets} (in the Appendix) provides a detailed distribution of the dataset sizes used in the literature according to their types (phishing, spam, and malware).  
Some research papers, e.g., \cite{lin2013malicious,xiong2015mird,verma2017s,ma2011learning}, and some others (Refer to Table \ref{tab:sizedatasets}), have used millions of URLs (both benign and malicious) for the evaluation of their systems. But these are usually \textit{static} datasets, i.e., data collected over time and stored for usage. 
Some researchers go a step further -- they  deployed and tested their classifiers in a real-time environment 
\cite{jeun2013collecting,alshboul2015detecting,dewan2015towards}. 
Moreover, malicious links can be shared via chat sessions and Twitter feeds -- therefore, including the papers that test such sources (under category \textit{Feed}) is essential.


Irrespective of size, the datasets {\em may not be diverse}. In the following subsection, we evaluate the diversity of some popular, publicly-available datasets.
\subsubsection{\textbf{Dataset Diversity}}~\label{sec:url-data-div}
We collected \hl{phishing} URLs from three different sources PhishTank (22,018 URLs), APWG (9,757 URLs) and OpenPhish (5,484 URLs) on Nov 2017\hl{,} Alexa \hl{as the basis for our}  legitimate dataset. The collected phishing URLs are \hl{those} that are live at the time of collection. We use the list of top domains from Alexa as a seed for our crawler\footnote{We limited the crawler to search up to three levels.} to generate a more realistic dataset. \hl{To}  increase the diversity of the legitimate websites, we retrieved the top 40 websites from each category on Alexa\footnote{https://www.alexa.com/topsites/category lists 17 domain categories.} excluding \hl{the} three categories:  \textit{Adult}, \textit{Regional}, and \textit{World}. We stopped the crawler after collecting 91,610 URLs.

\begin{figure}[h]
\centering
 \includegraphics[width=1\linewidth]{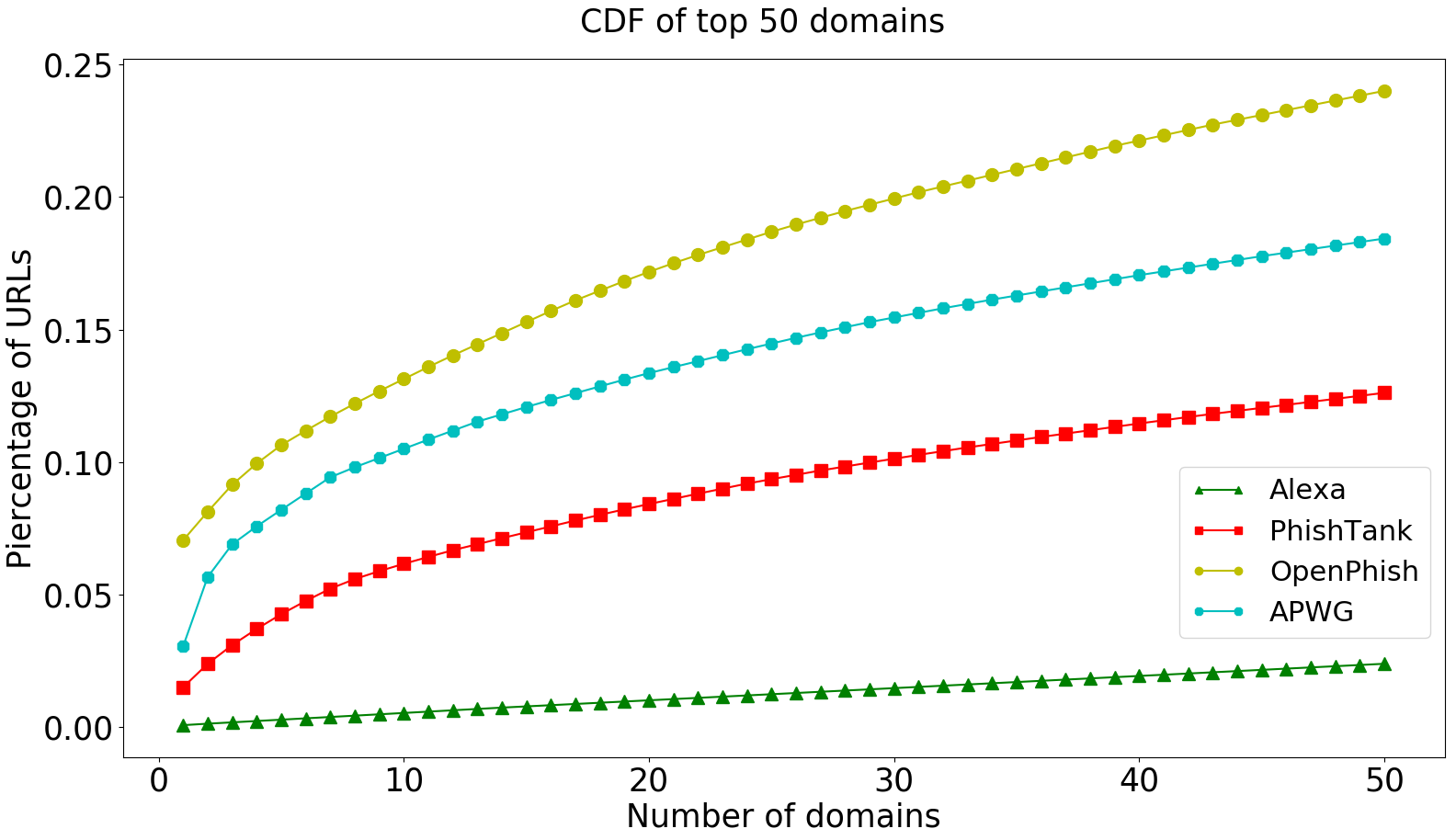}
 \caption{Cumulative distribution function of the top 50 domains in the phishing and legitimate dataset (Section \ref{sec:url-data-div})}  
 \label{fig:url_diversity}
 \end{figure}

Calculating the frequency of domains in each dataset (Figure~\ref{fig:url_diversity}) revealed that \hl{the top 50 domains} constitute 24\%, 12.6\% and 18\% of URLs extracted from OpenPhish, PhishTank, and APWG respectively. From these percentages, PhishTank seems to be a \hl{more} diverse dataset \hl{than the other two}.

For the Alexa dataset, the top 50 domains comprise 75\% of URLs. However, this can be mitigated by limiting the number of URLs crawled from each domain in the list. 
Thus, we changed our crawler's configuration to store at most 12 URLs for each domain. Following this modification, the percentage of URLs with the top 50 domains in Alexa dataset reduced to 2.3\% (out of 21,788 total URLs). When we limit the number of URLs for each domain,  the top \hl{$N$} domains \hl{variable becomes a linear function of dataset size}. \hl{Thus}, limiting the number of URLs collected \hl{from each domain}   can help build a more diverse legitimate dataset.

In the above experiment, we studied all datasets independently. We did not compare the phishing and legitimate datasets with each other since this would go into the realm of \hl{developing} detection techniques, which is \hl{beyond} the scope of this paper. 

The time \hl{period} of data\hl{sets} is also important since a classifier can overfit on some attributes common to instances from a specific time period. Such a classifier will fail miserably on newer kinds of attacks. \hl{So, we study the collection time periods of datasets in the next section.} 

\begin{table}[!htb]
\centering
\caption{Size ranges of phishing/legitimate URL datasets.}
\label{tab:sizedatasetscomb}
\resizebox{0.5\textwidth}{!}{
\begin{tabular}{|c|c|c|c|c|c|c|c|}
\hline
                           &         \multicolumn{7}{c|}{\textbf{Legitimate}}                                                                                       \\ \hline
                              \textbf{Phishing}       & \textbf{100s} & \textbf{1,000s}  & \textbf{10,000s} & \textbf{100,000s} & \textbf{$\geq$1 M} & \textbf{Feed} & \textbf{N/A}     \\ \specialrule{.2em}{.1em}{.1em}
 \textbf{100s }    & \cellcolor{yellow}{4} & \cellcolor{green}{1} &   &   &  &                         &    \\ \hline 
                           \textbf{1,000s}   & &\cellcolor{red}{6} & \cellcolor{green}{1} &  & \cellcolor{green}{1} & &                      \\ \hline   \textbf{10,000s}  &   & \cellcolor{green}{2}   & \cellcolor{red}{14} & \cellcolor{red}{7}   & \cellcolor{yellow}{4} &  & 
                           \\ \hline 
                           \textbf{100,000s} &   &  &  &\cellcolor{yellow}{3} & & & \\ \hline 
                           \textbf{$\geq$1 M}&   &  &  &  & \cellcolor{red}{8} & &  \cellcolor{green}{1}\\  \hline 
                           \textbf{Feed} &   &  &  &  & & \cellcolor{green}{2} & \\ \hline
                           \textbf{N/A} &   &  &  &  & & &  \cellcolor{red}{7}\\
\specialrule{.2em}{.1em}{.1em}                           
\end{tabular}}
\end{table}

\subsubsection{\textbf{Recency of data collection}}\label{sec:url-time}
\textit{Recency} of data collection is an important aspect of the evaluation datasets. \hl{Since attackers tend to launch their attacks for short time windows, malicious links can become dead quickly. So, URL datasets need to be collected close to attack launch time.} We specify \hl{recency} in Table \ref{tab:collecturls}. It shows the relative age of the dataset, which is $n - m$, where $m$ is the year \hl{during which the URLs were collected from the respective sources by the researchers}  and $n$ \hl{is} the year in which the paper was published.  
Some \hl{researchers} (e.g., \cite{dewan2015towards,lee2013warningbird,catakoglu2016automatic}) \hl{explicitly consider} the active  attacker and choose to train and retrain their systems on datasets collected across various time frames. This ensures better resiliency of the system to newer and probably more evolved attacks. In Table~\ref{tab:collecturls}, we mark such papers in bold. However, a large fraction (approx. 55\%) of the papers \hl{surveyed} do\hl{es} not report the collection time of datasets. 


\begin{table}[!htb]
\centering
\caption{Recency of URL datasets used in evaluation. \hl{Papers} in \textbf{bold} use data collected \hl{from} multiple \hl{periods}.}
\label{tab:collecturls}
\begin{tabular}{|c|c|l|}
\hline
\textbf{Recency in year(s)}  & \textbf{N}         & {\textbf{Literature}} \\ \hline \hline
\textbf{Real-time} & 5                  & ~\cite{burnap2015real}, \textbf{\cite{lee2013warningbird}}, \textbf{\cite{catakoglu2016automatic}}, \cite{jeun2013collecting},\textbf{\cite{dewan2015towards}},~\cite{alshboul2015detecting}         \\ \hline
 \textbf{1}    & 17        &  \begin{tabular}[c]{@{}l@{}} ~\cite{popescu2015study,ma2009beyond,gyawali2011evaluating,vu2016firstfilter,basnet2012mining,xiong2015mird,le2011phishdef,chhabra2011phi, eshete2014webwinnow} \\~\cite{marchal:phishscore,marchal:phishstorm,gupta2014bit,khonji2011novel,khonjiIJ11,vermaK15,nguyen2013detecting},\textbf{~\cite{catakoglu2016automatic}}   \end{tabular}    \\  \hline
                            \textbf{2-3}     & 7              &   \begin{tabular}[c]{@{}l@{}} ~\cite{verma2017s}, \textbf{\cite{lee2013warningbird}}, \cite{su2013suspicious,lee2014users}, \textbf{\cite{dewan2015towards}},~\cite{popescu2016practical,lin2013malicious}  \end{tabular}     \\ \hline
                             \textbf{$\geq$4}  & 5      &  ~\cite{basnet2015towards,vermaK15,dewan2017facebook,zhang2016url}, \textbf{\cite{catakoglu2016automatic}}       \\ \hline

\end{tabular}
\end{table}
We \hl{next} describe the different evaluation metrics used in phishing URL detection papers.

\subsection{Evaluation Metrics}\label{sec:url-metrics}

The choice of metric(s) is very important for sound  system evaluation. In Table~\ref{tab:urlmetrics1}, we report papers \hl{that} use one (along the diagonal) or a combination of two types of metrics. Papers that report more than two metrics are discussed but are not counted in this table. However, 10 out \hl{of} 61 papers do not report the evaluation metrics used.

Accuracy and Confusion Matrix are the most popular metrics. We group papers that use True Positive, False Positive, True Negative and False Negative values as metrics (both raw numbers as well as rates) into the Confusion Matrix (CMx) 
category. 
Next in popularity are: \textit{Precision} (P), \textit{Recall} (R), and \textit{$F_{1}$-score} (abbr. F, it is the harmonic mean of precision and recall) along with \textit{Error Rate} (ErR). Some papers report additional metrics, e.g., Receiver Operating Characteristics (ROC)~\cite{basnet2015towards}, Area Under Curve (AUC)~\cite{darling2015lexical, ma2009beyond}, and classification costs~\cite{zhao2013cost}. 
A few papers consider combinations of three or more metrics: \textit{Acc, CMx, Other}\footnote{Classification Cost} in~\cite{zhao2013cost}, \textit{Acc, PRF, CMx }in~\cite{popescu2015study}, \textit{Acc, PRF, AUC} in~\cite{bahnsen2017classifying,darling2015lexical}, \textit{Acc, PRF, CMx, ErR} in~\cite{zhang2016url} and \textit{Acc, PRF, CMx, AUC} in~\cite{marchal:phishstorm,marchal:phishscore}. 
More details on metrics \hl{used} along with \hl{relevant}  citations are reported in \hl{Appendix}  Table~\ref{tab:metricsurl}.

\begin{table}[!htb]
\centering
\caption{Distribution of evaluation metrics in phishing URL detection.}
\label{tab:urlmetrics1}
\begin{threeparttable}
\begin{tabular}{rccccc}
                                   & Acc                    & PRF                    & ErR                    & CMx                       & Other\tnote{1}                   \\ \cline{2-6} 
\multicolumn{1}{r|}{Acc}           & \multicolumn{1}{c|}{\cellcolor{yellow}{5}} & \multicolumn{1}{c|}{\cellcolor{red}{6}} & \multicolumn{1}{c|}{\cellcolor{green}{1}} & \multicolumn{1}{c|}{\cellcolor{yellow}{5}} & \multicolumn{1}{c|}{\cellcolor{green}{2}} \\ \cline{2-6} 
\multicolumn{1}{r}{PRF}           &  & \multicolumn{1}{|c|}{\cellcolor{yellow}{5}} & \multicolumn{1}{l|}{}  & \multicolumn{1}{c|}{\cellcolor{green}{1}} & \multicolumn{1}{l|}{}  \\ \cline{3-6} 
\multicolumn{1}{r}{ErR}           &               &           & \multicolumn{1}{|c|}{\cellcolor{red}{9}} & \multicolumn{1}{c|}{\cellcolor{green}{1}}  & \multicolumn{1}{l|}{}  \\ \cline{4-6} 
\multicolumn{1}{r}{CMx}           &  &   &   & \multicolumn{1}{|c|}{\cellcolor{yellow}{4}}  & \multicolumn{1}{c|}{\cellcolor{green}{2}} \\  \cline{5-6}
\multicolumn{1}{r}{Other}           &   &   &   &   & \multicolumn{1}{|c|}{\cellcolor{green}{1}} \\   \cline{6-6}
\end{tabular}
\begin{tablenotes}
\item[1] \textit{Balanced Success Rate, Root Mean Square Error, effective rules/patterns, Cost/Sum of classification, False Alarm rate}
\end{tablenotes}
\end{threeparttable}
\end{table}

Another important facet to consider is the use of \hl{appropriate} evaluation metrics when considering unbalanced datasets. Accuracy is not always the \hl{appropriate} for multiple reasons\hl{:} Base-rate fallacy \hl{in} Section \ref{sec-challenges}, unbalanced datasets, \hl{and} asymmetric costs. Similar is the case for \hl{the}  error rate metric. ROC, AUC, and confusion matrix values can \hl{be} better metrics  in such \hl{scenarios}.
Researchers should  also use metrics specifically proposed for unbalanced experiments~\cite{bekkar2013evaluation}, e.g., geometric mean, Matthews Correlation Coefficient (MCC), Balanced Accuracy, Balanced Detection Rate \cite{ELAetAl:IWSPA-AP2018}, etc.

However, of the papers that use highly unbalanced URL datasets:~\cite{ma2011learning,basnet2015towards} use error rate, \cite{sha2015limited} uses false positive and false negative rates\hl{,} which is better, since one can calculate other metrics \hl{(if dataset size and distribution of classes is also reported)}, \cite{su2013suspicious} uses malicious missing rate and detection rate, and \cite{dewan2017facebook} makes use of popular metrics:  accuracy, precision, recall, $F_{1}$-score and ROC. Thus, we observe \hl{mostly inappropriate} unbalanced dataset metrics in the studied literature.
\subsection{Selected Phishing URL Detection Papers}\label{sec:url-lit}
We now \hl{discuss}  phishing URL detection \hl{papers} that \hl{addressed at least a few} security challenges. The selection is based on \hl{the number of} challenges (Section~\ref{sec-techniques}) addressed. 


Zhao et al. \cite{zhao2013cost}  proposed a scalable cost-sensitive active learning (CSOAL) framework \hl{that tackled multiple} challenges \hl{listed} in Section \ref{sec-challenges}. To ensure efficiency and scalability of the proposed system in a web-based environment, the authors used online active learners for building the detection model. This also handled phishing attacks launched with short time windows. Moreover, an online learning algorithm ensured faster system retraining. To deal with the issue of data availability, the system used a fraction of the provided dataset during training. The model was trained using only 0.5\% of the data and was evaluated on a large (1M) and highly imbalanced (99 legitimate to 1 malicious) URL dataset. Thus, their method also \hl{considered} class imbalance. The authors proposed two novel cost-sensitive measures -- weighted cost and weighted sum of sensitivity and specificity for better system evaluation, but these were not compared to other metrics suitable for unbalanced data (see Section \ref{sec:url-metrics}). Other experiments include system scalability and study of training and testing times. However, the authors did not address the active attacker issue.

In \cite{darling2015lexical}, authors proposed a lightweight phishing detection system using lexical features for link analysis. Such a system ensures \hl{separation}  
between the victim and the attacker; \hl{and a lightweight scheme ensures speed and scalability}. The authors also study system performance by conducting ``generalization experiments'' across multiple evaluation datasets of varying ratios between benign and malicious classes in the dataset (ratios considered were 3:1, 1:1, and 1:3\hl{)}.
Researchers argue that their work is secure against an active attacker, but do not perform any real-time experiments to \hl{validate this claim}.  
However, the metrics used for evaluation of the proposed system are \hl{inappropriate for} unbalanced datasets. 

Blacklist dependent phishing URL detection techniques can be defeated
by phishers by simply launching new websites after short windows of time. Ma et. al., \cite{ma2009beyond}, addressed \hl{this} issue by proposing a URL classification technique that makes use of a wide variety of features\hl{:} lexical, host-based, blacklist-based and WHOIS properties. Although the authors used a relatively small dataset, they addressed the \textit{data diversity and generalization issue} by creating multiple datasets using  different publicly available sources. Other important attributes that stand out are: feature selection and comparison, \hl{classifier comparison}, and error analysis. However, the small and relatively balanced dataset \hl{is a serious limitation}.
Moreover, \hl{the active attacker issue was not considered directly}.

Novel distance-based features for phishing URLs based on Kullback-Leibler divergence \hl{(described in} \cite{kullback1951information} \hl{to measure the distance between distributions of normalized character frequencies in URLs and Standard English)} 
and Kolmogorov-Smirnov distance \hl{(measured using the two-sample Kolmogorov-Smirnov test statistic}~\cite{smirnov1948table} \hl{calculated from the distribution of character frequencies of the URLs versus that of standard English.)} were used in \cite{vermaK15}, along with modifications of some old features to make them more robust, e.g., the ratio of URL domain to URL length was used instead of just URL length. Phishing detection at URL level\hl{,} along with fast data structures used by the researchers\hl{,} ensures a fast and lightweight approach capable of handling attacks launched within short time windows. The authors used four different datasets collected from different real-world sources\footnote{Two datasets from researchers in China and two from publicly available sources - DMOZ, Alexa for legitimate and PhishTank for phishing} for evaluation, thus ensuring a diverse set of URLs. They also performed experiments to explicitly verify dataset diversity along with a cross-dataset experiment (training on one dataset and testing on another). The system performance using five supervised models was evaluated using accuracy, false positive rate, and classifier training/testing times. Thus, the system evaluation was setup to address some of the issues of earlier work. However, the proposed system was not evaluated using unbalanced datasets. The authors did not comment on system retraining for handling newer attacks.



Lee et al.~\cite{lee2013warningbird} evaluated malicious URLs collected from Twitter, a popular platform for sharing poisoned links. The \hl{authors} partially handled the active attacker challenge by proposing a system \hl{that} can identify correlations among URL redirect chains. The authors trained and evaluated the proposed system on a large dataset of malicious tweets using multiple metrics\hl{:} accuracy, AUC, false positive and negative rates.  The system \hl{was also evaluated} in a real-time environment using a sliding window to vary the span of the training data feed. The authors also \hl{studied} feature importance  using $F_1$-score\hl{s}. However, while classification time per URL  is \hl{reported}, the training time is not specified. 

We conclude this section with lessons learned and best practices based on our review of the literature.
\subsection{Lessons learned and best practices}\label{sec:url-summ}
We observed a number of useful practices in the literature as well as possibilities for improving future research. Use of  \hl{techniques} capable of handling data availability issues while providing reliable performance \cite{zhao2013cost} is a desirable practice. Also, the use of cost-sensitive metrics \cite{zhao2013cost,lee2013warningbird}\hl{,} which take into account the class-imbalance issue\hl{,} is an important criteri\hl{on} for evaluation.
Generalization experiments\hl{,} or cross-domain testing of systems\hl{,} is another recommended way for verifying the performance of the model on various types of attacks. We observed generalization experiments in \hl{a few} papers\hl{,e.g.,}~\cite{verma2017s,darling2015lexical}. While a number of papers evaluate\hl{d} on unbalanced  datasets \cite{zhao2013cost, sha2015limited, darling2015lexical, dewan2017facebook}, we did not \hl{find} any paper \hl{with} the \hl{appropriate} metrics (suggested in~\cite{bekkar2013evaluation}) to \hl{measure performance}. 

Feature selection can make a system faster and robust. Popular techniques for feature selection include information gain, \hl{$F_{1}$-scores}, etc. \hl{Phishing URL detection}  papers that used information gain \hl{include}  \cite{darling2015lexical,gupta2014bit,su2013suspicious}\hl{, whereas} \cite{sandracoordinator,lee2013warningbird} used $F_1$-score\hl{s} to determine \hl{feature} importance. Other papers that \hl{analyze} feature importance \hl{include}  \cite{eshete2014webwinnow}, \cite{mamun2016detecting}, \cite{dewan2017facebook,dewan2015towards}.

\section{Phishing Website Detection}
\label{sec-techniques-websites}
For phishing website detection, we review 67 papers based on the selection criteria in Section~\ref{sec:keyword}. We begin with the structure of a website.
\subsection{Website Structure}
\begin{figure}[h]
\centering
 \includegraphics[width=1\linewidth]{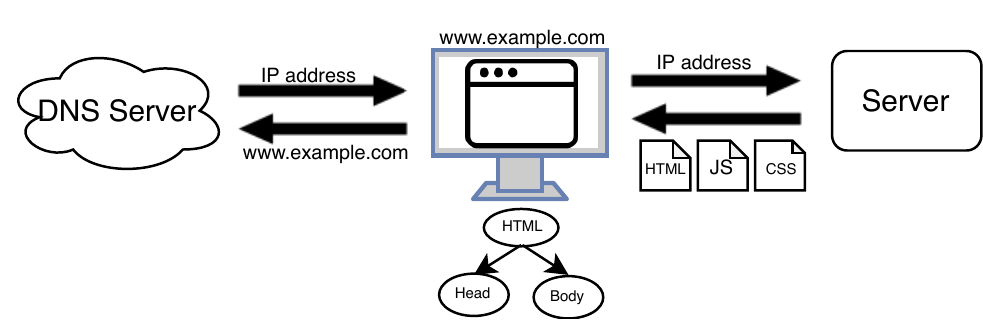}
 \caption{Structure of a typical website along with its network components}
 \label{fig:web_struct}
\end{figure}

A website refers to a collection of formatted documents that can be rendered and viewed using a web browser (e.g., Chrome, Firefox, Microsoft Edge). 
Websites consist of three elements: (1) URL pattern, (2) layout and (3) link structure. Figure \ref{fig:web_struct} shows the structure and elements of a typical website. 

As described in Section \ref{sec-techniques-urls}, each website needs an address (URL) for access. The syntax of a URL and its different components were described in Section \ref{sec-techniques-urls}. Besides coming up with a URL,  the domain must be registered with the DNS. When a user inputs the URL into a browser, it needs the IP address of that website, which it gets from the DNS server, to send HTTP (or HTTPS) requests. Figure \ref{fig:web_struct} shows the communications from the users' browser to the DNS server and the web server.

After the HTTP request, the browser receives the HTML content of the website. The \textit{layout} refers to the syntactic structure of the web page, i.e., the distribution of HTML elements. Website designers generally use the same layout for web pages with similar functionalities. Thus, users can easily identify the purpose of a web page by looking at its design. This can be a good source of features for detecting similarity between the websites since phishing websites try to mimic well-known target websites. Cascading Style Sheets (CSS) and JavaScript are the most common ways of specifying the layout. CSS is used to describe the position and other appearance properties of objects in the websites. JavaScript enables the content of a web page to dynamically change on the client side.

Each web page usually links to other web pages for easy navigation. These links may or may not be in the same domain as the original web page (Section~\ref{sec-techniques-urls}). Popular websites have lots of incoming links from other websites, whereas phishing websites do not, since they are short-lived. Web page rankings (e.g., PageRank) by search engines can be used to measure the importance of web pages. These rankings are based on the page  attributes including the link structure.



The following components of a website can be used as features to characterize it:
\begin{itemize}
    \item \textit{Network:} 
    \hl{We discussed above the syntactic features that are directly extracted from the URL.}
    Here, we include other features, which are indirectly related to the URL (e.g., DNS information, domain registration, and Secure Sockets Layer (SSL) certificate attributes).
    \item \textit{HTML content}: This includes features that are extracted from the HTML content of the website. Examples are:  \textit{number of links}, webpage content, tags, and \textit{number of hidden elements.}
    \item \textit{JavaScript}: Although Javascript features can be viewed as HTML features, we put them in a separate group because of their dynamic nature.
\end{itemize}

In the next section, we discuss the features used in the reviewed literature, with the above taxonomy. 

\subsection{Feature Extraction and Analysis}\label{website-feature-extraction}
\begin{table*}[!htb]
\centering
\caption{Features used in phishing website detection. For uncommon features, we cite the paper \hl{that} best describes it.}
\label{table:website-features-short}
\begin{tabular}{|l|c|c|lcclcc}
\hline
\multicolumn{1}{|c|}{\multirow{2}{*}{\textbf{HTML}}}     & \multicolumn{2}{c|}{Criteria}                 & \multicolumn{1}{c|}{\multirow{2}{*}{\textbf{Network}}}   &   \multicolumn{2}{c|}{Criteria\tnote{a}}               & \multicolumn{1}{c|}{\multirow{2}{*}{\textbf{Javascript}}} &\multicolumn{2}{c|}{Criteria}                  \\ \cline{2-3}\cline{5-6}\cline{8-9}
& FS & PT &\multicolumn{1}{l|}{} & \multicolumn{1}{c|}{FS}  & \multicolumn{1}{c|}{PT} &\multicolumn{1}{l|}{} &\multicolumn{1}{c|}{FS} & \multicolumn{1}{c|}{PT}\\
\specialrule{.2em}{.1em}{.1em}
\# of various tag types & M & M &\multicolumn{1}{l|}{Domain registration info}  &\multicolumn{1}{|c|}{S}& \multicolumn{1}{|c|}{H} & \multicolumn{1}{l|}{keywords to words ratio} &\multicolumn{1}{|c|}{S} &\multicolumn{1}{|c|}{H}
\\ \hline
                          HTML tag attributes   & M & M &\multicolumn{1}{l|}{Registrar ID}       &\multicolumn{1}{|c|}{S}&\multicolumn{1}{|c|}{M}   & \multicolumn{1}{l|}{\#  of suspicious strings \cite{canali2011prophiler}} & \multicolumn{1}{|c|}{S} & \multicolumn{1}{|c|}{M} 
                          \\ \hline 
%
                          Term Frequency       & L & S & \multicolumn{1}{l|}{\# of Nameservers}    &\multicolumn{1}{|c|}{S}&\multicolumn{1}{|c|}{M}                & \multicolumn{1}{l|}{\# of long strings (>40, >51)} & \multicolumn{1}{|c|}{S} & \multicolumn{1}{|c|}{M} \\ \hline
                         \# of elements out of place  & S & M & \multicolumn{1}{l|}{DNS record}         &\multicolumn{1}{|c|}{S} & \multicolumn{1}{|c|}{M}        & \multicolumn{1}{l|}{decoding routines} &\multicolumn{1}{|c|}{S} & \multicolumn{1}{|c|}{H} \\ \hline
                          \# of small/hidden elements & S & M &\multicolumn{1}{l|}{\# of DNS queries}     &\multicolumn{1}{|c|}{S}&\multicolumn{1}{|c|}{H}          & \multicolumn{1}{l|}{shellcode detection} &\multicolumn{1}{|c|}{S} & \multicolumn{1}{|c|}{H}    \\ \hline
                         \# of suspicious elements & S & M & \multicolumn{1}{l|}{HTTP header fields} &\multicolumn{1}{|c|}{M}&\multicolumn{1}{|c|}{M}   & \multicolumn{1}{l|}{\# of iframe strings} &  \multicolumn{1}{|c|}{S} & \multicolumn{1}{|c|}{M} \\ \hline
                          \# of internal/external links & S & M &\multicolumn{1}{l|}{Alexa rank} &\multicolumn{1}{|c|}{S}&\multicolumn{1}{|c|}{M} & \multicolumn{1}{l|}{\# of DOM modifying functions} &\multicolumn{1}{|c|}{S} &\multicolumn{1}{|c|}{H} \\ \hline
                          \begin{tabular}[c]{@{}l@{}}NULL links on \\site and footer\end{tabular}       & S & M &\multicolumn{1}{l|}{Gmail reputation \cite{whittaker2010large}}      &\multicolumn{1}{|c|}{S}& \multicolumn{1}{|c|}{M}     & \multicolumn{1}{l|}{\# of event attachment}  &\multicolumn{1}{|c|}{S}&\multicolumn{1}{|c|}{H} \\ \hline
                          \begin{tabular}[c]{@{}l@{}}More than one \\head tag/document\end{tabular} & S & M &\multicolumn{1}{l|}{\# of bytes/packets trans.}                        &\multicolumn{1}{|c|}{S}&\multicolumn{1}{|c|}{H}             &   \multicolumn{1}{l|}{\# of  \hl{suspicious objects} \cite{canali2011prophiler}} & \multicolumn{1}{|c|}{S} & \multicolumn{1}{|c|}{M}\\ \hline
                          invisible frames \cite{liang2009malicious} & S & M &\multicolumn{1}{l|}{Fake HTTP protocol}       &\multicolumn{1}{|c|}{S}&\multicolumn{1}{|c|}{M} &   \multicolumn{1}{l|}{\# of scripts} &\multicolumn{1}{|c|}{S} &\multicolumn{1}{|c|}{M}\\ \hline
                          \# of specific file type & S & M &\multicolumn{1}{l|}{\begin{tabular}[c]{@{}l@{}}\# of IP/port upon \\ complete download\end{tabular}}                              &\multicolumn{1}{|c|}{S}& \multicolumn{1}{|c|}{H}      &   \multicolumn{1}{l|}{\begin{tabular}[c]{@{}l@{}}\# of functions (eval, \\ setInterval, OnMouseOver...)\end{tabular}} &\multicolumn{1}{|c|}{S} & \multicolumn{1}{|c|}{M}\\ \hline
                          Visual \cite{bannur2011judging} & M & H &\multicolumn{1}{l|}{SSL Cert. attributes}                        &\multicolumn{1}{|c|}{S}& \multicolumn{1}{|c|}{M}\\ \cline{1-6}
                          \# of iframes & S & M \\ \cline{1-3} 
                          DOM-tree \cite{rosiello2007layout} & S & H \\ \cline{1-3}
                          ActiveX function &         S &  M& \\ \cline{1-3}
                          Right click disabled &  S & M                                 &                                           \\ \cline{1-3}
                          Server Form Handler &      S & M                                           \\ \cline{1-3}
                          \begin{tabular}[c]{@{}l@{}}Login-form identity\end{tabular} &                      S  & M           &                                                                     \\ \cline{1-3}
                          \begin{tabular}[c]{@{}l@{}}External term frequency \cite{marchal2016know} \end{tabular} &                      L  & H           &                                                                     \\ \cline{1-3}
\end{tabular}
\end{table*}

Table~\ref{table:website-features-short} shows the features used in the studied phishing website detection literature.\footnote{To avoid repeating URL features, we have added them to Table \ref{tab:urltypes} in the previous section.}
\hl{These features can improve detection performance. However, they also increase the training and classification times since most of them need moderate or high extraction and processing times.}
\textit{Feature size} for most of the website features is small except for \textit{Term Frequency,} which depends on the size of the vocabulary. We emphasize that  \textit{training \hl{of} models \hl{require} large-scale feature extraction, but for some of the website features, this may not be possible}. For example, retrieving the WHOIS information involves sending requests to a WHOIS \hl{server} which may start rejecting requests due to a rate-limiting policy.

There are several hard to evade website features \hl{-} \textit{Visual} similarity of websites is one such feature, because if a website is not visually similar to the target website, then the victim will not fall for it.
Another example is \textit{login-form identity}. The phisher can manipulate this feature by removing the login fields, but this would defeat the main purpose of a phishing attack, which is to steal  credentials.
In Table \ref{table-webpage-features} under \hl{the} Appendix, we list the  features along with citations to the papers that use them.

In the next section, we discuss the different phishing detection methods that use the aforementioned features.

\subsection{Detection Methods}\label{sub-website-detection-methods}
\begin{figure*}[b!]
\centering
 \includegraphics[width=0.9\linewidth]{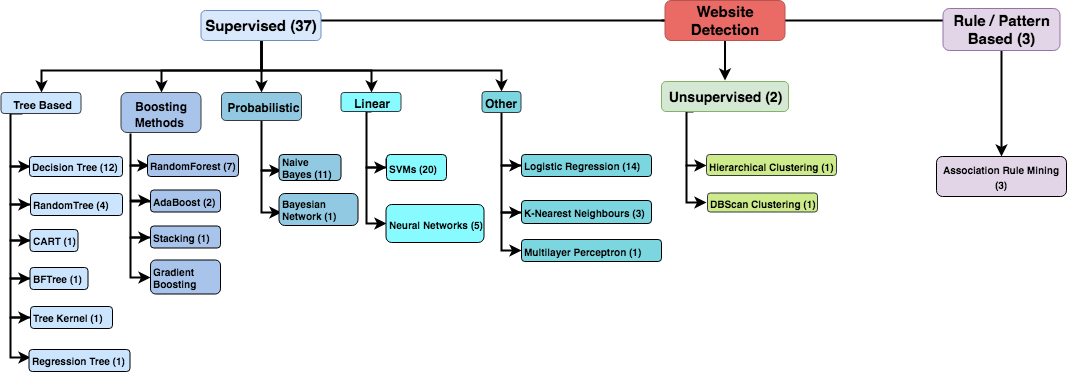}
 \caption{Algorithms used in phishing website detection. The number of papers using the method is in parentheses.}
 \label{fig:webmethodsdist}
\end{figure*}

Figure~\ref{fig:webmethodsdist} shows the different methods used in the phishing website \hl{detection} literature as well as the number of papers that use each method. Supervised techniques -- Support Vector Machines (SVM), logistic regression, decision tree, and Na\"{i}ve Bayes -- are the most popular phishing website detection methods. \hl{Only five papers} use unsupervised and rule-based methods. In contrast to phishing URL detection methods, we see that online learning and deep learning techniques are missing in phishing website detection. More detailed information about the detection methods used in each paper is in the Appendix (Table \ref{table-webpage-classification}).

While supervised learning is the preferred choice, such algorithms need a sizable ground truth or labeled data for training. Hence, we next describe the different datasets used in the phishing website detection literature.

\subsection{Dataset Properties}
Although the detection method is an important part of a proposed system, datasets that authors use to train and test their models  have a significant effect on the trustworthiness of their models.
\subsubsection{\textbf{Dataset sources and availability}}
Phishing website detection techniques have used almost the same datasets as phishing URL detection techniques, thus also exhibiting the lack of diversity issue.
 Although the papers are about phishing website detection, occasionally they use \textit{Malware} and \textit{Spam} websites to test/train their methods. Such papers are grouped together in the  \textit{Malicious} group (\textit{spam} dataset refers to URLs extracted from the body of spam emails). Tables~\ref{table-website-dataset} and~\ref{table-website-dataset_2} list the papers that use phishing and malicious sources along with the legitimate ones. Papers in which researchers used their own data sources under are in \textit{Author's} column. If they reveal the details of their sources publicly, they are listed as \textit{Author's public}, otherwise as \textit{Author's private}.

\begin{table*}[h]
\centering
\caption{\hl{Legitimate and phishing} website datasets: \hl{\textit{sources and availability}}}
\label{table-website-dataset}
\resizebox{2\columnwidth}{!}{
\begin{threeparttable}
\begin{tabular}{|c|c|c|c|c|c|c|c|c|}
\hline
\multirow{2}{*}{\diagbox[width=2.5cm]{\textbf{Phish}}{\textbf{Legit.}}}
& \multirow{2}{*}{\begin{tabular}[c]{@{}c@{}}\textbf{DMOZ}\\(dep)\end{tabular}} & \multirow{2}{*}{\begin{tabular}[c]{@{}c@{}}\textbf{Alexa}\\(pub)\end{tabular}} & \multirow{2}{*}{\begin{tabular}[c]{@{}c@{}}\textbf{Yahoo Dir.}\\(pub)\end{tabular}} & \multirow{2}{*}{\begin{tabular}[c]{@{}c@{}}\textbf{Twitter}\\(pub)\end{tabular}} & \multicolumn{2}{c|}{\textbf{Author's}}  & \multirow{2}{*}{\begin{tabular}[c]{@{}c@{}}\textbf{Misc.}\tnote{a}\\(pub)\end{tabular}} & \multirow{2}{*}{\textbf{N/A}}  \\ \cline{6-7}
                                                                         & \multicolumn{1}{c|}{}                            & \multicolumn{1}{c|}{}                             & \multicolumn{1}{c|}{}                                       & \multicolumn{1}{c|}{}                               & \multicolumn{1}{c|}{\textbf{public}} & \multicolumn{1}{c|}{\textbf{private}} &  &                                    \\ \specialrule{.2em}{.1em}{.1em}
PhishTank (pub)        & \begin{tabular}[c]{@{}l@{}}\cite{xiang2010hierarchical,choi2011detecting, feroz2015examination, nguyen2013detecting, lee2016phishing, chiew2015utilisation,geng2015combating,thakur2014catching}\end{tabular}           & \begin{tabular}[c]{@{}c@{}}\cite{thakur2014catching,miyamoto2008evaluation, Xiang:2011:CFM:2019599.2019606, huang2010mitigate,xu2014gemini, marchal2012proactive,tan2016phishwho, manek2014demalfier, gowtham2014comprehensive, he2011efficient, varshney2016phish, maurer2012using, ramesh2014efficacious, dong2015beyond, zhang2017two, raodetection}\\\cite{xiang2010hierarchical,chiew2015utilisation}\end{tabular}            & \begin{tabular}[c]{@{}c@{}}\cite{miyamoto2008evaluation, xiang2010hierarchical, Xiang:2011:CFM:2019599.2019606}\\\cite{pao2012malicious,yue2013fine, moghimi2016new, abdelhamid2014phishing, mohammad2014intelligent, rajab2017new, Mohammad2014, liu2010automatic,choi2011detecting}\end{tabular}                      & \cite{thomas2011design}              & \cite{gowtham2014comprehensive, dong2010defending}                       & \cite{tan2014phishing, Gowtham2014, mao2017phishing, fang2015proactive, corona2017deltaphish, zhang:textual,wenyin2012antiphishing, barraclough2013intelligent}                       & \begin{tabular}[c]{@{}c@{}}\cite{xiang2010hierarchical, pao2012malicious}\\\cite{marchal2016know,rosiello2007layout} \end{tabular} 	& \begin{tabular}[c]{@{}c@{}}\cite{liang2016cracking, aburrous2010intelligent, abbasi2015enhancing, shahriar2012trustworthiness, mao:alarm,zhang2017phishing} \end{tabular} \\ \hline
APWG\tnote{b} (pub)             & \cite{geng2015combating}           & \multicolumn{1}{c|}{}            & \multicolumn{1}{c|}{}                      & \multicolumn{1}{c|}{\cite{thomas2011design}}              & \multicolumn{1}{c|}{}                       & \multicolumn{1}{l|}{}                       &                   &  \cite{aburrous2010intelligent, abbasi2015enhancing}  \\ \hline
UAB Phishing (pvt)     & & & & & & & & \cite{britt2012clustering, wardman2009identifying, wardman2011high}                      \\ \hline
Millersmiles (pub)     & \multicolumn{1}{c|}{}           & \multicolumn{1}{c|}{\cite{xiang2010hierarchical, he2011efficient}}            & \multicolumn{1}{c|}{\cite{xiang2010hierarchical, mohammad2014intelligent, rajab2017new, Mohammad2014}}                      & \multicolumn{1}{c|}{}              & \multicolumn{1}{c|}{\cite{dong2010defending}}                       & \multicolumn{1}{l|}{}                       &  \cite{xiang2010hierarchical}              &      \\ \hline
Twitter (pub) & & & & \cite{aggarwal2012phishari, thomas2011design} & & & &\\ \hline
Author's (pvt) & & & & & & \cite{zhuang2012intelligent,zhang2014domain, vargas:enemies, whittaker2010large,kosba2014adam} & &\\ \hline
Miscellaneous\tnote{c} (pub)    & & \cite{tan2016phishwho, zhang2017two} & & & & \cite{ramesh2017intelligent}                       &  \cite{thabtah2016dynamic,el2017detection}  & 
\\ \hline
\end{tabular}
\begin{tablenotes}
\item[a] \textit{Intelsecurity.com (defunct) \cite{marchal2016know}, UCI repository \cite{thabtah2016dynamic,el2017detection}, TrendMicro \cite{pao2012malicious}, Google Safe Browsing \cite{xiang2010hierarchical}, URoulette \cite{rosiello2007layout}}
\item[b] \textit{Anti-Phishing Working Group}
\item[c] \textit{OpenPhish \cite{ramesh2017intelligent}, UCI repository \cite{thabtah2016dynamic}}
\end{tablenotes}
\end{threeparttable}
}
\end{table*}

\begin{table}[h]
\centering
\caption{\hl{Legitimate and malicious} website datasets: \hl{\textit{sources and availability}}}
\label{table-website-dataset_2}
\resizebox{1\columnwidth}{!}{
\begin{threeparttable}
\begin{tabular}{|c|c|c|c|c|c|}
\hline
\begin{tabular}[l]{@{}l@{}}\diagbox[width=2.5cm]{\textbf{Malicious}}{\textbf{Legit.}} \end{tabular}   & \begin{tabular}[c]{@{}c@{}}\textbf{DMOZ}\\(dep)\end{tabular} & \begin{tabular}[c]{@{}l@{}}\textbf{Alexa}\\(pub)\end{tabular} & \begin{tabular}[c]{@{}c@{}}\textbf{Yahoo}\\(pub)\end{tabular} & \begin{tabular}[c]{@{}c@{}}\textbf{Twitter}\\(pub)\end{tabular} & \begin{tabular}[c]{@{}c@{}}\textbf{Author}\\(pvt)\end{tabular}  \\ \specialrule{.2em}{.1em}{.1em}
\begin{tabular}[c]{@{}c@{}}DNS-BlackHole\\(pub)\end{tabular}   & \begin{tabular}[c]{@{}l@{}}\cite{choi2011detecting}\\\cite{geng2015combating}    \end{tabular}       & \cite{marchal2012proactive}            & \cite{choi2011detecting}                      & \multicolumn{1}{c|}{}              & \multicolumn{1}{c|}{}                           \\ \hline
Author's (pub) & \multicolumn{1}{c|}{}           & \cite{xu2014evasion}            & \multicolumn{1}{c|}{}                      & \multicolumn{1}{c|}{}              & \multicolumn{1}{c|}{}                    \\ \hline
Author's (pvt) & \multicolumn{1}{c|}{}           & \cite{canali2011prophiler}            & \multicolumn{1}{c|}{}                      & \begin{tabular}[c]{@{}c@{}}\cite{thomas2011design}\\\cite{bannur2011judging}\end{tabular}              &  \begin{tabular}[c]{@{}c@{}}\cite{stringhini2013shady}\\\cite{ mohaisen2015towards}\end{tabular}                       \\ \hline
Misc.\tnote{a} (pub)    & \cite{choi2011detecting}           & \cite{ manek2014demalfier}            & \cite{yue2013fine,choi2011detecting}                      &               & \cite{liang2009malicious}                            \\ \hline
\end{tabular}
\begin{tablenotes}
\item[a] \textit{Wepawet \cite{canali2011prophiler}, MalwareURL \cite{manek2014demalfier}, host-files.net \cite{manek2014demalfier}, stopbadware.org \cite{liang2009malicious}, jwSpamSpy \cite{choi2011detecting} and webspam-uk07 \cite{yue2013fine}}
\end{tablenotes}
\end{threeparttable}
}

\end{table}

\begin{figure}
\centering
 \includegraphics[width=0.75\linewidth]{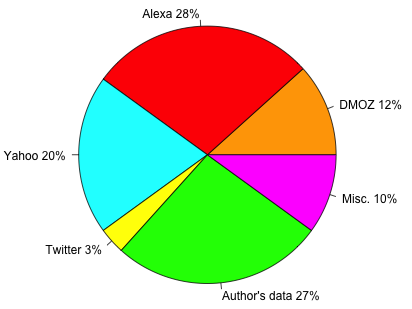}
 \caption{Distribution of papers \hl{based on} legitimate website source}  
 \label{fig:legiturldist_website}
 \end{figure}

\begin{figure}
\centering
 \includegraphics[width=0.7\linewidth]{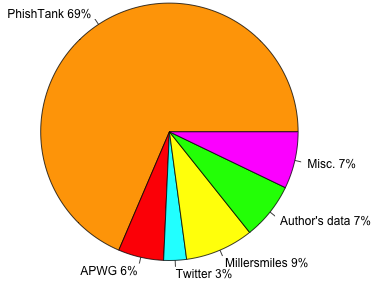}
 \caption{Distribution of papers \hl{based on}  phishing website source}  
 \label{fig:phishurldist_website}
 \end{figure}

Figures~\ref{fig:legiturldist_website} and~\ref{fig:phishurldist_website} show the distribution of phishing and legitimate sources respectively. They show similar trends as we saw before in phishing URL detection (Figures \ref{fig:legiturldist} and \ref{fig:phishurldist}).
For the phishing dataset, 69\% of the studies used PhishTank as their source of data and 9\% used Millersmiles  and 6\% used \hl{APWG. For} legitimate datasets, the sources are more diverse than phishing ones. Alexa (28\%) is the most widely used dataset, but DMOZ (12\%) and Yahoo Directory (20\%) are also used in several studies (unfortunately, they are now deprecated). 


\subsubsection{\textbf{Dataset Size}}
Table~\ref{table-webpage-dataset-size-summarized} shows the categorized sizes of the legitimate, phishing, or malicious website datasets used by each reviewed phishing website detection paper. 
We observe that more than 60\% of the research papers report experiments on balanced datasets (red cells along the diagonal line of Table \ref{table-webpage-dataset-size-summarized}). These systems do not consider the base-rate fallacy in their evaluation, hence their methods may perform quite differently in the real world. Also, only about one-third of the reviewed papers used at least 10,000 legitimate, phishing and/or malicious websites for evaluation. The rest suffer from the small dataset concern (eight papers do not mention their dataset sizes). More details on the sizes of datasets used by each paper can be found in the Appendix (Table \ref{table-webpage-dataset-size}). 

\begin{table}[h]
\centering
\caption{Size ranges of legitimate/phishing website datasets.}
\label{table-webpage-dataset-size-summarized}
\begin{tabular}{|c|c|c|c|c|c|c|}
\hline
                           \textbf{Type}      & \multicolumn{6}{c|}{\textbf{Legitimate}} \\ \hline
\textbf{\hl{Phishing}} & 100s & 1,000s & 10,000s & 100,000s                                       & \textgreater1M     & N/A \\ \specialrule{.2em}{.1em}{.1em}
100s     & \cellcolor{red}{6} & \cellcolor{yellow}{5}                                                                                                                                       &                                                                                                                                 &                                                &                           &  \\ \hline 
                           1,000s   & \cellcolor{yellow}{4}                        & \cellcolor{red}{13} & \cellcolor{green}{2}                                                                                       & \cellcolor{green}{2} &                        &  \cellcolor{yellow}{4}  \\ \hline
                           10,000s  &                                                                                                                                     &                                                                                                                                                                                                                      & \cellcolor{red}{7} & \cellcolor{green}{1}                        & \cellcolor{green}{1}  & \cellcolor{yellow}{3} \\ \hline
                           100,000s &                                                                                                                                     &                                                                                                                                                                                                                                                  &                                                                                                                                 &                                                & \cellcolor{green}{1} & \cellcolor{green}{1} \\ \hline 
                           N/A &                                                                                                                                     & \cellcolor{green}{1}                                                                                                                                                                                                                                                 &                                                                                                                                 &                                                & & \cellcolor{red}{13}\\ \hline 

\end{tabular}
\end{table}

\subsubsection{\textbf{Dataset Diversity}}\label{sec:web-data-div}
As in the URL section, we analyze the diversity of PhishTank, OpenPhish, APWG, and Alexa, since these are also the most widely used sources for websites. However, instead of checking the diversity of URLs, we analyze the diversity found in the websites' content. We extracted all the text from the HTML body using the Python library called Beautiful Soup\footnote{https://www.crummy.com/software/BeautifulSoup/bs4/doc/} to compare the websites' content. We also added a filter to remove all stopwords. Another preprocessing step is \hl{the} removal of all the dead and non-English links. Then we converted them into vectors using the TFIDF representation. After converting each website into a vector, we use Cosine Similarity to measure the similarity between websites.

 After preprocessing, we ended up with 5,751, 1,665, 1861 and 63,670 URLs for PhishTank, OpenPhish, APWG, and Alexa, respectively.
Table~\ref{table-similarity-distribution} shows the percentage of datasets with a specific range of similarities. For example, column ``0-10'' shows the fraction of each dataset that has cosine similarity values in the  0\% to 10\% range.
The results showed that all four datasets except APWG are diverse from a website content point of view (1.3\% of websites had more than 50\% similarity for PhishTank and Alexa, and 6.2\% for OpenPhish, compared to 21.6\% for APWG). 


\begin{table}[h]
\centering
\caption{Percentage of websites with different ranges of similarities in each dataset. Columns specify the ranges of the Cosine Similarity in percentage.}
\label{table-similarity-distribution}
\resizebox{1\columnwidth}{!}{
\begin{tabular}{|c|c|c|c|c|c|c|}
\hline
\multirow{2}{*}{\textbf{Dataset}} & \multicolumn{6}{c|}{\textbf{Ranges}}                              \\ \cline{2-7} 
                         & [0-10]  & (10-20] & (20-30] & (30-40] & (40-50] & \textgreater{}50 \\ \hline
APWG                     & \textbf{72.78}  & 3.0   & 0.68  & 0.64  & 0.33  & 21.59            \\ \hline
OpenPhish                & \textbf{77.27} & 11.28 & 1.32  & 0.29  & 0.46  & 6.26             \\ \hline
PhishTank                & \textbf{81.59} & 13.55 & 2.21  & 0.81  & 0.5   & 1.31             \\ \hline
Alexa                    & \textbf{86.06} & 10.34 & 1.55  & 0.48  & 0.21  & 1.32             \\ \hline
\end{tabular}
}
\end{table}

\subsubsection{\textbf{Recency of data collection}}
Table~\ref{table-webpage-time-period} describes the recency of the dataset \hl{collection time} with respect to the paper's publication date. Unfortunately, only a few papers mentioned the time period during which their datasets were collected (25 out of 67). The importance of features may change over time, some features which were not important during the past year may now be useful, or the other way around. The use of newer data for evaluation also captures the capacity of a system to detect the latest attacks.
Among the papers that mentioned the recency, a majority of the work is published less than a year after data collection.

There are only a handful of research methods that propose real-time website-based detection systems \cite{thomas2011design, dong2015beyond, aggarwal2012phishari}. Feature extraction time is a major obstacle for real-time website-based detection systems since they must extract the content from the website.


\begin{table}[h]
\centering
\caption{Recency of website datasets used in evaluation. \hl{Papers} in \textbf{bold} use data collected \hl{from} multiple \hl{periods}.}
\label{table-webpage-time-period}
\begin{tabular}{|c|c|l|}
\hline
\textbf{Recency in year(s)}         & \textbf{N}  & \textbf{Literature}                                                                                              \\ \hline \hline
\textbf{Real time}           & 3  & \begin{tabular}[c]{@{}l@{}}\cite{thomas2011design, dong2015beyond, aggarwal2012phishari} \end{tabular}\\
\hline
\textbf{1}        & 17     & \begin{tabular}[c]{@{}l@{}}\cite{el2017detection,marchal2016know, miyamoto2008evaluation, lee2016phishing, xu2014evasion,stringhini2013shady}, \textbf{\cite{pao2012malicious}}\\\cite{thakur2014catching,vargas:enemies,whittaker2010large,Gowtham2014,zhuang2012intelligent,xiang2010hierarchical,aggarwal2012phishari,wardman2011high,britt2012clustering, wardman2009identifying} \end{tabular}\\ \hline
\textbf{2-3} & 2 &  \textbf{\cite{pao2012malicious}},\cite{corona2017deltaphish}                                                          \\ \hline
\textbf{3-4}     &  4      & \cite{moghimi2016new,shahriar2012trustworthiness,geng2015combating,aburrous2010intelligent}                                                                           \\ \hline
\end{tabular}
\end{table}

\subsection{Evaluation Metrics}
As mentioned in Section \ref{sec:url-metrics}, evaluation metrics should be chosen carefully when the dataset is unbalanced.
We only found two papers that use a metric specifically suited to unbalanced datasets (MCC) \cite{el2017detection,zhang:textual}; the rest reported commonly used metrics, which are more suitable for balanced datasets. Among these two papers, only \cite{zhang:textual} actually used an unbalanced dataset.

Table~\ref{tab:metricsmap} lists the metrics used for evaluating proposed methods in the literature. It only includes the papers that used up to two metrics (seven of 67 papers do not mention metrics). For the complete list of papers and their metrics refer to Table~\ref{table-webpage-metrics} in the Appendix.

\begin{table}[!htb]
\centering
\caption{Distribution of evaluation metrics in phishing website detection.}
\label{tab:metricsmap}
\begin{threeparttable}
\begin{tabular}{rcccccc}
                                   & Acc                    & PRF                    & ErR                    & CMx   & AUC                    & Other\tnote{a}                    \\ \cline{2-7} 
\multicolumn{1}{r|}{Acc}           & \multicolumn{1}{c|}{\cellcolor{yellow}{4}} & \multicolumn{1}{c|}{\cellcolor{yellow}{3}} & \multicolumn{1}{c|}{\cellcolor{yellow}{3}} & \multicolumn{1}{c|}{\cellcolor{red}{9}} & \multicolumn{1}{c|}{} & \multicolumn{1}{c|}{}\\ \cline{2-7} 
\multicolumn{1}{r}{PRF}           & \multicolumn{1}{l|}{}  & \multicolumn{1}{c|}{\cellcolor{yellow}{5}} & \multicolumn{1}{c|}{\cellcolor{green}{1}}  & \multicolumn{1}{c|}{\cellcolor{green}{2}} & \multicolumn{1}{l|}{}  & \multicolumn{1}{c|}{\cellcolor{green}{1}} \\ \cline{3-7} 
\multicolumn{1}{r}{ErR}           & \multicolumn{1}{c}{}         & \multicolumn{1}{c|}{} & \multicolumn{1}{c|}{\cellcolor{red}{9}}  & \multicolumn{1}{c|}{\cellcolor{green}{2}}  & \multicolumn{1}{c|}{} & \multicolumn{1}{c|}{}\\ \cline{4-7} 
\multicolumn{1}{r}{CMx}           & \multicolumn{1}{l}{}  & \multicolumn{1}{l}{}  & \multicolumn{1}{l|}{}  & \multicolumn{1}{c|}{\cellcolor{red}{11}}  & \multicolumn{1}{c|}{} & \multicolumn{1}{c|}{}\\ \cline{5-7} 
\multicolumn{1}{r}{AUC} &\multicolumn{1}{c}{} &\multicolumn{1}{c}{} &\multicolumn{1}{c}{} & \multicolumn{1}{c}{} &\multicolumn{1}{c|}{\cellcolor{green}{1}} &\multicolumn{1}{c|}{} \\ \cline{6-7} 
\multicolumn{1}{r}{Other}           & \multicolumn{1}{l}{}  & \multicolumn{1}{l}{}  & \multicolumn{1}{l}{}  & \multicolumn{1}{l}{}   & \multicolumn{1}{c|}{} & \multicolumn{1}{c|}{\cellcolor{green}{2}} \\ \cline{7-7}  
\end{tabular}
\begin{tablenotes}
\item[a] \textit{MCscore and Kappa statistic}
\end{tablenotes}
\end{threeparttable}
\end{table}

As shown in Table~\ref{tab:metricsmap}, Confusion Matrix is the most used metric in the literature\hl{. The Error Rate and PRF (Precision, Recall and $F_{1}$-score) are also popular among researchers, even though they are not appropriate metrics,} when the dataset is unbalanced since they depend upon which class is considered positive.
Among the reported metrics, AUC and Confusion Matrix (we can calculate all other metrics from the matrix) are \hl{better} choices. Of the papers that use an unbalanced dataset, several did not use the \hl{appropriate} metrics \cite{zhang:textual, aggarwal2012phishari, geng2015combating, dong2015beyond, stringhini2013shady, mohammad2014intelligent, canali2011prophiler, moghimi2016new}. The following nine reported the Confusion Matrix \cite{dong2010defending, Gowtham2014, canali2011prophiler, zhang2017two, yue2013fine, nguyen2013detecting, xu2014evasion, pao2012malicious, whittaker2010large}. One reported AUC \cite{geng2015combating}, and three reported both AUC and Confusion Matrix   \cite{Xiang:2011:CFM:2019599.2019606,gowtham2014comprehensive, marchal2016know}. 

\subsection{Selected Phishing Website Detection Literature}

Next, we present a detailed analysis of the reviewed literature that addressed one or more of the challenges described in Section~\ref{sec-challenges}.

\hl{Using unbalanced dataset is important because of the base-rate fallacy challenge}. While many papers \cite{marchal2016know, dong2010defending, pao2012malicious, xu2014evasion, lee2016phishing, popescu2016practical} used unbalanced datasets, papers  \cite{bannur2011judging,thomas2011design} addressed this issue differently. Researchers in \cite{bannur2011judging} evaluated their algorithm by changing the ratio of malicious-to-benign examples (1:2, 1:5, 1:10, 1:15 and 1:20). The results show that both precision (97.6\% to 94.2\%) and recall (96.6\% to 81.6\%) decrease with an increase in the number of legitimate websites i.e., performance drops with \hl{increase in degree of imbalance}. 
The malicious links in the dataset were collected only from spam emails, thus making it less diverse. They also did not mention the time period of the dataset they collected to train and test their method. \hl{Researchers in} \cite{thomas2011design} \hl{also observed similar results by changing the non-spam to
spam ratios (1:1, 1:4, and 1:10).}

A robust detection system should be able to thwart zero-day attacks \hl{(an attack that occurs on the same day a vulnerability is discovered)} -- thus, testing a system in a real-time environment is indispensable. The authors in \cite{thomas2011design} proposed `Monarch' as a live phishing website detection system that uses a Logistic Regression classifier trained offline using blacklists and annotation schemes -- the \textit{ground truth}. The system's training URLs were aggregated from two streams: links in emails from a mail service provider as well as shared on Twitter. These two sources capture a widely different set of features in terms of lexical properties of URLs, hosting infrastructure, and page content (HTML and links). The training setup was done using varying ratios (1:1, 1:4, 1:10) of non-spam to spam (links caught using  spam traps and/or blacklists were marked as spam, rest were labeled as non-spam). The system was tested in real-time using links extracted from the same two sources. The system was tested on 15 million URLs/day with classifier retraining every 4 days. However, the authors do not clearly discuss the ratio of spam and non-spam in their testing data stream. They note the deficiencies in their ground truth data - but do not provide any ideas for improvement.  

The active attacker poses a critical threat to phishing website detection systems. The framework proposed in \cite{xu2014evasion} analyzed the possible interaction between attacker (evasion) and defender (counter-evasion). They demonstrated that an active attacker can evade machine learning based detection models (J48 classifier trained on 124 application/network-layer features) by manipulating the features of the malicious website. Their results validated that the proposed active method can increase the false negative rate to 89.1\% by only changing 5\% of features on average. They also showed that proactive training can reduce the success rate of the attacker. The authors also claimed that their method can make the decision for a given website after a few seconds but they did not report the classification time specifically.


Periodic system retraining is another way to \hl{identify} newer types of attacks (active attacker challenge). Researchers in \cite{whittaker2010large} used a model trained (with daily retraining) on Google's phishing blacklist to update the blacklist automatically. The authors addressed multiple goals in their paper, decreasing false positives while increasing recall, handling noise in training data (inaccurate labels), and decreasing the detection latency (to smoothly process millions of web pages every day). They achieved the low latency goal by creating a pipeline system which tries to drop obvious non-phishing websites during the early stages. Their pipeline starts with extracting lightweight URL features followed by the content and hosting features. They use a novel machine learning algorithm similar to Random Forest and online learning algorithms for their system. Their model achieved 90\% precision with only 0.1\% false positive rate on a testing data of 75 million URLs collected over two weeks. However, the authors do not share details about \hl{the} steps they took to overcome the dataset's noise.
 
Marchal et al. \cite{marchal2016know} addressed three security challenges in their method: the base-rate fallacy, lack of data availability, and active attackers. They proposed a phishing detection system that uses a relatively small training data (1036 phishing and 4531 legitimate) and tested on 101,553 websites (100,000 legitimate and 1553 phishing). 
The system uses features that cannot be fully manipulated by the attacker (reinforcement against active \hl{attackers}) divided into four categories based on whether the attacker has full or partial control on the feature. For example, distribution of the terms in external links of a phishing website is completely out of the control of the attacker. 
Their method had a 99\% AUC and less than 1\% false positive rate using more than 200 features. They further grouped the features into five different categories (URL, term usage, starting and landing main level domain, registered domain name usage, and webpage content) and tested the effectiveness of each of them separately. Based on their results, URL features contributed the most (0.88 $F_1$-score) compared to others.   
Additionally, their method is language independent and was tested on a dataset of six different languages (German, English, French, Portuguese, Italian, and Spanish) to show that the performance is similar for websites with different languages. However, they did not report the training and detection times of their system. 


Applying machine learning techniques on extracted features from different sections of the website (URLs, HTML content, domains, etc.) is not enough to distinguish legitimate websites from their phishing counterparts.
Using the user's interactions with websites, SpiderWeb \cite{stringhini2013shady} builds a graph from the redirect chain of the websites and distinguishes legitimate and malicious websites by defining a threshold. Since the redirect chains are collected from a diverse set of web users, the attacker needs to make the redirect chain of the malicious website similar to legitimate websites to evade detection. This can only be done by exposing the malicious website to a larger audience which increases the chance of getting caught. The main limitation of this work lies in the complexity of redirection chains - it requires a minimum number of different redirect chains (visits from different users) that lead to the same website to make a decision. They did an experiment with 10,914 users and \hl{found} that their system detected the phishing websites after 723 users got compromised (93\% successfully blocked).

Among all the systems \hl{reviewed} here, none addressed all of the security challenges. The closest system \cite{marchal2016know} addressed three out of all the challenges mentioned in Section~\ref{sec-challenges}. A more thorough system is required to fill this gap. As a starting point for a better system, we should extract the best practices from existing works. 
Before moving to phishing email detection techniques, we summarize the lessons learned and best practices of phishing website detection techniques based on our review of the literature.

\subsection{Lessons learned and best practices}

Among the systems that used a large and unbalanced dataset in their training and testing phases, we realized that ensemble methods (bagging or boosting) are the best fit. They resulted in detection rates as high as 99\% and error rates of less than 1\%. DMOZ and Phishtank are the main datasets used in these systems as well as the author collected datasets. This can be seen \hl{as evidence} that researchers should focus more on ensemble methods \cite{liu2009exploratory}. 

Using relatively diverse dataset can make the model more robust and general. There are several works that used Alexa as their source for legitimate websites. Although Alexa's list of top websites is diverse in term of having different categories of \hl{websites, it} only contains domain names. So, there is a clear distinction between legitimate and phishing websites. One way to remove this bias is to use Alexa's list as a basis for crawling \cite{canali2011prophiler} and use crawled \hl{websites} as a dataset.

\section{Phishing/Spear Phishing Email Detection} \label{sec-tech-emails}
Email is a popular vector of phishing and spear phishing attacks. Thus, detection of these attacks is critical. 
In this section, we review 37 phishing and eight \hl{spear phishing} email detection papers. 
\subsection{Email Structure}\label{subsec-intro-email}


An email \hl{consists} of two main parts: Header fields (abbr. \textbf{Header}), and a \textbf{Body}, which is optional.\footnote{\url{https://tools.ietf.org/html/rfc2822}} Simple Mail Transfer Protocol or SMTP\footnote{\url{https://tools.ietf.org/html/rfc5321}} is the Transmission Control Protocol/Internet Protocol (TCP/IP) protocol used in sending and receiving emails. Originally, emails were only able to handle plain text messages. Multi-Purpose Internet Mail Extensions (MIME)\footnote{\url{https://tools.ietf.org/html/rfc2049}} supported new file types like JPEG and HTML.

The Header is a sequence of lines that are composed of a field name, followed by a colon (``:''), then \hl{some information}, and ends with a newline (Example VIII.1). 
Though mostly hidden from the reader, the header is very important since it contains sender's and recipients' information, an optional message-ID field, and the route (hops and SMTP relays) taken by the email. 
\textit{Authentication-results} is an important header field for spam and phishing detection and displays the results of authentication protocols such as Sender Policy Framework (SPF). SPF is an open standard specifying a method to prevent sender address forgery.\footnote{\url{http://www.openspf.org/Introduction}} Another widely used protocol is DomainKeys Identified Mail (DKIM) for email integrity.\footnote{\url{http://www.dkim.org/}} 

The email body is optional and usually contains greetings, the actual message, and a possible signature field. It also includes images and other types of attachments. The MIME extension enabled the body to be coded with non-ASCII elements like HTML or Base64. 
An email body formatted with MIME content has a \textit{MIME-Version} field  in the header.

The header and body are the main parts for extracting features from an email, so we refer to them as Header and Body Features respectively. We describe the different types of features used in email detection research in the next section.  We present the other types of features (External and Attachment related) and the papers that \hl{use} them respectively in  Appendix Table~\ref{table-email-features}.

\begin{MyBox} \label{email_example}
\scriptsize
\textbf{From:} onlinebanking@regions.com  Fri Nov 23 11:01:49 2007\\
\textbf{Return-Path:} <onlinebanking@regions.com>\\
\textbf{X-Original-To:} jose@login.monkey.org\\
\textbf{Delivered-To:} jose@login.monkey.org\\
\textbf{Received:} from mail1.monkey.org (spammy.monkey.org [152.160.49.220])
	by funky.monkey.org (Postfix) with ESMTP id 1577F469B2
	for <jose@login.monkey.org>; Fri, 23 Nov 2007 11:01:49 -0500 (EST)\\
\textbf{Received:} from mail2.monkey.org (mail2.monkey.org [204.181.64.8])\\
	by mail1.monkey.org (Postfix) with ESMTP id 06788131FF92\\
	for <jose@monkey.org>; Fri, 23 Nov 2007 11:01:49 -0500 (EST)\\
\textbf{Received:} from 80.216.6.214 by ; Fri, 23 Nov 2007 10:59:28 -0600\\
\textbf{Message-ID:} <200710511.012001@mail.monkey.org>\\
\textbf{Date:} Fri, 23 Nov 2007 11:01:47 -0500 (EST)\\
\textbf{From:} onlinebanking@regions.com\\
\textbf{To:} undisclosed-recipients:;\\
\textbf{Content-Length:} 0\\
\end{MyBox}

\subsection{Email Feature Extraction and Analysis}\label{subsec-feature-email}
 


\textbf{Header features:}
\hl{The research literature} focuses on sender related fields -- e.g., \textit{From}, \textit{Sender}, \textit{Mail From}, \textit{Reply-to}, etc., --
especially the Fully Qualified Domain Name (FQDN) 
in those entries and verifying their consistency throughout the header.
\hl{The FQDN is the complete domain name for a specific host on the internet.\footnote{https://kb.iu.edu/d/aiuv} It is made of the hostname and domain name. In the following example: \textit{``mymail.somecollege.edu"}, the hostname is \textit{mymail}, and the domain is \textit{somecollege.edu}.}

\hl{To illustrate}, an email supposedly from Amazon should not have a \textit{Reply-to} entry with an Outlook domain name. \hl{However, we caution that the sender can spoof all the header fields pertaining to sender domain and email address.} 
The \textit{Subject} is commonly analyzed \hl{(Table} ~\ref{table-email-features}\hl{)}, it being an important lure used by phishers. So researchers use different types of lexical analysis relying on Natural Language Processing (NLP) \hl{techniques} to extract features and search for blacklisted words.
Another interesting field is the \textit{Received} field. It allows emails to be tracked down because every hop the emails passes through will register its own entry (specifying the domains of the sending and receiving hosts, timestamp, etc.) \cite{hong:analysis}.
An email received from a coworker in the same building using the company's mailing server should not be coming from a host in another country.

\textbf{Body features:} Researchers use lexical analysis to detect common phrasing patterns or words, e.g., \textit{Click here}, \textit{Urgent}, \textit{Warning}, \textit{Account closure notice}, etc.
\hl{NLP techniques}  \cite{vermaH13} \hl{were} used to \hl{design}: 1) Semantic features e.g., detecting language that elicits urgency or threat \cite{verma:detecting}. 2) Stylometric features e.g., detecting writing habits and profiling users \cite{Stringhini2015,Duman:emailprofiler}. 
Researchers also extract important features from URLs included in the email body (as covered in Section \ref{sec-techniques-urls}).

In Table \ref{table:email-features-short}, we list the various features that appeared in phishing email detection research papers, as well as their size and required processing time. 
 URL features have longer extraction times (as seen in Section~\ref{url-features}) compared to ``shallow'' lexical features \hl{(e.g., word frequencies) extracted from the body}. The size of most email features is relatively small, except for features that rely on the vocabulary of the whole dataset like TFIDF \cite{salton1986introduction}. 
Weighing the features with algorithms like Information Gain (IG) makes it possible to get subsets of features that give better classification results 
\cite{Andronicus:classification,masoumeh:feature,isredzaRahmi:profiling,Ammar:online,Isredza:phishing,fergus:feature,DBLP:journals/corr/YasinA16}.

\begin{table*}[!htb]
\centering
\caption{Features used in phishing email detection. For uncommon features, we cite the paper \hl{that} best describes it.} 
\label{table:email-features-short}
\begin{threeparttable}
\begin{tabular}{|l|c|c|lcc}
\hline
\multicolumn{1}{|c|}{\multirow{2}{*}{\textbf{Body}}}     & \multicolumn{2}{c|}{Criteria}                 & \multicolumn{1}{c|}{\multirow{2}{*}{\textbf{Header}}}   &   \multicolumn{2}{c|}{Criteria}               \\ \cline{2-3}\cline{5-6}
& FS & PT &\multicolumn{1}{l|}{} & \multicolumn{1}{c|}{FS} & \multicolumn{1}{c|}{PT} \\
\specialrule{.2em}{.1em}{.1em}
\begin{tabular}[c]{@{}l@{}} Lexical Features (\# of words, \# of characters,\\
Function words, Tokens\tnote{a}, Regular expressions, etc.)\end{tabular} & M & M &\multicolumn{1}{l|}{\textit{``Message-Id''}}  &\multicolumn{1}{|c|}{S} & \multicolumn{1}{|c|}{S} 
\\ \hline
                       Style metrics (number of paragraphs in the email, Yule metric, etc.)& S & S &\multicolumn{1}{l|}{\textit{``Received''} Fields}       &\multicolumn{1}{|c|}{S}&\multicolumn{1}{|c|}{S} 
                          \\ \hline 
\begin{tabular}[c]{@{}l@{}} Topic in the body\tnote{b}     \end{tabular}  & S  & S   &\multicolumn{1}{l|}{\begin{tabular}[c]{@{}l@{}} \textit{``From''}\end{tabular} }   &\multicolumn{1}{|c|}{S}& \multicolumn{1}{|c|}{S} 
\\ \hline
                        \begin{tabular}[c]{@{}l@{}} Latent Semantic Indexing\tnote{c}   \end{tabular}    & S & S & \textit{``Mail from''}    &\multicolumn{1}{|c|}{S}& \multicolumn{1}{|c|}{S} \\ \hline
       Readability indexes\tnote{d}    & S & S  &\multicolumn{1}{l|}{\begin{tabular}[c]{@{}l@{}} \textit{``Sender''}\end{tabular} }   &\multicolumn{1}{|c|}{S}& \multicolumn{1}{|c|}{S}  \\ \hline
        \begin{tabular}[c]{@{}l@{}}NLP (Part-of-Speech tags, Named entities,\\ Wordnet properties, etc.)  \end{tabular}         & M & S &\multicolumn{1}{l|}{\begin{tabular}[c]{@{}l@{}} \textit{``Mail-to''}\end{tabular} }   &\multicolumn{1}{|c|}{S}& \multicolumn{1}{|c|}{S}  \\ \hline
       Semantic Network Analysis\tnote{e}          & S & H  &\multicolumn{1}{l|}{\begin{tabular}[c]{@{}l@{}} \textit{``Delivered To''}\end{tabular} }   &\multicolumn{1}{|c|}{S}& \multicolumn{1}{|c|}{S}  \\ \hline
                          \begin{tabular}[c]{@{}l@{}}  Vocabulary Richness\tnote{f} \end{tabular}       & S & M  & \multicolumn{1}{l|}{Authentication-Results (SPF, DKIM, etc.)}       &\multicolumn{1}{|c|}{S}& \multicolumn{1}{|c|}{S}\\ 
                          \hline
                          \begin{tabular}[c]{@{}l@{}}  Urgency, reward, threat language in the body    \end{tabular} & S & S&\multicolumn{1}{l|}{\begin{tabular}[c]{@{}l@{}} \textit{``Subject''} features (length of subject, \# of words,\\\# of characters, vocabulary richness, etc.) \end{tabular}}    &\multicolumn{1}{|c|}{S}&\multicolumn{1}{|c|}{S} \\ 
                          \hline
                          \begin{tabular}[c]{@{}l@{}} Greeting, signature, farewell in the message  \end{tabular} & S  & S & \multicolumn{1}{l|}{Blacklisted words in \textit{``Subject''}} &\multicolumn{1}{|c|}{S}&\multicolumn{1}{|c|}{S}\\ 
                          \hline
         Presence of both ``From:'' \& ``To:'' in email body &                    S  & S &\multicolumn{1}{l|}{{\# of words and/or characters in the \textit{``Send''} field}} &\multicolumn{1}{|c|}{S}&\multicolumn{1}{|c|}{S} \\
                         \hline
       \# of linked to domains &      S  & S     &\multicolumn{1}{l|}{\textit{``Sender''} domain $\neq$ \textit{``Reply-to''} domain}      &\multicolumn{1}{|c|}{S}&\multicolumn{1}{|c|}{S} \\
             \hline
           URL features      & M & M &\multicolumn{1}{l|}{ \begin{tabular}[c]{@{}l@{}}\textit{``Sender''} domain $\neq$ \textit{``Message-Id''} domain\end{tabular}}                        &\multicolumn{1}{|c|}{S}&\multicolumn{1}{|c|}{S}\\ \hline
                        HTML features  & M & S&\multicolumn{1}{l|}{ \begin{tabular}[c]{@{}l@{}}\textit{``Sender'' / ``From''} $\neq$  email's modal domains \end{tabular}}                        &\multicolumn{1}{|c|}{S}&\multicolumn{1}{|c|}{S} \\ 
     \hline 
                      Presence and/or \# of forms in email body& S & S &\multicolumn{1}{l|}{Timestamp, \textit{``Sent date''}}       &\multicolumn{1}{|c|}{S}&\multicolumn{1}{|c|}{S} \\
                       \hline
                          Blacklisted words in the message &  M & M  &\multicolumn{1}{l|}{\begin{tabular}[c]{@{}l@{}} Source IP, Autonomous System Number\end{tabular}}                              &\multicolumn{1}{|c|}{S}& \multicolumn{1}{|c|}{M} \\ \hline
            Scripts/JavaScripts features in email body&  M  & S   &\multicolumn{1}{l|}{\textit{``Subject''}: Fwd, Reply}        &\multicolumn{1}{|c|}{S}& \multicolumn{1}{|c|}{S} \\ \hline         
                \# of \textit{onClick} events in email body &      S  & S  &\multicolumn{1}{l|}{Interaction habits\tnote{g}}                        &\multicolumn{1}{|c|}{S}& \multicolumn{1}{|c|}{M}                             \\ \hline
                          
                          \begin{tabular}[c]{@{}l@{}}Feature from images/logos in the message\end{tabular} &      S & S       &\multicolumn{1}{l|}{\textit{``Cc''}, \textit{``BCc''} fields}                        &\multicolumn{1}{|c|}{S}& \multicolumn{1}{|c|}{S} \\ \hline                                                             
                          \begin{tabular}[c]{@{}l@{}}Mention of the sender \end{tabular} &              S & S                 &\multicolumn{1}{l|}{  \begin{tabular}[c]{@{}l@{}}  \textit{``X-Mailer''} \end{tabular} }                        &\multicolumn{1}{|c|}{S}& \multicolumn{1}{|c|}{S}
                          \\ \hline

                          \begin{tabular}[c]{@{}l@{}}\# of links \& images links in email body  \end{tabular} &                      S   & S   &\multicolumn{1}{l|}{  \begin{tabular}[c]{@{}l@{}}  \textit{``X-Originating-IP''} \end{tabular} }                        &\multicolumn{1}{|c|}{S}& \multicolumn{1}{|c|}{S}
                          \\ \hline
               
                          \begin{tabular}[c]{@{}l@{}}\# of tables in email body \end{tabular} &                      S   & S           & 
                       \multicolumn{1}{l|}{\textit{``X-Originating-hostname''}}                        &\multicolumn{1}{|c|}{S}& \multicolumn{1}{|c|}{S} 
                            \\ \hline
                          \begin{tabular}[c]{@{}l@{}}Recipient's email address in email body \end{tabular} &                    S & S            &\multicolumn{1}{l|}{  \begin{tabular}[c]{@{}l@{}}  \textit{``X-spam-flag'} \end{tabular} }                        &\multicolumn{1}{|c|}{S}& \multicolumn{1}{|c|}{S}  
                            \\ \hline
                           \begin{tabular}[c]{@{}l@{}} Phishing terms weight\tnote{h} \end{tabular} &       L & M            &\multicolumn{1}{l|}{  \begin{tabular}[c]{@{}l@{}}  \textit{``X-virus-scanned'} \end{tabular} }                        &\multicolumn{1}{|c|}{S}& \multicolumn{1}{|c|}{S}      
                         \\ \hline
                          \begin{tabular}[c]{@{}l@{}} TFIDF \end{tabular} &     L & S   
                          \\ \cline{1-3}
                          Email size &                      S  & S           &                            \\ \cline{1-3}
                          \begin{tabular}[c]{@{}l@{}}\# of email body parts\end{tabular} &                    S  & S           &                           
                            \\ \cline{1-3}
                          \begin{tabular}[c]{@{}l@{}} MIME Version or Content-Type\tnote{i} \\ \end{tabular} &                    S  & S           &    
                           \\ \cline{1-3}
                          \begin{tabular}[c]{@{}l@{}}JavaScript PopUp Windows\end{tabular} &                    S & S           &              
                            \\ \cline{1-3}
                          \begin{tabular}[c]{@{}l@{}} Link displayed $\neq$ Link in destination  \end{tabular} &                    S  &M           &     
                          \\ \cline{1-3} 
                          \begin{tabular}[c]{@{}l@{}} Img links $\neq$ spoofed target address\end{tabular} &                    S & M           &     
                          \\ \cline{1-3} 
                          \begin{tabular}[c]{@{}l@{}} Hidden text in the email, Salting techniques\tnote{j} \end{tabular} &                    S & S           &     
                          \\ \cline{1-3}  
                           \begin{tabular}[c]{@{}l@{}}\# of internal/xxternal links \end{tabular} &                    S  & S           &     
                          \\ \cline{1-3}  
\end{tabular}
\begin{tablenotes}
\item[a] Semantic relations can also be used to add tokens
\item[b] Words that tend to appear together in emails \cite{bergholz2008improved}
\item[c] Statistical technique for the analysis of two-mode and co-occurrence data.
\item[d] quantitative description of the text writing and organisation style \cite{Han:2016:ASP:2851613.2851801}.
\item[e] Use Semantic Network Analysis to extract and create \hl{the} feature vector as explained in \cite{kim:semantic}
\item[f] The ratio of the number of words to the number of
characters in the document. \cite{fergus:feature}
\item[g] User's history of conversations and list of contacts \cite{Stringhini2015}
\item[h] Terms that have the highest term frequency in the phishing dataset \cite{DBLP:journals/corr/YasinA16}.
\item[i] Multipart/Alternative, text/plain, text/html
\item[j] Adding or distorting content not perceivable by the reader \cite{adre:new}
\end{tablenotes}
\end{threeparttable}
\end{table*}

\begin{figure*}[!htb]
\centering
 \includegraphics[width=\linewidth]{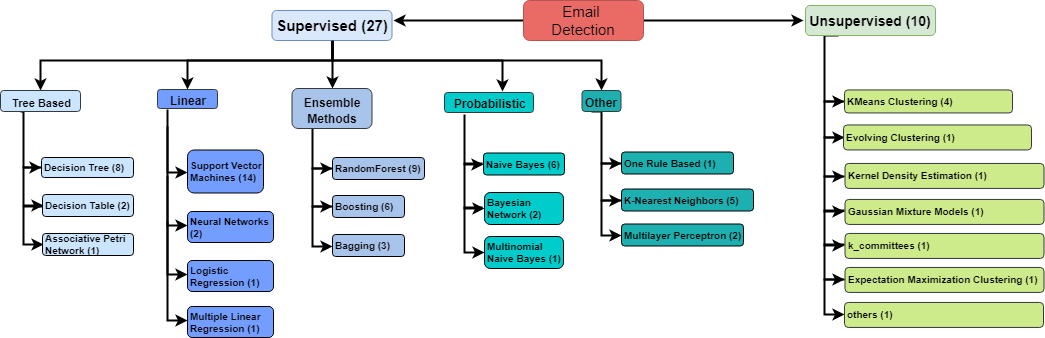}
 \caption{Algorithms used in phishing email detection. The number of papers using the method is in parentheses.}  
 \label{fig:emailmethodsdist}
\end{figure*}

\subsection{Detection Methods}
Figure~\ref{fig:emailmethodsdist} shows the classification and clustering methods used in the phishing email detection literature and the number of papers that employed them. \hl{It} shows \hl{a preference for} Support Vector Machine (SVM) and Random Forest (RF). The number of research papers \hl{using} supervised algorithms is higher than the number \hl{using}  unsupervised algorithms. 
More details and information about the methods used by each paper can be found in Table~\ref{table-email-classification} in the appendix.
As previously mentioned in Section \ref{sub-website-detection-methods}, the lack of diverse and new datasets gives a false impression or interpretation of supervised classifiers' results. Some researchers  have tackled the lack of labeled datasets by using unsupervised algorithms to cluster or create profiles of users' writing habits\cite{Duman:emailprofiler,Stringhini2015}. 

\begin{table*}[h]
\centering
\caption{\hl{Legitimate and phishing} email datasets: \hl{\textit{sources and availability}}}
\label{table-email-dataset}
\begin{threeparttable}
\begin{tabular}{|c|c|c|c|c|c|c|c|}
\hline
\diagbox[width=2.0cm]{\textbf{Phish}}{\textbf{Legit.}}    & \begin{tabular}[c]{@{}c@{}}\textbf{SpamAssassin}\\(pub)\end{tabular} &  \begin{tabular}[c]{@{}c@{}}\textbf{Enron}\\(pub)\end{tabular} & \begin{tabular}[c]{@{}c@{}}\textbf{Author's}\\(pvt)\end{tabular}  &\begin{tabular}[c]{@{}c@{}}\textbf{Misc}\tnote{a}\end{tabular} & \begin{tabular}[c]{@{}c@{}}\textbf{N/A}\end{tabular} \\

\specialrule{.2em}{.1em}{.1em}
Nazario (pub)        & \begin{tabular}[c]{@{}c@{}}\cite{Isredza:phishing,verma:phish,fette2007learning,khonji:lexial,Figueroa2017,isredzaRahmi:profiling,Ammar:evolving,Ammar:online,DBLP:journals/corr/YasinA16,bergholz2008improved,toolan:ensembles, fergus:feature,masoumeh:feature,Andronicus:classification}
\end{tabular}           &\cite{vermaH13}    & \cite{Ammar:online,verma:detecting}
& &        
\\ \hline
APWG (pub)             &          & \multicolumn{1}{c|}{}                           & \multicolumn{1}{c|}{}   &  & \cite{kim:semantic}\\ \hline
Author's (pvt) & \cite{Ammar:online} & \cite{Stringhini2015} & \begin{tabular}[c]{@{}c@{}}\cite{chowdhury:multilayer,adre:new,Stringhini2015,orman:semantics,Olivo:2013,chandrasekaran2006phishing,huang2017gossip,Dazeley2010,Chiang:2015:AMC:2778890.2779103}\end{tabular} 
& & 
\\ \hline
Miscellaneous\tnote{b} & \cite{Islam:2013:MPD:2405859.2406232,ramanathan} & &  & \cite{DBLP:conf/ecir/GanstererP09}
& \cite{Seifollahi2017,yearwood:profiling,john:applying} 
\\ \specialrule{.2em}{.1em}{.1em}
Malware/Spam          &\multicolumn{1}{l}{}  &\multicolumn{1}{l}{}  &\multicolumn{1}{l}{}  &\multicolumn{1}{l}{}  &        \\
\hline

Author's (pvt) & 
& \multicolumn{1}{c|}{\cite{Stringhini2015}}                      & \cite{adre:new,Stringhini2015,Chiang:2015:AMC:2778890.2779103}
& &     \\ \hline
Miscellaneous\tnote{c}   &           &                       &  &     \cite{DBLP:conf/ecir/GanstererP09}  & \begin{tabular}[c]{@{}l@{}} 
\end{tabular}  \\ \hline
\specialrule{.2em}{.1em}{.1em}
\hl{Spear Phishing}     &\multicolumn{1}{l}{}  &\multicolumn{1}{l}{}  &\multicolumn{1}{l}{}  &\multicolumn{1}{l}{}  &
\\ \hline
Author's (pvt) & \multicolumn{1}{c|}{}                    &\cite{Stringhini2015}                      & \cite{Duman:emailprofiler,Stringhini2015,Han:2016:ASP:2851613.2851801,Ho2017DetectingCS}
&    &  \\ \hline
UCI repo & \multicolumn{1}{c|}{}                    &\cite{Laszka:2016:MSF:3015812.3015893,Laszka:2015:OPF:2887007.2887140}                      & 
&    &  \\ \hline
N/A   & \multicolumn{1}{l|}{}           &              &  & &\cite{Mengchen:optimizing}
\\ \hline
\specialrule{.2em}{.1em}{.1em}
\end{tabular}
\begin{tablenotes}
\item[a] \textit{Authors gathering emails from various sources like companies, Phishery (dead source) TREC \cite{DBLP:conf/ecir/GanstererP09}}
\item[b] \textit{Major Australian Bank \cite{Seifollahi2017,yearwood:profiling,john:applying}, PhishingCorpus \cite{Islam:2013:MPD:2405859.2406232,ramanathan}, SPAM Archive \cite{ramanathan}}
\end{tablenotes}
\end{threeparttable}
\end{table*}

\subsection{Dataset properties}\label{subsubsec-dataset-email}
\subsubsection{\textbf{Dataset sources and availability}}

\begin{figure}[!htb]
\centering
 \includegraphics[width=0.7\linewidth]{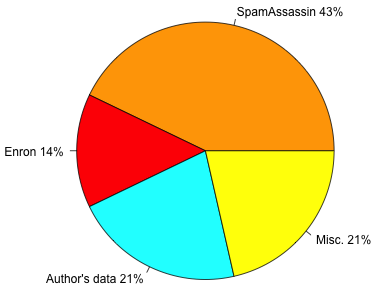}
 \caption{Distribution of papers \hl{based on} legitimate email source}  
 \label{fig:legiturldist_email}
 \end{figure}

\begin{figure}[!htb]
\centering
 \includegraphics[width=0.75\linewidth]{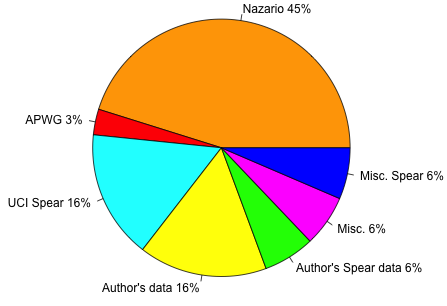}
 \caption{Distribution of papers \hl{based on} phishing email source}  
 \label{fig:phishurldist_email}
 \end{figure}

We describe the different dataset sources used in phishing email detection literature in Table~\ref{table-email-dataset}. Figures~\ref{fig:legiturldist_email} and~\ref{fig:phishurldist_email} show the distribution of these sources. It is clear that the most popular datasets for phishing and legitimate emails are Nazario\footnote{\url{https://monkey.org/~jose/phishing/}} and SpamAssassin\footnote{\url{http://www.csmining.org/index.php/spam-assassin-datasets.html}} respectively. 
We also include sources for malware, spam, and \hl{spear phishing} datasets in Table~\ref{table-email-dataset}.


Researchers need to be aware that publicly available datasets can be sanitized. For example, the Enron dataset\footnote{\url{https://www.cs.cmu.edu/~enron/}} and the SpamAssassin dataset have some obfuscated addresses and domains. Moreover, many of the emails in the Enron dataset only have partial headers (missing a number of fields) instead of full headers. 


An increasing number of authors used their own private datasets or worked with companies to get their email logs and archives.
It has the benefit of training and testing models  on realistic and relatively new emails. However, it raises the issue of reproducibility and comparisons of systems. Although the objective is to release a new system with \hl{a} high detection rate, it would also help the field of research if authors \hl{shared} new datasets. That being said, we are aware of issues related to sensitive information and \hl{realize} that sharing can be difficult.

\subsubsection{\textbf{\hl{Dataset Size}}}
As mentioned above, the ratio of benign to malicious samples in phishing detection is very important. Therefore, we report the \hl{number} of the papers that used different combinations of dataset sizes in Table~\ref{tab-sizedatasets-emails}. 
 \hl{It} also shows the sizes of the phishing and legitimate dataset  sizes that are commonly used together. 
We notice from this table that most authors use relatively balanced ratios (e.g., 4 to 6 or 4.5 to 5.5). 
Examples of research in the literature that used unbalanced datasets with different ratios are \cite{Andronicus:classification,Isredza:phishing, isredzaRahmi:profiling,adre:new,vermaH13}.

\subsubsection{\textbf{Dataset Diversity}}~\label{sec:email-data-div}
Another concern is the dataset diversity. Therefore, we analyze the email body content diversity of the most used datasets: SpamAssassin, Enron, and Nazario.
With this aim, we extract all the text from the email body then we remove all the HTML tags, CSS, and header of the emails. We also filter out all stop words to remove uninformative words. Then we convert the extracted text into vectors using TFIDF representation. The SpamAssassin dataset includes both spam and ham emails. The Inverse Document Frequency was computed using all the emails. After getting the vectors, we use Cosine Similarity (Equation \hl{(}\ref{eq:cosine}\hl{)}) to measure the similarities. The Dataset sizes were 10,745 for SpamAssassin, 156,070 for Enron (Sent folder), and 8,551 for Nazario. Table~\ref{table-similarity-distribution-emails} shows the percentage of datasets with a specific range of similarities 
\hl{It shows} that the datasets have diverse email body content\hl{:} 90.91\% of emails in Nazario dataset and more than 97\% in SpamAssassin and Enron datasets have less than 10\% similarity. 


\begin{table}[!htb]
\centering
\caption{Size ranges of phishing/legitimate email datasets.} 
\label{tab-sizedatasets-emails}
\begin{tabular}{|c|c|c|c|c|c|}
\hline
                \textbf{Type}           &         \multicolumn{5}{c|}{\textbf{Legitimate}}                                                                                       \\ \hline
                             \textbf{ Phishing }      & \textbf{100s} & \textbf{1,000s}  & \textbf{10,000s} & \textbf{100,000s} & \textbf{N/A}      \\ \specialrule{.2em}{.1em}{.1em}
   \textbf{100s }    & \cellcolor{yellow}{5} & \cellcolor{yellow}{5} &     & 
   &\\ \hline 
                           \textbf{1,000s}   & &\cellcolor{red}{9} & \cellcolor{yellow}{4} &  &\cellcolor{red}{7}  
                           \\ \hline   \textbf{10,000s}  &   & \cellcolor{green}{1}   & \cellcolor{green}{2} &  & \\
                          \hline
                          \textbf{N/A} &  & \cellcolor{green}{2} & \cellcolor{green}{1}&\cellcolor{green}{1}  &\\ 
                          \hline

\end{tabular}
\end{table}


\begin{table}[h]
\centering
\caption{Percentage of emails with different ranges of similarities in each dataset. Columns specify the ranges of the Cosine Similarity in percentage.}
\label{table-similarity-distribution-emails}
\begin{threeparttable}
\begin{tabular}{|c|c|c|c|c|c|c|}
\hline
\multirow{2}{*}{\textbf{Dataset}} & \multicolumn{6}{c|}{\textbf{Ranges}}                              \\ \cline{2-7} 
                         & [0-10]  & (10-20] & (20-30] & (30-40] & (40-50] & \textgreater{}50 \\ \hline
SA\tnote{a}                    & \textbf{98.15} & 1.45  & 0.24  & 0.061  & 0.02 & 0.04         

\\ \hline
Nazario                & \textbf{90.91} & 6.60 & 1.32 & 0.53  & 0.26 & 0.33           

\\ \hline
Enron   & \textbf{97.16} & 2.56 & 0.18  & 0.04 & 0.01 & 0.02             \\ \hline

\end{tabular}
\begin{tablenotes}
\item[a] \textit{SA} - Spam Assassin 
\end{tablenotes}
\end{threeparttable}
\end{table}

\subsubsection{\textbf{Recency of data collection}}
Nazario, SpamAssassin, Enron, \hl{and} UCI datasets are made of old 
emails (early 2000s), except for Nazario who released new emails annually from 2015-17. This could explain why some authors prefer to rely on private sources, e.g., personal emails or companies' datasets.
Table~\ref{table-email-time-period} describes the recency of the datasets in relation to the papers' publication dates.
From the table, we conclude that almost 70\% of the papers \hl{used} old datasets. Thus\hl{,} we recommend  \hl{using} recent sources of emails if possible. For example, the Wikileaks website holds archives with thousands of recent legitimate emails from varied sources e.g., Clinton emails, Hacking Team, Sony, etc. An interesting source for phishing emails are Phishbowls, which are a frequently updated collection of emails with full or partial header maintained by universities e.g., Cornell IT Phishbowl.\footnote{\url{https://it.cornell.edu/phish-bowl}} We refer to \cite{ELAetAl:IWSPA-AP2018} where the authors built a dataset with emails from Wikileaks archives and Phishbowls from different universities.

\begin{table}[h]
\centering
\caption{Recency of email datasets used in evaluation. \hl{Papers} in \textbf{bold} use data collected from multiple \hl{periods}.}
\label{table-email-time-period}
\begin{tabular}{|c|c|l|}
\hline
\textbf{Recency in year(s)}      & \textbf{N}     & \textbf{Literature}                                                                                              \\ \hline \hline
\textbf{Real time}   &       3   & \begin{tabular}[c]{@{}l@{}}\cite{Duman:emailprofiler,husak:phigaro}
\textbf{\cite{Ho2017DetectingCS}}
\end{tabular}\\
\hline
\textbf{1}       &   5   & \begin{tabular}[c]{@{}l@{}}\textbf{\cite{Ammar:online,verma:phish}} \cite{yearwood:profiling,chowdhury:multilayer,kim:semantic} 
\end{tabular}\\ \hline
\textbf{2-3}& 5 & \begin{tabular}[c]{@{}l@{}}\cite{adre:new,DBLP:conf/ecir/GanstererP09,bergholz2008improved} \textbf{\cite{toolan:ensembles,huang2017gossip}}\\  
%
\end{tabular}                                                       \\ \hline
\textbf{3-4}      &   7  & \begin{tabular}[c]{@{}l@{}}\textbf{\cite{khonji:lexial,Isredza:phishing}}\cite{john:applying}\\\textbf{\cite{Han:2016:ASP:2851613.2851801}} \textbf{\cite{toolan:ensembles,huang2017gossip,fergus:feature}} \end{tabular} 
\\ \hline
\textbf{$\geq$ 4}  &      26   & \begin{tabular}[c]{@{}l@{}}\cite{Figueroa2017,fette2007learning,isredzaRahmi:profiling,Ammar:evolving}
\\\cite{Stringhini2015,Andronicus:classification,verma:detecting,masoumeh:feature,DBLP:journals/corr/YasinA16,Olivo:2013}
\\\cite{ramanathan,Laszka:2015:OPF:2887007.2887140,Laszka:2016:MSF:3015812.3015893,Seifollahi2017,vermaH13,Chiang:2015:AMC:2778890.2779103,Islam:2013:MPD:2405859.2406232}\\
\textbf{\cite{huang2017gossip,Ho2017DetectingCS,Han:2016:ASP:2851613.2851801,khonji:lexial,fergus:feature,Ammar:online,Isredza:phishing}}
\\
\textbf{\cite{verma:phish,toolan:ensembles}}
\end{tabular}                                                          \\ \hline
\end{tabular}
\end{table}



\subsection{Evaluation Metrics} \label{sec:email-metric}
We list in Table~\ref{tab:metricsmap-email} the metrics and the number of papers that used \hl{up to two of} these metrics in their \hl{evaluations}.
Information about papers that cover more than two metrics can be found in Table \ref{table-email-metrics} of the Appendix. We note that 7 papers out of 37 did not mention the metrics \hl{they used}.
The most common evaluation metrics used in the phishing email detection literature are Accuracy (Acc), Precision, Recall, and $F_1$-score (PRF), and the confusion matrix (CMx). 
As mentioned before, Precision, Recall, $F_{1}$-score and Accuracy are unsuitable for evaluating unbalanced datasets. Hence we recommend the use of metrics like AUC and the confusion matrix values instead.

In the \hl{surveyed} phishing email detection literature, only one paper used G-mean \cite{Chiang:2015:AMC:2778890.2779103}, 13 reported the confusion matrix, and seven reported AUC.


\begin{table}
\centering
\caption{Distribution of evaluation metrics in phishing email detection.} 
\label{tab:metricsmap-email}
\begin{threeparttable}
\begin{tabular}{rcccccc}
             & Acc    & PRF   & ErR    & CMx   & AUC                    & Other\tnote{a}                    \\ \cline{2-7} 
\multicolumn{1}{r|}{Acc}           & \multicolumn{1}{c|}{\cellcolor{red}{14}} & \multicolumn{1}{c|}{\cellcolor{yellow}{4}} & \multicolumn{1}{c|}{\cellcolor{green}{1}} & \multicolumn{1}{c|}{\cellcolor{yellow}{3}} & \multicolumn{1}{c|}{\cellcolor{yellow}{3}} & \multicolumn{1}{c|}{\cellcolor{green}{2}}
\\ \cline{2-7} 
\multicolumn{1}{r}{PRF}           & \multicolumn{1}{l|}{}  & \multicolumn{1}{c|}{\cellcolor{red}{14}} & \multicolumn{1}{c|}{\cellcolor{green}{1}}  & \multicolumn{1}{c|}{\cellcolor{red}{8}} & \multicolumn{1}{c|}{\cellcolor{yellow}{4}}  & \multicolumn{1}{c|}{\cellcolor{green}{1}} 
\\ \cline{3-7} 
\multicolumn{1}{r}{ErR}           & \multicolumn{1}{c}{}         & \multicolumn{1}{c|}{} & \multicolumn{1}{c|}{\cellcolor{green}{2}}  & \multicolumn{1}{c|}{\cellcolor{green}{1}}  & \multicolumn{1}{c|}{} & \multicolumn{1}{c|}{}
\\ \cline{4-7} 
\multicolumn{1}{r}{CMx}           & \multicolumn{1}{l}{}  & \multicolumn{1}{l}{}  & \multicolumn{1}{l|}{}  & \multicolumn{1}{c|}{\cellcolor{red}{14}}  & \multicolumn{1}{c|}{\cellcolor{yellow}{3}} & \multicolumn{1}{c|}{\cellcolor{green}{1}}\\ \cline{5-7} 
\multicolumn{1}{r}{AUC} &\multicolumn{1}{c}{} &\multicolumn{1}{c}{} &\multicolumn{1}{c}{} & \multicolumn{1}{c}{} &\multicolumn{1}{c|}{\cellcolor{red}{7}} &\multicolumn{1}{c|}{\cellcolor{green}{1}} \\ \cline{6-7} 
\multicolumn{1}{r}{Other\tnote{a}}           & \multicolumn{1}{l}{}  & \multicolumn{1}{l}{}  & \multicolumn{1}{l}{}  & \multicolumn{1}{l}{}   & \multicolumn{1}{c|}{} & \multicolumn{1}{c|}{\cellcolor{yellow}{4}} \\ \cline{7-7}  
\end{tabular}
\begin{tablenotes}
\item[a] \textit{One-Error, Coverage and Average Precision \cite{yearwood:profiling}, G-Mean \cite{Chiang:2015:AMC:2778890.2779103}, Rand Index (RI) \cite{Seifollahi2017}, Root Mean Square Error (RMSE) \cite{Ammar:evolving}}
\end{tablenotes}
\end{threeparttable}
\end{table}





\subsection{Selected Email Detection Literature}
We now take a deeper look at the phishing email detection literature that addressed at least some of the security challenges mentioned in Section~\ref{sec-techniques}. Then we summarize the lessons learned and recommend some good practices.

In a realistic phishing scenario, the number of phishing emails is considerably lower than legitimate ones. Therefore, \hl{the} base-rate fallacy is a critical issue that needs to be considered. We notice that researchers stopped using a 1:1 ratio. 
 However, the ratios used are still not realistic. We cite for example \cite{isredzaRahmi:profiling}, where the authors used a dataset made of 4,150 legitimate and 2,687 phishing emails from SpamAssassin and Nazario respectively. Masoumeh et al. \cite{masoumeh:feature} relied on the same sources but used different ratios: 1,700 ham and 1,000 phishing emails. An instance of a realistic ratio is \cite{adre:new}, where the authors collected  36,364 ham emails and 3,636 phishing emails from private sources. 

The lack of new datasets is an important security challenge. As we concluded previously from Table~\ref{table-email-time-period}, researchers still employ old data. If we \hl{consider} the \hl{evolution} of phishing attacks, then using emails from old sources to train and test classifiers is a problem. Authors who work with companies use their private email logs collected from recent months or years for their research. An example of such papers is \cite{yearwood:profiling} where the authors gathered \hl{2,048} phishing emails from an Australian bank in the span of 5 months. Another example would be \cite{Ammar:online} where the authors collected from their own private server 2,000 legitimate and 300 phishing emails in the same year the paper was published. We also cite \cite{chowdhury:multilayer} where the authors collected 1,492 (986 legitimate 506 \hl{phishing}) emails in the span of 3 months, almost a year before the paper was published. However, the use of private datasets exacerbates the problem of comparing systems.

Testing the robustness of the phishing email detection system is a good prevention mechanism against the active attacker.
Authors test for robustness in different ways. One way would be to train the phishing email detector on one subset of the dataset, but test it on another subset that was collected in a later time span, e.g., \cite{adre:new}. The overall performance was a bit lower but they still achieved an $F_1$-score of 98.66\% compared to the original $F_1$-score of 99.89\%. The same authors also added 20,000 emails to their dataset and considered spam and phishing as an unwanted class, and then tested an unchanged model. They achieve an  $F_1$-score of 99.48\%.
Another way to test for robustness and how the classifier would perform against \hl{zero-day attacks} is to train the model on one dataset and test it on emails from a different dataset. This was done, for example, in \cite{Ammar:online}. The authors trained their classifiers on Nazario and SpamAssassin datasets, then tested on emails collected from their private servers. The performance only dropped by almost 1\% from 99\% to 98\%. Legitimate datasets were varied in \cite{vermaH13}. They trained only on a subset of Enron Inbox emails and then tested on a different subset of Enron Inbox emails as well as a subset of Enron Sent emails. 
A few papers also leverage semantics to increase the robustness of the features and increase the performance of classifiers. The following \hl{papers} \cite{vermaH13,verma:detecting,Seifollahi2017,ramanathan,DBLP:journals/corr/YasinA16} reported the use of WordNet\footnote{\url{https://wordnet.princeton.edu/}} to enrich the textual features. Wordnet is a large lexical dataset for English and is used to identify the semantic relationships of the tokens used as features.

An issue that we encountered in the reviewed literature was the ability to perform  \hl{a} comparative analysis of systems. Indeed, few researchers report the training and testing times of their models and/or the \hl{details} of the machines used to run those models. \hl{Exceptions include} \cite{Ammar:online,Ammar:evolving,adre:new}. 


In the next subsection, we focus on the research done \hl{in spear phishing} detection. The readers will notice that it is \hl{shorter} compared to \hl{email detection}. \hl{The reason is the paucity} of research \hl{on this topic}.

\subsection{Spear phishing}\label{subsec:spearphishing}
Spear phishing emails are malicious emails that are targeted towards an individual, \hl{or} a specific group of people\hl{,} or a specific organization. It is called targeted because it contains information sensitive to the intended victim.
Hackers use spear phishing attacks not only to steal money or personal information but also to install malware or obtain credentials of an account with higher privileges.\hl{\footnote{\hl{https://www.csoonline.com/article/3334617/what-is-spear-phishing-why-targeted-email-attacks-are-so-difficult-to-stop.html}}}

Research covering spear phishing emails is still small compared to \hl{research on} phishing emails, URLs, and websites. Based on the attributes mentioned in Section~\ref{sec-techniques}, we analyzed eight papers. Here, we highlight the \hl{spear phishing email literature that} addressed \hl{at least some of the} security challenges mentioned in Section~\ref{sec-challenges}\hl{.}
The base-rate fallacy is even more of an issue for spear phishing emails since they are \hl{scarce} compared to even normal phishing emails. 
An example of research that used \hl{an imbalanced} dataset is Han et al. \cite{Han:2016:ASP:2851613.2851801}. The emails were collected from Symantec's enterprise email scanning services and contained 1,467 spear phishing emails from eight different known campaigns and 14,043 benign emails collected between 2011 and 2013. Another example is Stringhini et al.\cite{Stringhini2015}, where they used the Enron Corpus and three email datasets from a large security company (that were not detected by spam-filtering software) to collect and work on 43,274 spam, 17,473 phishing, 546 \hl{spear phishing} emails. \hl{The} last example is Trang Ho et al. \cite{Ho2017DetectingCS}. The authors worked on a dataset from Lawrence Berkeley National Lab (LBNL), which consisted of 372 million emails and event logs gathered since 2013. The security team of LNBL and the method developed by the researchers only discovered 21 spear phishing emails combined. This \hl{is} an example of a realistic ratio of benign to spear phishing emails.

The time-scale of \hl{spear phishing} attacks \hl{are} short, thus the need for real-time detection systems.
Trang Ho et al. \cite{Ho2017DetectingCS} exploited domain and sender reputation features and created a new anomaly detection technique. 
Their method scores a click-in-email event by generating the feature vector and comparing it with previous events using a time window (e.g., three or six months). Their method was able to detect 17 out of 19 spear phishing campaigns in addition to 
two others that were previously undetected by LBNL. 
Another live detection system was made by Stringhini et al. \cite{Stringhini2015}. They built a system that is based on senders' writing habits. User profiles were built using writing habits (occurrence of characters, functional and special words, style metrics, etc.), composition and sending habits and interaction habits (social network of the user). After the system is trained using SVM with SMO (Sequential Minimal Optimization), it launches an identity-verification process when \hl{a received} email is considered suspicious. Han et al. \cite{Han:2016:ASP:2851613.2851801} attempt to improve upon the \hl{time-consuming} task of manually analyzing datasets of emails to detect which are spear phishing. They used a graph-based semi-supervised learning framework to differentiate between benign, known and unknown spear phishing campaigns. The dataset contained sets of labeled spear phishing campaigns and unlabeled emails. 
They achieved a 0.9 $F_1$-score with a 0.01 false positive rate in detecting spear phishing emails that belong to previously known campaigns. However, $F_1$-score may not be the best metric to use for imbalanced datasets. We encourage authors to evaluate their system using the metrics suggested in Section \ref{sec:email-metric}.

We summarize \hl{next} the lessons learned from reviewing the phishing and \hl{spear phishing} email detection literature.

\subsection{Lessons learned and best practices}\label{sub-email-best-practices}
There is still much to be done to enhance the research practices in the \hl{phishing and spear phishing email} detection fields. Researchers should give more attention to the reproducibility of their work by specifying: the \hl{architecture} of the machines that run the training and testing models; the training and testing times; and sharing the datasets used or possible ways to collect them (if not available already). 
Research using balanced datasets is decreasing. 
However, the ratios of \hl{legitimate  to phishing emails} used in the literature are still not realistic. Researchers can start by using data with skewed ratios between legitimate and malicious classes and evaluate their systems using metrics made for unbalanced datasets. 
To detect \hl{zero-day attacks} 
and reinforce detection systems against active attackers, \hl{more research is needed on periodic retraining} on new datasets and \hl{on} real-time detection mechanisms.

\section{Opportunities for Future Research}~\label{sec:gik-uw}
A careful review of the literature on phishing URL, website and email detection from the security challenges perspective reveals multiple opportunities for future contribution, which are discussed below.
\begin{itemize}
\item \textit{Dataset Issues:} We observe a \textit{lack of good dataset sources and \hl{a} dearth of diversity in the data used for evaluation of detection systems.} PhishTank and APWG are popular sources of phishing URLs used in URL and website detection; DMOZ (now deprecated) and Alexa's list of top websites are common sources of legitimate links. However, Alexa \textit{only} provides legitimate domain names, which do not capture the true nature and variety of complete URLs that a user can come across in a real-world scenario. Only a few authors go the extra mile to include a detailed description of how the data was collected~\cite{amrutkardetecting}. 


Along with its diversity, \textit{recency of the data} must also be taken into account during evaluation. Newer types of phishing URLs are submitted to crowd-sourced sites like APWG, PhishTank. But \hl{the} use of deprecated sources \hl{such as} DMOZ, Yahoo Directory, etc., renders the proposed system useless to newer attacks. 
In email detection, we notice a clear affinity towards using old data sources like Nazario, SpamAssassin, and/or Enron. To ensure diversity, authors also use private emails or emails collected from companies. Although, a better option (more recent and realistic), using private sources does not help in the clear shortage of publicly available datasets. 

Another major issue observed is the \textit{availability of quality data}. Publicly available phishing URLs (PhishTank) may have dead links, presence of duplicates, incomplete links, etc. 
Organizations and individuals may encounter more sophisticated attacks in \hl{the} form of emails or websites, which could evade detection by humans and by classifiers. A number of papers also use phishing as well as \hl{spear phishing} emails collected by them, which are not publicly available. Such URLs or emails are usually not submitted to crowd-sourced sites and could represent harder to detect attacks.  



\item \textit{Evaluation metrics:} Use of \textit{accuracy} as an evaluation metric is not desirable. It does not capture the true
performance of the classifier in the case of unbalanced data. Moreover, it gives a false sense of optimism (base-rate fallacy). A better evaluation would include metrics of performance on both classes or cost-weighted scores of classification like Matthew's Correlation Coefficient (MCC)~\cite{bekkar2013evaluation}, balanced detection rate~\cite{ELAetAl:IWSPA-AP2018}, etc. Using appropriate metrics like MCC, balanced detection rate, precision, recall, etc., for evaluating performance on unbalanced datasets  can also tackle the base-rate fallacy issue.

\item \textit{Retraining the systems:} System retraining is  important for adapting detection models based on machine learning methods against an active attacker. However, much of the relevant literature does not mention \hl{the} retraining of classification models trained on features extracted from older websites, URLs, and/or emails (malicious and benign). This is a major concern since the pre-trained detection models are most likely to fail when phishers change the type and nature of the malicious websites/URLs and URLs embedded in emails,  thus changing the attributes that the classifier \hl{depends upon for its judgment}. We refer readers to  papers, e.g.,~\cite{whittaker2010large,liang2016cracking,ma2011learning}, which describe techniques/periodicity for retraining. 

\item \textit{Generalization experiment:} Common techniques for training and testing the data involve five-fold or ten-fold cross-validation or in some cases, splitting the original dataset into training (usually 70\%-60\%) and testing (usually 30\%-40\%) data. But in such cases, the source of the data remains the same and the test set may have attributes which are common to the training set -- thus ensuring better performance of the classifier. This, however, may not be the case in a real-world scenario. Therefore a proposed methodology should be \hl{cross-evaluated} (trained on one dataset and tested on a different dataset) on attack types from a varied range of data sources. Deploying a system in a real-world setup (as part of a mail server, or web browser) also tests the system's performance against newer attacks.

\item \textit{Supervised Machine Learning:} We observed an important gap in the detection techniques employed for malicious \hl{URLs}, websites \hl{and emails}: a majority of the implementations use supervised learning methods. The major necessity for training and evaluating the performance of such supervised systems is the considerable amount of labeled data. However, this brings us back to a major security challenge -- \textit{data availability}. 

\item \textit{Evaluation times:} Phishing attacks are usually launched for a short window of time (a few days or weeks). Therefore, for \hl{the} detection of malicious websites, the mechanism used should be fast as well as capable of zero-day attack detection. However, only a few papers (e.g., \cite{verma2017s}) report run times for training the classifiers. In our opinion, training and detection times should be emphasized while discussing the performance of a classifier. 


\end{itemize}

\section{The Weakest Link - User Studies}~\label{sec-userstudies}

Regardless of detectors' performance, users play an important role in preventing attacks, since determined attackers can find ways to bypass detection techniques. Till date, users are considered the most vulnerable link in the phishing ecosystem. According to a study by Intel, 97\% of people are not able to detect phishing emails.\footnote{\hl{https://www.marketwatch.com/press-release/97-of-people-globally-unable-to-correctly-identify-phishing-emails-2015-05-12}} There are several studies on users' susceptibility to phishing attacks, e.g., \cite{goel2017got, halevi2013pilot}, and on how to improve their knowledge about them, e.g.,  \cite{kumaraguru2010teaching, lin2011does}. 
In this section, we start with an overview of different deception techniques used by attackers. We then systematically review the following attributes of 63 selected studies:
\begin{itemize}
\item \textit{Participants and Environment:} The number of participants recruited in the study and whether they were recruited from a specific population (e.g.,\ university students). Running the experiment in the lab usually forces the researchers to reduce the number of participants, so we also consider the environment of the study as an attribute (Lab/Real). \hl{Researchers can ask participants to come to the lab to do the experiment (Lab) or send them the emails (or link to an online survey) so they can do the experiment wherever they want (Real)}.
\item \textit{Evaluation:} The variables that were measured in the study: detection rate, background knowledge, personality, and cognitive process. \hl{Personality is used to distinguish qualities or characteristics of individuals. It consists of five major traits: Openness, Conscientiousness, Extraversion, Agreeableness, and Neuroticism} \cite{mccrae1999five}. Cognitive process refers to four variables related to humans' decision making process: 1) neural activity: \hl{to measure individuals effort in detection phishing attacks}, 2) eye movement: \hl{to find parts of the email/website that people focus on}, 3) cognitive impulsivity \cite{barratt1994impulsiveness}: Tendency to act without much thought, and 4) heuristic/systematic processing \cite{chaiken1980heuristic}: Explains how people receive and process a message, either by using their background knowledge or by investigating and analyzing the message. Besides these quantitative measures, qualitative analysis of participants' reasoning can be used to understand their thought processes and performance.
\item \textit{Context:} \hl{Whether the study is about phishing webpages or phishing emails.}
\item \textit{Experiment Design:} Important aspects of the design of the experiment. It includes: whether the experiment involves deception (i.e., participants are not informed that the experiment is about phishing) and   whether participants' behavior changes with time (long-term study). Design of the \hl{questionnaire}  is also important, some studies only ask participants about their vulnerabilities \hl{(Q \& A)} while more complex studies show the participants phishing/legitimate emails or webpages (experimental).
\item \textit{Goal of the study:} We found two different goals for user studies on phishing attacks. 1) Detection: studying the quantitative and qualitative performance of participants on detecting different types of attacks. 2) Training: studying the effectiveness of a training method. 
Although most training studies are followed by a detection study, we make a distinction between the pure \textit{detection} studies and \textit{training} studies.
\end{itemize}

Categorizing the previous research into different groups based on the aforementioned attributes helps us identify the areas that are well studied and also the gaps in existing research. After reviewing the studies based on the aforementioned attributes, we point out their strengths and weaknesses and conclude with a discussion on the gaps in existing research and some possible future directions.



\subsection{Participants and Environment}
Table~\ref{table-user-study-participants} shows the number and type of participants along with the environment of each study. We categorized the participants into three groups: students, employees of a company, and unrestricted population (i.e., participants sampled from a general population). Empty cells in the table show the areas that remain unexplored. Although there are lots of arguments in the literature about the lab versus real world experiment \cite{brewer2000research, winkler1973experiments, jakobsson2006designing}, based on Table~\ref{table-user-study-participants}, it is clear that both types of experiments are popular among researchers. 
No study on employees has been conducted in the lab environment. \hl{Although the real world experiment is more general in comparison to the lab experiment, the lab experiment helps the researcher analyze human behavior in more detail and understand the factors that affect their responses.}

Apart from the type and number of participants, the variables that are being studied play an important role on the quality of the experiment. While on one hand, studying several variables helps to look at different aspects at the same time; on the other, it makes the experiment more complicated and probably requires more participants. In the next section we discuss variables that have been studied in the reviewed literature.

\begin{table*}[h]
\centering
\caption{Type of  participants and the environment of phishing user studies}
\label{table-user-study-participants}
\begin{threeparttable}
\begin{tabular}{|c|c|l|l|l|}
\hline
\multirow{2}{*}{\textbf{Environment}} & \multirow{2}{*}{\textbf{Type of Participants}} & \multicolumn{3}{c|}{\textbf{Number of Participants}}                                                         \\ \cline{3-5} 
                             &                                       &  \multicolumn{1}{c|}{\textless50}                                                                                                                                                   &  \multicolumn{1}{c|}{$\geq$50 and \textless200}                                                                                                                          &  \multicolumn{1}{c|}{$\geq$200}                                                                                                             \\ \hline \hline 
\multirow{3}{*}{Lab}         & Student                               & \cite{neupane2015multi, lin2011does, arachchilage2016phishing, xiong2017aiping, neupane2016neural, baslyman2016smells, miyamoto2015eye, moreno2017fishing, kirlappos2012security,alsharnouby2015phishing} & \cite{pattinson2012some, parsons2016users, parsons2013phishing, jensen2017combating, harrison2015examining, sun2017exploring, vishwanath2015examining,ramesh2017intelligent}, \cite{sun2016teaching}\tnote{a}, \cite{sun2017effects}\tnote{a} & \cite{wang2012research, alseadoon2013typology, blythe2011f, jansson2013phishing, lastdrager2017effective, moody2011phish}, \cite{sun2016mediating}\tnote{a} \\ \cline{2-5} 
                             & Employee                              &                                                                                                                                                               & \cite{balozian2017managers}\tnote{a}                                                                                                                   &                                                                                                                             \\ \cline{2-5} 
                             & Unrestricted                      & \cite{kumaraguru2010teaching, wu2006web}, \cite{conway2017qualitative}\tnote{a}                                                                    & \cite{martin2009phishing, zhao2017design}, \cite{arachchilage2014security}\tnote{a}, \cite{arachchilage2013game}\tnote{a}                     & \cite{wang2016overconfidence}, \cite{junger2017priming, alqarni2016toward}\tnote{a}                              \\ \hline
\multirow{3}{*}{Real}        & Student                               & \cite{yang2015effectiveness}                                                                                                                                & \cite{halevi2013pilot, harrison2016individual, harrison2016user}                                                               & \cite{dodge2006using, goel2017got,vishwanath2011people,perrault2018using,benensonunpacking2017}                                                                  \\ \cline{2-5} 
                             & Employee                              & \cite{halevi2015spear}                                                                                                                                      & \cite{silic2016dark,xin:phishing,holm2014empirical}                                                                                                                                   & \cite{caputo2014going, coronges2012influences, kearney2013phishing}                                      \\ \cline{2-5} 
                             & Unrestricted                      &                                                                                                                                                               & \cite{park2014comparing, oliveira2017dissecting, canfield2016quantifying, moreno2017fishing}                                                           & \cite{xiong2017aiping, vidas2013qrishing, sheng2010falls, nicholson2017can, liu2011smartening, wright2010influence,rajivan2018creative}                            \\ \hline
\end{tabular}
\begin{tablenotes}
\item[a] \textit{Studies that only asked participants about their vulnerability (Q/A) instead of showing them the email/website}
\end{tablenotes}
\end{threeparttable}
\end{table*}

\subsection{Experimental Attributes}
Table \ref{table-user-study-context-variables} groups the literature based on the context, evaluation, and whether or not deception was involved in the studies. Based on the table, it is obvious that there is more research on emails than webpages. 
Another observation is the lack of a long-term study. People's behavior can change over time, so it is important to redo the experiment after a long period of time to see how their behavior changes. We also mentioned the variables that each work studied in Table \ref{table-user-study-context-variables}. Although there are a few works that study several attributes together, most of the existing works study the effect a single variable (e.g. personality) on participants' phishing detection ability. 


\begin{table*}[h]
\centering
\caption{Parameters of phishing user studies: length, long-term or not, context, deception, and variables used. \textit{DR} - Detection Rate, \textit{BK} - Background Knowledge, \textit{CP} - Cognitive Process, \textit{PR} - Personality}
\label{table-user-study-context-variables}
\begin{tabular}{|c|c|c|c|l|c|c|c|c|}
\hline
\multirow{2}{*}{\textbf{Long-term}} & \multirow{2}{*}{\textbf{context}} & \multirow{2}{*}{\textbf{deception}} & \multirow{2}{*}{\textbf{N}} &  \multicolumn{1}{c|}{\multirow{2}{*}{\textbf{Literature}}}                 & \multicolumn{4}{c|}{\textbf{Variables}}         \\ \cline{6-9} 
                                      &                                     &                                       &                                                                                                                                                                                                                                                                                                       &  & DR         & BK         & CP         & PR         \\ \hline
                                      \hline
No                                    & Email                               & Yes                                   & 13 & \begin{tabular}[c]{@{}l@{}}\cite{parsons2016users, wang2012research, harrison2016individual, parsons2013phishing, benensonunpacking2017, oliveira2017dissecting}\\\cite{goel2017got,caputo2014going,dodge2006using,kearney2013phishing,jansson2013phishing, harrison2015examining,holm2014empirical}\end{tabular} & \checkmark &            &            &            \\ \cline{4-9} 
                                      &                                     &                                       & 3 & \cite{halevi2013pilot, coronges2012influences, sheng2010falls}                                                                                                                                                                                                                                      & \checkmark & \checkmark &            &            \\ \cline{4-9} 
                                      &                                     &                                       & 2 & \cite{jensen2017combating,xin:phishing}                                                                                                                                                                                                                                                             & \checkmark &            & \checkmark &            \\ \cline{4-9}                                      
                                      &                                     &                                       & 1 & \cite{rajivan2018creative}                                                                                                                                                                                                                                                             & \checkmark &            &  & \checkmark            \\ \cline{4-9}
                                      &                                     &                     &   4              & \cite{alseadoon2013typology, harrison2016user, vishwanath2015examining, wright2010influence}                                                                                                                                                                                                        & \checkmark & \checkmark & \checkmark &            \\ \cline{4-9} 
                                      &                                     &                &  2                      & \cite{halevi2015spear, moody2011phish}                                                                                                                                                                                                                                                              & \checkmark & \checkmark &            & \checkmark \\ \cline{4-9} 
                                      &                                     &             &  1                         & \cite{pattinson2012some}                                                                                                                                                                                                                                                                            & \checkmark & \checkmark & \checkmark & \checkmark \\ \cline{3-9} 
                                      &                                     & \multirow{2}{*}{No}     & 6              & \cite{martin2009phishing, park2014comparing, blythe2011f, caputo2014going, vishwanath2011people,perrault2018using}                                                                                                                                                                                                    & \checkmark &            &            &            \\ \cline{4-9} 
                                      &                                     &                   & 3                    & \cite{wang2016overconfidence, canfield2016quantifying, nicholson2017can}                                                                                                                                                                                                                            & \checkmark &            & \checkmark &            \\ \cline{2-9} 
                                      & \multirow{5}{*}{Webpage}            & Yes                     & 5              & \cite{yang2015effectiveness, lin2011does, silic2016dark, vidas2013qrishing, zhao2017design}                                                                                                                                                                                                         & \checkmark &            &            &            \\ \cline{3-9} 
                                      &                                     & \multirow{4}{*}{No}       & 4            & \cite{wu2006web, arachchilage2016phishing, kumaraguru2010teaching, liu2011smartening}                                                                                                                                                                                                               & \checkmark &            &            &            \\ \cline{4-9} 
                                      &                                     &                & 3                       & \cite{neupane2015multi, xiong2017aiping, miyamoto2015eye}                                                                                                                                                                                                                                           & \checkmark &            & \checkmark &            \\ \cline{4-9} 
                                      &                                     &             & 3                          & \cite{moreno2017fishing, kirlappos2012security,alsharnouby2015phishing}                                                                                                                                                                                                                             & \checkmark & \checkmark & \checkmark &            \\ \cline{4-9} 
                                      &                                     &            & 1                           & \cite{neupane2016neural}                                                                                                                                                                                                                                                                            & \checkmark &            & \checkmark & \checkmark \\ \hline
Yes                                   & Email                               & Yes              & 1                     & \cite{lastdrager2017effective}                                                                                                                                                                                                                                                                      & \checkmark & \checkmark &            &            \\ \hline
\end{tabular}
\end{table*}

\textbf{Multiple Comparisons Problem}:  We noticed that several  papers, e.g.,\ \cite{sun2016teaching, sun2016mediating, vishwanath2015examining, moody2011phish, arachchilage2013game, vishwanath2011people, lastdrager2017effective, nicholson2017can, sun2017exploring, harrison2016user, moreno2017fishing, jensen2017combating, wang2016overconfidence,harrison2016individual, martin2009phishing, alsharnouby2015phishing}, made their statistically significant conclusion by running multiple tests on a single dependent variable. Multiple comparisons problem causes incorrect rejection of null hypothesis, if not accompanied by p-value adjustment (increases type 1 error) \cite{coombs1996univariate}. 
Bonferroni \cite{armstrong2014use} is the simplest way to adjust the p-value, which is too conservative in rejecting the null hypothesis (increases type 2 error). More complex methods like Holm-Bonferroni and Benjamini-Hochberg \cite{benjamini1995controlling} can be used to lower type 1 and 2 errors simultaneously (compared to Bonferroni).

In the next section, we give a brief summary of the works selected based on how thoroughly they covered the aforementioned attributes, involving emails or websites (some with deception and some without).

\subsection{Selected User Study Literature}
Studies on users' behaviour are not completely aligned with the challenges we described in Section \ref{sec-challenges}. Therefore, instead of selecting the papers based on those challenges, we evaluate the literature based on the following: 1) how many different variables were studied; 2) context (email or website); 3) environment; and 4) whether or not it includes deception.

Researchers in \cite{halevi2013pilot} studied the effect of psychological traits on people's ability to detect email phishing attacks and online information sharing. They sent a ``prize scam'' email to 100 students from a psychology class that requested an immediate response. They also asked about the privacy settings of participants on Facebook, the frequency of posting, type of data they posted, etc. The results show that for women, certain personality traits (neuroticism and openness) are more likely to be associated with vulnerability to phishing attacks as well as with online information sharing on a social network.

The effect of priming people on detection of phishing emails has been studied in \cite{pattinson2012some}. They divided a total of 117 participants into two groups. One group was not informed that they are participants in a phishing study, but the other group was informed. As expected, their results indicate, that informed participants perform better than not-informed participants. Also, they found that background computer knowledge improves participants' phishing detection rate.

A few studies investigated the relationship of brain activity with phishing and other security-related tasks. Researchers in \cite{neupane2015multi} and \cite{neupane2016neural} used Functional Magnetic Resonance Imaging (fMRI) \hl{and Electroencephalogram (EEG) to measure brain's electrical activity} and an eye-tracker to have a better understanding of users' decision-making process. They used both existing phishing websites (by downloading and hosting them on their own network) and manually created ones. The EEG results show that at a subconscious level, there might be hidden differences in how users detect real and fake websites. The fMRI results show an increase in brain activity (on regions related to decision-making, attention and problem solving) during the detection task. Although participants tried to process the existing information to differentiate between real and fake webpages, their performance still low. This suggests that people may not have enough knowledge about the cues that really need their attention. The eye-gaze pattern captured by the eye-tracker revealed that participants did not pay enough attention to the cues that show whether a webpage is real or fake (``less time on URL; more time on login field or website logo'').

Training is one of the techniques to help people avoid phishing attacks. Attackers may choose a domain name very similar to the domain name of the target company, e.g.,  bank\textbf{0}famerica.com (using `0' instead of the letter `o'). So, can we say that paying attention to the URL will help detect some of the phishing websites? To answer this question, researchers did a study on the effectiveness of highlighting the domain in helping people detect phishing webpages \cite{lin2011does}. A group of 22 people were showed 16 webpages (8 legitimate and 8 fraudulent) and asked how safe these websites are. Then they asked participants to re-evaluate the same 16 webpages by focusing on the address bar. They categorized the participants into three types based on which clues they observed: A) those who pay attention to the content of the webpage, B) those who pay attention to the information found in the address bar and AB) those who rely primarily on the content and sometimes use the address bar information. Their results show that the domain name highlighting is not effective for type A and AB participants. So, domain name highlighting cannot be used as a single method for detecting phishing webpages. A similar study has been done on a larger scale (320 participants), which resulted in a similar conclusion \cite{xiong2017aiping}.

\hl{So far, we only discussed user studies on regular phishing attacks. Training people to detect spear phishing attacks is also important, since these attacks are harder to detect automatically due to their targeted nature, thus we} review them in the next section.

\subsection{Spear phishing}


Despite the importance of spear phishing attacks, we found only two user studies on this type of attack \cite{caputo2014going, halevi2015spear}. In both works, spear phishing emails were sent to the employees of a company. The goal in \cite{caputo2014going} is to study the effect of training on employees (1369 employees), while the goal of \cite{halevi2015spear} is to study the employees' (40 employees) vulnerability to spear phishing attack. It has been shown in \cite{caputo2014going} that immediate feedback is not enough for increasing people's awareness. \hl{They sent a series of carefully crafted spear phishing emails to the employees and gave them immediate training whenever they fell for an attack.} The study in \cite{halevi2015spear} found that there is a correlation between users' conscientiousness and their responses to the spear phishing. Another interesting finding is the negative correlation between participants' subjective estimate of their own vulnerability and the likelihood of being phished. 

We conclude this section by reviewing the unexplored areas in the existing research and some suggestion for future works.

\subsection{\hl{Opportunities for Future Research}}
Based on  Table~\ref{table-user-study-participants}, the important missing studies are: a deep analysis of participants' reasoning on (1) real-world studies with unrestricted participants and (2) in-lab experiments with employees. There is typically an inverse relationship between the depth of behavioral analysis and the number of participants. The fewer the participants, the better researchers can analyze the responses of each participant individually. An employee-only study is also important because many of the spear phishing studies use employees of a company as their target. 

Another missing study is a comparison between conducting the same experiment in the lab or in a real-world situation. There is only one study in which the experiment was conducted in both real and lab environment \cite{moreno2017fishing}. Researchers first ran the experiment in the lab to have more control, and then, to increase confidence and power (by increasing the number of participants), they did the experiment in a real scenario. Although they showed that results from real and lab environment are similar, their goal was not to compare these two environments. Also, there is only one study that considers long-term effects (Table~\ref{table-user-study-context-variables}) in their experiment design \cite{lastdrager2017effective}. They re-tested the participants (primary school students) after \hl{two and four weeks} of conducting a phishing training (interactive presentation). Surprisingly, their results showed that after a gap of four weeks, the performance of the participants diminished to the level before the training had occurred. The main limitation of their work is that they used a paper-based questionnaire to record participants performance instead of showing them the email/webpage on the computer. More studies with diverse populations are needed to make a general conclusion about the temporal effect on people's performance.

As we mentioned, only two \hl{groups} studied users' behaviour  on spear phishing \hl{attacks}. More studies are required to understand the different aspects of spear phishing attacks as well as how people respond to \hl{them}. Then training techniques can be developed based on the results of these studies to increase people's awareness about the risk of spear phishing attacks.



\section{Related Work} \label{sec-related}

Phishing has been the topic of several surveys as well as books. We organize the related work along the major themes: emails, websites, URLs, user studies, and general surveys. The last category includes those surveys that examine more than one dimension of phishing detection. We conclude this section by mentioning surveys on two related areas: spam detection and web page classification. 

{\bf Emails.} There is one survey that \textit{focuses primarily} on phishing email literature \cite{almomaniGAMA13}. In this 2013 survey, 
the researchers review mostly on machine-learning techniques for email filtering, and their relative advantages and disadvantages. They also classify the approaches proposed in the literature according to  different stages of the attack flow as: network level protection, authentication, client side tool, user education, and server side filters and classifiers. We discuss other email surveys under General surveys. 



\textbf{URLs.} Recently a survey on phishing URL detection has become available \cite{sahooLH17}. In this survey, features for phishing URL detection are covered, and different kinds of machine learning techniques are surveyed: batch, on-line, representation, and ``other.'' 
Also,   \cite{sataneD13,sharmaP15} survey a few papers on phishing URL detection, and \cite{gupta17TJ,guptaAP17,aleroudZ17} discuss some of the issues and literature associated with phishing URL detection.

\textbf{Websites.} Three teams of researchers have  surveyed phishing website detection \cite{mohammadTM15,varshneyMA16,dou17KK}. In \cite{mohammadTM15}, phishing website detection papers were studied along five dimensions: blacklist/whitelist, instantaneous protection, decision support tools, community rating tools such as Web of Trust, and intelligent heuristics, for the first time. Subsequently, Varshney et al. \cite{varshneyMA16} added search-based, visual-similarity, DNS-based, and proactive phishing URL based techniques.  They also discussed pros and cons for each type of detection technique and some selected papers, for which they  summarized results and dataset sizes. In \cite{dou17KK}, three criteria were used to select papers for discussion (novelty, attention -- measured by a number of citations, and completeness). 
\hl{In a more recent study of phishing techniques} \cite{qabajeh2018recent}\hl{, the authors give a detailed description of how content-based methods have been used for phishing detection in previous literature. The researchers provide a brief history of phishing along with an analysis of automatic detection methods and how they help fight phishing website attacks.}
 
\textbf{User studies.} To the best of our knowledge, there is no comprehensive survey of user studies of phishing or spear phishing, although some surveys, e.g., ~\cite{guptaAP17,gupta17TJ,aleroudZ17,purkait12}, discuss it as part of their taxonomy, training, or study of selected solutions. In \cite{khonji2013phishing}, the authors survey a few papers on user studies. 

\begin{table*}[!htb]
\centering
\caption{Comparing earlier phishing surveys with the present survey. \textit{Year range} - publication years of papers reviewed}
\label{table:comparison}
\begin{tabular}{|l|c|c|c|c|c|c|c|c|c|c|c|}
\hline
                                                                                                            \multirow{2}{*}{\begin{tabular}[c]{@{}c@{}}Survey\end{tabular}} & \multirow{2}{*}{\begin{tabular}[c]{@{}c@{}}Year\\Range\end{tabular}} & \multicolumn{3}{c|}{Detection Techniques} & \multirow{2}{*}{\begin{tabular}[c]{@{}c@{}}User \\ Studies\end{tabular}} &\multicolumn{4}{c|}{Aspects}                                      & \multirow{2}{*}{\begin{tabular}[c]{@{}c@{}}Security\\ Challenges\end{tabular}} \\ \cline{3-5}\cline{7-10}
    &  & URL   & Web  & Email  &  & \begin{tabular}[c]{@{}c@{}}Dataset\\ Diversity\end{tabular} & Features & \begin{tabular}[c]{@{}c@{}}Detection \\ Methods\end{tabular} & \begin{tabular}[c]{@{}l@{}}Evaluation\\ Metrics\end{tabular}  &                                                                                \\ \specialrule{.2em}{.1em}{.1em}
\begin{tabular}[c]{@{}c@{}}Gupta et. al. \cite{gupta17TJ}\end{tabular}      & 2005-2016     &     &\checkmark        & \checkmark      &             &   & \checkmark        &                                                             &                                                               &                 \\ \hline
\begin{tabular}[c]{@{}l@{}}Gupta et. al. \cite{guptaAP17}\end{tabular}  & 2004-2016         &      & \checkmark        & \checkmark    &             &                                                           & \checkmark        &                                                             &                                                                      &                                                                              \\ \hline
 \begin{tabular}[c]{@{}l@{}} \hl{Chiew et. al.} \cite{chiewYT18}\end{tabular}    & 2004-2017       &      \checkmark     &     \checkmark    & \checkmark       &   &          &       \checkmark  &    \checkmark       &            &                                                                                \\ \hline
\begin{tabular}[c]{@{}l@{}}Dou et. al. \cite{dou17KK}\end{tabular} & 2005-2016             &     & \checkmark        &       &              &      &         &   &            &         \\ \hline
\begin{tabular}[c]{@{}l@{}}Almomani et. al. \cite{almomaniGAMA13}\end{tabular} & 2004-2012  &       & \checkmark         & \checkmark       &               &         &    \checkmark      & \checkmark     &    \checkmark                   &                                                                                \\ \hline
\begin{tabular}[c]{@{}l@{}}Satane and Dasgupta \cite{sataneD13}\end{tabular} & 2007-2014    &   \checkmark    &    \checkmark      &        &               &              &  \checkmark        &                 &                 &             \\ \hline
\begin{tabular}[c]{@{}l@{}}Sharma and Parveen \cite{sharmaP15}\end{tabular}   & 2012-2015   &   \checkmark    &          &        &               &         & \checkmark         &    \checkmark    &  \checkmark                 &                                \\ \hline
\begin{tabular}[c]{@{}l@{}}Aleroud and Zhou \cite{aleroudZ17}\end{tabular}    & 2005-2015   &   \checkmark    &  \checkmark        &        &      \checkmark         &                                                              &          &                                                        \checkmark      &         &                                                                                \\ \hline
\begin{tabular}[c]{@{}l@{}}Mohammad et. al. \cite{mohammadTM15}\end{tabular} & 2004-2012    &       &   \checkmark       &        &  \checkmark             &      &          &                \checkmark   &                                       &                \\ \hline
\begin{tabular}[c]{@{}l@{}}Varshney et. al. \cite{varshneyMA16}\end{tabular}  & 2004-2016   &       &\checkmark          &        &               &         &    \checkmark      &       \checkmark    &               &                                                                                \\ \hline
\begin{tabular}[c]{@{}l@{}}Purkait \cite{purkait12}\end{tabular}   & 2004-2011                &       & \checkmark    & \checkmark       &  \checkmark             &            &       &         &                       &          \\ \hline
\begin{tabular}[c]{@{}l@{}}Khonji et. al. \cite{khonji2013phishing}\end{tabular} & 2006-2011 &   \checkmark    &      \checkmark    &    \checkmark    &     \checkmark          &          &          &                       &                                                          \checkmark            &                      \\ \hline
  \begin{tabular}[c]{@{}l@{}} \hl{Qabajeh et. al.} \cite{qabajeh2018recent}\end{tabular}    & 2005-2017       &         & \checkmark        &     &                                                  &          &        \checkmark           &          &            &                                                                                \\ \hline

\begin{tabular}[c]{@{}l@{}}This Survey\end{tabular} & 2004-2018 &   \checkmark    &      \checkmark   &    \checkmark    &     \checkmark          &     \checkmark     &       \checkmark   &                \checkmark       &                                                          \checkmark            &        \checkmark              \\ \hline

\end{tabular}
\end{table*}

\textbf{General surveys.} In a 2012 survey, the author surveyed a range of countermeasures for phishing, ranging from email detection to training and legal studies\cite{purkait12}. In a 2013 survey, \cite{khonji2013phishing}, researchers examined 20 proposed solutions in detail. These included both software detection techniques as well as some training approaches for human awareness and vulnerability. 
In \cite{gupta17TJ}, researchers have given: a nice taxonomy of phishing attacks and defenses, identified features for phishing email detection, and compared 15 phishing email detection techniques and 15 phishing website detection techniques over the period 2000-2016. In \cite{guptaAP17}, some dataset sources are listed, taxonomies are discussed with more examples, features for phishing email detection are listed and 18 solutions are discussed over the period 2000-2016. 

In a 2017 phishing survey \cite{aleroudZ17}, researchers have proposed a new, multi-dimensional phishing taxonomy and classified phishing countermeasures in five categories: machine learning, text mining, human users,
profile matching, and others, with the last category further subdivided into:
search engines, ontology, client-server authentication, and honeypot countermeasures. They compare anti-phishing tools, both commercial and research prototypes, and identify gaps in literature in terms of unstudied or little-studied attack vectors and communication channels. 
\hl{A 2018 survey} \cite{chiewYT18} \hl{on phishing reviews popular techniques and vectors or channels for  phishing operations. It presents in detail several techniques and propagation methods for social engineering attacks as well as evaluate how such existing methods can be combined to launch more sophisticated attacks in future.}

Two related areas worth mentioning are \textit{spam detection} and \textit{web page classification}. Although the goals of spam and phishing are different, and phishing detection is more challenging because of deception, there are similarities in the approaches, algorithms, and features that have been employed. We refer the reader to the survey of spam detection for more details \cite{blanzieriB08}. By comparing the work on spam detection and on phishing detection, we observe that phishing work has not utilized text mining and natural language processing techniques as much as spam detection. Web page classification also exploits text classification techniques. We refer to \cite{qiD09} for an excellent survey on web page classification.

We summarize our observations about the phishing surveys in Table~\ref{table:comparison}.  No previous work, to our knowledge, has attempted to systematize and evaluate the phishing literature from the perspective of security challenges or dataset diversity,  and we did not find any previous survey of spear phishing research literature. 





\section{Conclusions} \label{sec-concl}
In this paper we have covered phishing and spear phishing detection techniques and user studies from the perspective of security challenges. To our knowledge, this is the first systematic survey of these \hl{important topics} from this viewpoint, and the only comprehensive survey of spear phishing research and user studies of phishing/spear phishing.  We identified several \hl{opportunities for future research}. 
\hl{Specifically}, we found that deep learning techniques have not been exploited much in phishing detection. A large number of papers use a wide variety of features, but do not include any information about feature importance or feature selection -- this is important since using \hl{many} features can slowdown system performance. 
Other gaps include: the lack of proper datasets\footnote{\hl{We recently created email datasets as part of the IWSPA-AP Shared task} \cite{VermaD18}\hl{, which are available by request to the authors.}} and metrics for evaluation,  and lack of diverse population in user studies and long-term experiments.
\hl{Our systematization reviews and outlines the common security challenges prevalent in cybersecurity, which can be utilized} for the evaluation of future research on these topics and may lead to improved and more practical schemes. 

\section*{Acknowledgments}
We thank Omprakash Gnawali, Luis F.T. De Moraes and \hl{Sima Shafaei} for their helpful comments and suggestions on this paper. \hl{This research was supported in part by NSF grants CNS 1319212, DGE 1433817, DUE 1241772, and DUE
1356705. This material is also based upon work supported by, or in part by, the U. S. Army Research Laboratory and the U. S. Army Research Office under contract/grant number W911NF-16-1-0422.
}


%


\ifCLASSOPTIONcaptionsoff
  \newpage
\fi



%
\bibliographystyle{IEEEtran}
\bibliography{sok-refs}

\section*{Appendix}
In the Appendix, we give tables with bibliographic numbers of the papers, so that interested readers can easily find the original sources for each entry in our tables in the main part of the paper. 
\setcounter{table}{0}
\renewcommand{\thetable}{A\arabic{table}}

\begin{table}[!htb]
\centering
\caption{Distribution of metrics used for evaluating phishing URL detection methods}
\label{tab:metricsurl}
\resizebox{\columnwidth}{!}{
\begin{threeparttable}
\begin{tabular}{|c|c|l|}
\hline
\textbf{Eval. Metrics} &\textbf{N}         & \textbf{Literature} \\ \hline \hline
Accuracy         &28           & \begin{tabular}[c]{@{}l@{}}~\cite{cao2016detection,nepali2016you,rathod2015comparative,darling2015lexical,vermaK15,egan2011evaluation,nguyen2014novel,popescu2015study,gupta2014bit,zhao2013cost,lee2013warningbird}\\ ~\cite{khonji2011novel,feroz2014examination,huang2014malicious,feroz2015phishing,chu2013protect,eshete2014webwinnow,zhang2016url,bahnsen2017classifying,alshboul2015detecting}\\~\cite{verma2017s,garera2007framework,dewan2015towards,marchal:phishscore,marchal:phishstorm,pradeepthi2014performance,nguyen2013detecting,patil2016malicious} \end{tabular} \\ \hline
\begin{tabular}[c]{@{}c@{}}Precision\\Recall\\F-score \end{tabular}&20&  \begin{tabular}[c]{@{}l@{}}~\cite{mamun2016detecting,popescu2015study,rathod2015comparative,darling2015lexical,gupta2014bit,sandracoordinator, basnet2012mining,sananse2015phishing,bahnsen2017classifying,alshboul2015detecting}\\~\cite{burnap2015real, cao2016detection, marchal:phishscore,marchal:phishstorm, lee2013warningbird,nepali2016you,verma2017s,lee2014users,pradeepthi2014performance,zhang2016url} \end{tabular}          \\ \hline
Error Rate &13         &  \begin{tabular}[c]{@{}l@{}}~\cite{rathod2015comparative,vermaK15,ma2009beyond,ma2011learning,xiong2015mird,le2011phishdef,basnet2015towards}\\~\cite{gyawali2011evaluating,blum2010lexical,zhang2016url,lin2013malicious,su2013suspicious,gabriel2016detecting} \end{tabular}          \\ \hline
\begin{tabular}[c]{@{}c@{}}Confusion\\Matrix\tnote{a}\end{tabular}  &20                 & \begin{tabular}[c]{@{}l@{}}~\cite{feroz2014examination,sha2015limited,astorino2016malicious,darling2015lexical,guan2009anomaly,popescu2015study,ma2009beyond,zhao2013cost,vu2016firstfilter,khonjiIJ11,patil2016malicious} \\~\cite{chu2013protect,lee2013warningbird,sananse2015phishing,eshete2014webwinnow,zhang2016url,marchal:phishscore,marchal:phishstorm,popescu2016practical,nguyen2013detecting} \end{tabular}           \\ \hline
\begin{tabular}[c]{@{}c@{}}Area Under\\Curve\end{tabular} & 4 & ~\cite{darling2015lexical,marchal:phishscore,marchal:phishstorm,bahnsen2017classifying}\\ \hline
Others\tnote{b}      &6               &  \begin{tabular}[c]{@{}l@{}}~\cite{nguyen2014novel,zhao2013cost,khonjiIJ11},~\cite{jeeva2016intelligent,garera2007framework,popescu2016practical} \end{tabular}          \\ \hline
\end{tabular}
\begin{tablenotes}
\item[a] \textit{Consists of the raw True Positive, False Positive, True Negative, False Negative values as well as rates like TPR, FPR, TNR, FNR, specificity and sensitivity calculated from the confusion matrix}.
\item[b] \textit{Balanced Success Rate~\cite{gyawali2011evaluating}, Root Mean Square Error~\cite{nguyen2014novel}, effective rules/patterns~\cite{jeeva2016intelligent}, Cost/Sum of classification~\cite{zhao2013cost}, False Alarm rate~\cite{popescu2016practical}}
\end{tablenotes}
\end{threeparttable}}
\end{table}


\begin{table}[!htb]
\centering
\caption{Distribution of metrics used for evaluating phishing website detection methods}
\label{table-webpage-metrics}
\resizebox{\columnwidth}{!}{
\begin{threeparttable}
\begin{tabular}{|c|c|l|}
\hline
\textbf{Eval. Metrics}  & N        & \textbf{Literature} \\ \hline \hline
Accuracy                & 26      & \begin{tabular}[c]{@{}l@{}}\cite{manek2014demalfier, choi2011detecting, yue2013fine, lee2016phishing, abdelhamid2014phishing, zhang2017two, tan2016phishwho,rajab2017new,feroz2015examination,moghimi2016new} \\\cite{varshney2016phish,ramesh2014efficacious, liu2010automatic,zhang2014domain,raodetection,marchal2016know, wardman2011high,thomas2011design} \\ \cite{aggarwal2012phishari, liang2009malicious, tan2014phishing, abbasi2015enhancing, wenyin2012antiphishing,thabtah2016dynamic,el2017detection,xu2014evasion} \end{tabular} \\ \hline
\begin{tabular}[c]{@{}c@{}}Precision\\Recall\\F-score\end{tabular} & 27  & \begin{tabular}[c]{@{}l@{}}\cite{manek2014demalfier,el2017detection,bannur2011judging, stringhini2013shady, mohaisen2015towards, kosba2014adam, feroz2015examination,whittaker2010large,mao2017phishing} \\\cite{zhang2017phishing, lee2016phishing,zhuang2012intelligent,abbasi2015enhancing,aggarwal2012phishari,zhang2014domain,dong2015beyond,gowtham2014comprehensive} \\\cite{thakur2014catching,miyamoto2008evaluation,geng2015combating,mao:alarm,zhang:textual,marchal2016know,thabtah2016dynamic,moghimi2016new,raodetection, Xiang:2011:CFM:2019599.2019606} \end{tabular}          \\ \hline
Error Rate       & 21   &  \begin{tabular}[c]{@{}l@{}}\cite{liang2009malicious,stringhini2013shady, rosiello2007layout,thomas2011design, pao2012malicious, mohaisen2015towards,raodetection,aburrous2010intelligent,wardman2011high,canali2011prophiler,bannur2011judging} \\\cite{kosba2014adam,mohammad2014intelligent,wardman2009identifying,barraclough2013intelligent,wenyin2012antiphishing,miyamoto2008evaluation,moghimi2016new,liu2010automatic,el2017detection, zhang:textual}\end{tabular}          \\ \hline
\begin{tabular}[c]{@{}c@{}}Confusion\\Matrix\end{tabular}      &      35         & \begin{tabular}[c]{@{}l@{}}\cite{thakur2014catching,choi2011detecting,miyamoto2008evaluation,geng2015combating,xiang2010hierarchical,chiew2015utilisation,marchal2016know, stringhini2013shady, lee2016phishing, huang2010mitigate,mohaisen2015towards,bannur2011judging,Xiang:2011:CFM:2019599.2019606,xu2014gemini} \\\cite{pao2012malicious,he2011efficient,gowtham2014comprehensive, yue2013fine,varshney2016phish,zhang2017two,raodetection,fang2015proactive,moghimi2016new,shahriar2012trustworthiness,wardman2009identifying} \\\cite{el2017detection,xu2014evasion,thabtah2016dynamic,Gowtham2014,tan2014phishing,ramesh2017intelligent,whittaker2010large,tan2016phishwho,ramesh2014efficacious,dong2010defending,thomas2011design}\end{tabular}           \\ \hline
\begin{tabular}[c]{@{}c@{}}Area Under\\Curve \end{tabular}& 8 & \begin{tabular}[c]{@{}l@{}}\cite{marchal2016know, feroz2015examination, miyamoto2008evaluation, lee2016phishing,geng2015combating,corona2017deltaphish,gowtham2014comprehensive,Xiang:2011:CFM:2019599.2019606} \end{tabular} \\ \hline
Others\tnote{a}               & 7         &  \begin{tabular}[c]{@{}l@{}}\cite{zhang:textual,el2017detection, marchal2012proactive, whittaker2010large,vargas:enemies, moghimi2016new, dong2015beyond} \end{tabular}          \\ \hline
\end{tabular}
 \begin{tablenotes}
\item[a] \textit{Matthews Correlation Coefficient (MCC) \cite{el2017detection,zhang:textual}, Precision-recall curve \cite{whittaker2010large}, Kappa statistic \cite{moghimi2016new, dong2015beyond}, MCscore \cite{marchal2012proactive} and Text Similarity \cite{vargas:enemies}}
\end{tablenotes}
\end{threeparttable}
}
\end{table}

\begin{table}[!htb]
\centering
\caption{Distribution of metrics used for evaluating phishing email detection methods}
\label{table-email-metrics}
\resizebox{1\columnwidth}{!}{
\begin{threeparttable}
\begin{tabular}{|c|l|l|}
\hline
\textbf{Eval. Metrics}     & N     & \textbf{Literature} \\ \hline \hline
Accuracy                 &14    & \begin{tabular}[c]{@{}l@{}}\cite{Duman:emailprofiler,Andronicus:classification, Islam:2013:MPD:2405859.2406232,chandrasekaran2006phishing}\\ 
\cite{Figueroa2017, Isredza:phishing, yearwood:profiling,masoumeh:feature,Ammar:evolving, bergholz2008improved} \\ 
\cite{Stringhini2015,DBLP:conf/ecir/GanstererP09,Dazeley2010, orman:semantics}
\end{tabular}\\
\hline
 \begin{tabular}[c]{@{}c@{}}Precision\\ Recall \\F-score\end{tabular} & 12 &  \begin{tabular}[c]{@{}l@{}}\cite{Figueroa2017,Mengchen:optimizing,chandrasekaran2006phishing}
 \\
 \cite{adre:new,DBLP:journals/corr/YasinA16,bergholz2008improved,chowdhury:multilayer,toolan:ensembles,Ammar:online,Islam:2013:MPD:2405859.2406232,ramanathan,Chiang:2015:AMC:2778890.2779103}
 
 \end{tabular}   \\ \hline
\begin{tabular}[c]{@{}c@{}}Confusion \\ Matrix\tnote{a} \end{tabular}          & 14           & \begin{tabular}[c]{@{}l@{}}\cite{chandrasekaran2006phishing,huang2017gossip,Figueroa2017,Ammar:online}
\\
\cite{Isredza:phishing,adre:new,Laszka:2016:MSF:3015812.3015893,Chiang:2015:AMC:2778890.2779103,Laszka:2015:OPF:2887007.2887140,Mengchen:optimizing}\\ 
\cite{Stringhini2015,ramanathan,DBLP:journals/corr/YasinA16,vermaH13}
\end{tabular} \\
\hline
\begin{tabular}[c]{@{}c@{}} Area Under\\ Curve  \end{tabular} & 7       & \begin{tabular}[c]{@{}c@{}} \cite{Chiang:2015:AMC:2778890.2779103,masoumeh:feature,Andronicus:classification,Islam:2013:MPD:2405859.2406232,DBLP:journals/corr/YasinA16, Olivo:2013,ramanathan} \end{tabular}    
\\ \hline
Error Rate          & 2& \cite{Figueroa2017,isredzaRahmi:profiling}
\\ \hline

Other\tnote{b}  &4      & \begin{tabular}[c]{@{}l@{}} \cite{yearwood:profiling,Chiang:2015:AMC:2778890.2779103,Seifollahi2017,Ammar:evolving} 
\end{tabular}            \\ \hline
\end{tabular}
\begin{tablenotes}
\item[a] \textit{Consists of the raw True Positive, False Positive, True Negative, False Negative values as well as rates like TPR, FPR, TNR, FNR, specificity and sensitivity calculated from the confusion matrix}
\item[b] \textit{Miscellaneous metrics used by different authors. One-Error, Coverage and Average Precision \cite{yearwood:profiling}, G-Mean \cite{Chiang:2015:AMC:2778890.2779103}, Rand Index (RI) \cite{Seifollahi2017}} 
, Root Mean Square Error (RMSE) \cite{Ammar:evolving}
\end{tablenotes}
\end{threeparttable}}
\end{table}
\begin{table}[!htb]
\centering
\caption{\hl{Methods used} in phishing URL detection: \textit{Types, Algorithms and Literature}}
\label{tab:urlmethods}
\resizebox{1\columnwidth}{!}{
\begin{threeparttable}
\begin{tabular}{|c|l|l|l|}
\hline
\textbf{Types} & \textbf{Algorithms} & \textbf{N} &\textbf{Literature} \\ \hline \hline
\multirow{22}{*}{Supervised}     & Decision Tree  & 15&\begin{tabular}[c]{@{}l@{}}~\cite{gupta2014bit,dewan2015towards,dewan2017facebook,rathod2015comparative,cao2016detection,eshete2014webwinnow}\\~\cite{pradeepthi2014performance,basnet2015towards,vermaK15,marchal:phishscore,burnap2015real,marchal:phishstorm,mamun2016detecting,darling2015lexical,feroz2014examination}\end{tabular}          \\ 
\cline{2-4} 
                                 & Bayesian Classifier &1&  ~\cite{rathod2015comparative}          \\ \cline{2-4} 
                                 & Logistic Regression   &11&   \begin{tabular}[c]{@{}l@{}} ~\cite{darling2015lexical,ma2009beyond,feroz2014examination,basnet2012mining,su2013suspicious,eshete2014webwinnow}\\~\cite{basnet2015towards,vermaK15,ma2011learning,nepali2016you,garera2007framework} \end{tabular}       \\ \cline{2-4} 
                                 & Na\"ive Bayes    & 11&  \begin{tabular}[c]{@{}l@{}}                     ~\cite{darling2015lexical,ma2009beyond,feroz2014examination,pradeepthi2014performance,nepali2016you,dewan2015towards}\\~\cite{basnet2015towards,vermaK15,gupta2014bit,burnap2015real,dewan2017facebook} \end{tabular}       \\ \cline{2-4} 
                                 & Support Vector Machines &12 &    \begin{tabular}[c]{@{}l@{}} ~\cite{ma2009beyond,le2011phishdef,basnet2015towards,marchal:phishscore,marchal:phishstorm,vermaK15,dewan2015towards}\\~\cite{ma2011learning,chu2013protect,lee2013warningbird,nepali2016you,dewan2017facebook}  \end{tabular}     \\ \cline{2-4} 
                                 & RandomForest  & 14&  \begin{tabular}[c]{@{}l@{}}
                                 ~\cite{feroz2014examination,pradeepthi2014performance,dewan2017facebook,vermaK15,basnet2015towards,marchal:phishscore,marchal:phishstorm,cao2016detection}\\~\cite{mamun2016detecting,dewan2015towards,gupta2014bit,nepali2016you,eshete2014webwinnow,sananse2015phishing}   \end{tabular}      \\ \cline{2-4} 
                                 & BayesNet     & 4& ~\cite{feroz2014examination,cao2016detection,burnap2015real,eshete2014webwinnow}          \\ \cline{2-4} 
                                 & RandomTree  & 4& ~\cite{pradeepthi2014performance,marchal:phishscore,marchal:phishstorm,eshete2014webwinnow}           \\ \cline{2-4} 
                                 & AdaBoost    & 2& ~\cite{vermaK15,dewan2015towards}          \\
                                 \cline{2-4} 
                                 & Stacking    & 1& ~\cite{vermaK15}   \\ 
                                  \cline{2-4} 
                                 & RNN-LSTM    & 1& ~\cite{bahnsen2017classifying}   \\ 
                                 \cline{2-4} 
                                 & LMT  & 3&   ~\cite{pradeepthi2014performance,marchal:phishscore,marchal:phishstorm}        \\  \cline{2-4} 
                                 & k-Nearest Neighbors   & 3& ~\cite{darling2015lexical,pradeepthi2014performance,mamun2016detecting}          \\
                                 \hline
                                 
\multirow{1}{*}{Unsupervised}                      & k-Means Clustering     & 2& \begin{tabular}[c]{@{}l@{}} ~\cite{catakoglu2016automatic,feroz2015phishing}\end{tabular}           \\ \hline
\multirow{8}{*}{\begin{tabular}[c]{@{}c@{}}Online\\Learning\end{tabular}} & Perceptron  & 5& ~\cite{le2011phishdef,verma2017s,ma2011learning,zhao2013cost,gyawali2011evaluating}          \\ \cline{2-4} 
                                 & Multilayer Perceptron      & 3&~\cite{pradeepthi2014performance,basnet2015towards,burnap2015real}           \\ \cline{2-4} 
                                 & Confidence Weighted &8  & \begin{tabular}[c]{@{}l@{}}~\cite{le2011phishdef,verma2017s,egan2011evaluation,ma2011learning}\\~\cite{zhao2013cost,blum2010lexical,huang2014malicious,lin2013malicious}   \end{tabular}        \\ \cline{2-4} 
                                 & AROW\tnote{a} &2  & ~\cite{le2011phishdef,verma2017s}           \\ \cline{2-4} 
                                 & Passive Aggressive&4 & ~\cite{ma2011learning,verma2017s,zhao2013cost,lin2013malicious} 	\\	\cline{2-4} 
                                 & Cost sensitive\tnote{b}&1     & ~\cite{zhao2013cost}           \\\cline{2-4} 
                                 & Not mentioned & 1& ~\cite{feroz2015phishing}\\	\hline
\multirow{4}{*}{\begin{tabular}[c]{@{}c@{}}Rule/Pattern \\based\end{tabular}} 	                     & PART&3 & ~\cite{marchal:phishscore,marchal:phishstorm,vermaK15}\\ \cline{2-4} 
								 & JRip &2& ~\cite{marchal:phishscore,marchal:phishstorm}	\\\cline{2-4} 
								 & Association Rule Mining &1 & ~\cite{jeeva2016intelligent}	\\ \cline{2-4}	
                                 & Greedy Pattern Selection &1& ~\cite{huang2014malicious}  \\		\cline{2-4}	
                                 & TFD Pattern Matching &1& ~\cite{yuan2013tfd}  \\\hline         
Others                           & \begin{tabular}[c]{@{}l@{}}MD5 Hash Algorithms,\\ HMM, BFTree\end{tabular} &3&  ~\cite{feroz2014examination,sananse2015phishing,lee2014users}          \\ \hline
\end{tabular}
\begin{tablenotes}
\item[a] \textit{Adaptive Regularization of Weights}
\item[b] \textit{Learners (CS-OAL, CS-Passive Aggressive, Label Efficient perceptron etc.)\\ mentioned in the paper} \cite{zhao2013cost}
\end{tablenotes}
\end{threeparttable}}
\end{table}


\begin{table}[h]
\centering
\caption{\hl{Methods used} in \hl{reviewed} phishing website detection \hl{literature}}
\label{table-webpage-classification}
\resizebox{1\columnwidth}{!}{
\begin{tabular}{|l|c|l|}
\hline
\multicolumn{1}{|c|}{\textbf{Method}}          & \textbf{N}   & \multicolumn{1}{c|}{\textbf{Literature}}                                                                                                                                                                                                                     \\ \hline \hline
SVM                & 20  & \begin{tabular}[c]{@{}l@{}}\cite{manek2014demalfier, pao2012malicious, xu2014evasion, choi2011detecting, yue2013fine,he2011efficient, moghimi2016new,gowtham2014comprehensive,raodetection}\\\cite{mohaisen2015towards, stringhini2013shady, bannur2011judging, Gowtham2014,corona2017deltaphish,abbasi2015enhancing}\\\cite{miyamoto2008evaluation,zhang2014domain,kosba2014adam,el2017detection,zhang2017phishing}\end{tabular} \\ \hline
Na\"ive Bayes                    & 12          & \begin{tabular}[c]{@{}l@{}}\cite{manek2014demalfier, canali2011prophiler, xu2014evasion, feroz2015examination,thabtah2016dynamic,abbasi2015enhancing}\\ \cite{miyamoto2008evaluation,zhang2014domain,dong2015beyond,zhuang2012intelligent,zhang:textual,aggarwal2012phishari}  \end{tabular}                                                                                    \\ \hline
Logistic Regression               & 14       & \begin{tabular}[c]{@{}l@{}}\cite{manek2014demalfier, canali2011prophiler, bannur2011judging, feroz2015examination,zhang2014domain,abbasi2015enhancing,xu2014evasion,thabtah2016dynamic,el2017detection}\\ \cite{miyamoto2008evaluation,dong2015beyond,Mohammad2014,thomas2011design,raodetection} \end{tabular}                                                                                \\ \hline
Random Tree           & 2                   & \cite{canali2011prophiler},\cite{vargas:enemies}                                                                                                                                                                                                                       \\ \hline
Random Forest                   & 7         & 
\begin{tabular}[c]
{@{}l@{}}\cite{canali2011prophiler, feroz2015examination, miyamoto2008evaluation, lee2016phishing,aggarwal2012phishari, zhang2014domain,raodetection}  \end{tabular}                                                                                                                                                        \\ \hline
Decision Tree              &    14       & \begin{tabular}[c]{@{}l@{}}\cite{canali2011prophiler, xu2014evasion, feroz2015examination, miyamoto2008evaluation,thabtah2016dynamic,mohammad2014intelligent,el2017detection, thakur2014catching} \\ \cite{abbasi2015enhancing,abdelhamid2014phishing,aggarwal2012phishari,dong2015beyond,aburrous2010intelligent,raodetection}\end{tabular}                                                                                                                                                          \\ \hline
CART    & 1& ~\cite{el2017detection} \\\hline
Bayesian Network    & 4                     & \cite{canali2011prophiler, feroz2015examination,raodetection, zhang:textual}      
\\ \hline
Multilayer Perceptron & 1 & \cite{raodetection} \\ \hline
BFTree             & 1            & \cite{feroz2015examination}
                                                                                       \\ \hline
Gradient Boosting     & 1                   & \cite{marchal2016know}                                                                                                                                                                                                                            \\ \hline
Tree Kernel & 1 & \cite{abbasi2015enhancing} \\ \hline

\begin{tabular}[c]{@{}l@{}}Neural Network     \end{tabular}          & 6           & \cite{thabtah2016dynamic, zhang2017two,barraclough2013intelligent,miyamoto2008evaluation,el2017detection,zhang2017phishing}                                                                                     
\\ \hline
\begin{tabular}[c]{@{}l@{}}Regression Trees (RT) \\Bayesian Additive RT\\ Bagging      \end{tabular}   & 1                  & \cite{miyamoto2008evaluation}                         \\ \hline
AdaBoost & 3 & \cite{miyamoto2008evaluation,raodetection,el2017detection} \\ \hline
Bagging & 1 & \cite{el2017detection} \\ \hline
k-Nearest Neighbor             & 4                       &\begin{tabular}[c]
{@{}l@{}} \cite{yue2013fine,dong2015beyond, choi2011detecting, el2017detection}
\end{tabular}                                                                                                                                                                                                                                \\ \hline
Hierarchical Clustering & 1 & \cite{zhuang2012intelligent} \\ \hline
DBSCAN Clustering & 1 & \cite{liu2010automatic} \\ \hline
Association rule & 3 & \cite{abdelhamid2014phishing,mohammad2014intelligent,aburrous2010intelligent} \\ \hline
\end{tabular}}
\end{table}


\begin{table}[h]
\centering
\caption{\hl{Methods used} in reviewed phishing email detection literature. EM- Expectation Maximization}
\label{table-email-classification}
\resizebox{1\columnwidth}{!}{
\begin{threeparttable}
\begin{tabular}{|l|c|l|}
\hline
\multicolumn{1}{|c|}{\textbf{Method}}     &\textbf{N}        & \multicolumn{1}{c|}{\textbf{Literature}}                                                                                                                                                                                                                                                                                                                                                 
\\ \hline

Bagging &3    & \cite{masoumeh:feature,chowdhury:multilayer,DBLP:conf/ecir/GanstererP09}

			\\ \hline
Boosting        &6  & \begin{tabular}[c]
{@{}l@{}} \cite{isredzaRahmi:profiling}\cite{khonji:lexial,Isredza:phishing,Islam:2013:MPD:2405859.2406232,ramanathan}\\\cite{DBLP:conf/ecir/GanstererP09} 
\end{tabular}        			
						
                        \\ \hline
Bayesian Network& 2                        & \cite{Isredza:phishing,chowdhury:multilayer}
                      \\ \hline  
                        
     Decision Table   &2                              & \cite{Isredza:phishing,isredzaRahmi:profiling}
\\ \hline

Decision Tree       &8             & \begin{tabular}[c]{@{}l@{}}
\cite{Chiang:2015:AMC:2778890.2779103,masoumeh:feature,DBLP:conf/ecir/GanstererP09,toolan:ensembles}\\
\cite{chowdhury:multilayer,Islam:2013:MPD:2405859.2406232, DBLP:journals/corr/YasinA16,fergus:feature}
\end{tabular}  



\\ \hline
k-means              &4       &\cite{Seifollahi2017,Dazeley2010,Ammar:online,john:applying}   

\\ \hline
\begin{tabular}[c]
{@{}l@{}}
k-Nearest Neighbor\\ (kNN) \end{tabular}&5                                    &\begin{tabular}[c]
{@{}l@{}} \cite{Ho2017DetectingCS,chowdhury:multilayer,DBLP:conf/ecir/GanstererP09,Dazeley2010,toolan:ensembles}
\end{tabular}                                                                          \\ \hline
\begin{tabular}[c]{@{}l@{}}Logistic Regression\end{tabular}& 1 & \cite{toolan:ensembles}
\\ \hline
\begin{tabular}[c]
{@{}l@{}}
Neural Network    \end{tabular}  &2                         & \cite{Ammar:evolving,Ho2017DetectingCS} 
\\ \hline
Na\"ive Bayes               &6              & \begin{tabular}[c]{@{}l@{}}\cite{Chiang:2015:AMC:2778890.2779103,isredzaRahmi:profiling, DBLP:conf/ecir/GanstererP09,Ammar:online}\\
\cite{Islam:2013:MPD:2405859.2406232,DBLP:journals/corr/YasinA16}\end{tabular} 

\\ \hline




Random Forest           &9                & 
\begin{tabular}[c]
{@{}l@{}}\cite{isredzaRahmi:profiling,khonji:lexial,chowdhury:multilayer,Andronicus:classification,DBLP:conf/ecir/GanstererP09,Ammar:online,Isredza:phishing} \\\cite{DBLP:journals/corr/YasinA16,Islam:2013:MPD:2405859.2406232}
\end{tabular}                                                                       
\\
\hline


SVM          &13         & \begin{tabular}[c]{@{}l@{}}
\cite{Stringhini2015,Duman:emailprofiler,Ammar:online,DBLP:conf/ecir/GanstererP09,Islam:2013:MPD:2405859.2406232,adre:new,chandrasekaran2006phishing}\\
\cite{bergholz2008improved,Chiang:2015:AMC:2778890.2779103,Figueroa2017,DBLP:journals/corr/YasinA16,Olivo:2013,toolan:ensembles}
\end{tabular}
\\ \hline 
\begin{tabular}[c]{@{}l@{}}EM Clustering \end{tabular}& 1 & \cite{orman:semantics} \\ \hline
Multi-layer Perceptron & 2 & \cite{DBLP:journals/corr/YasinA16,Ammar:online} \\ \hline





Other\tnote{a} &7                            &  \begin{tabular}[c]
{@{}l@{}}
\cite{Chiang:2015:AMC:2778890.2779103,Seifollahi2017,Ho2017DetectingCS,isredzaRahmi:profiling,chowdhury:multilayer,Dazeley2010,Ammar:evolving}\\ 
\end{tabular}
\\ \hline
\end{tabular}
\begin{tablenotes}
\item[a] \textit{OneR \cite{isredzaRahmi:profiling}, Multiple Linear Regression (MLR)\cite{Ammar:evolving}, SMO\cite{chowdhury:multilayer}, Associative Petri network \cite{Chiang:2015:AMC:2778890.2779103},
multinomial Naive Bayes (MNB),  DCClust, MS-MGKM, k-committees \cite{Dazeley2010}, INCA \cite{Seifollahi2017}, Kernel Density Estimation (KDE), Gaussian Mixture Model (GMM) \cite{Ho2017DetectingCS}}
\end{tablenotes}
\end{threeparttable}}
\end{table}

\begin{table*}[!htb]
\centering
\caption{Range of dataset size used by each work for evaluation, separately for legitimate, spam and malware URLs}
\label{tab:sizedatasets}
\resizebox{2\columnwidth}{!}{
\begin{tabular}{|c|c|c|c|c|c|c|c|c|}
\hline
                           &          & \multicolumn{7}{c|}{\textbf{Legitimate}}                                                                                       \\ \hline
                           &          & \textbf{100s} & \textbf{1,000s}  & \textbf{10,000s } & \textbf{100,000s} & \textbf{$\geq$1M} & \textbf{Feed} & \textbf{N/A}    \\ \specialrule{.2em}{.1em}{.1em}
\multirow{5}{*}{\textbf{Phishing}}  & \textbf{100s }    & \begin{tabular}[c]{@{}l@{}}~\cite{guan2009anomaly,astorino2016malicious}\\~\cite{eshete2014webwinnow,jeeva2016intelligent}\end{tabular} & ~\cite{jeeva2016intelligent} &   &   &  &      &                        \\ \cline{2-9} 
                           & \textbf{1,000s}   & &\begin{tabular}[c]{@{}c@{}}~\cite{le2011phishdef,pradeepthi2014performance,burnap2015real} \\~\cite{gupta2014bit,garera2007framework} \end{tabular} & ~\cite{ma2009beyond} &  & ~\cite{sha2015limited} &                 &     \\ \cline{2-9}   & \textbf{10,000s}  &   & ~\cite{nguyen2014novel,cao2016detection,nguyen2013detecting}   & \begin{tabular}[c]{@{}c@{}}~\cite{dewan2017facebook,ma2009beyond,basnet2012mining}\\~\cite{lee2014users,basnet2015towards,vermaK15,marchal:phishscore,marchal:phishstorm}\\~\cite{chu2013protect,zhao2013cost}\\~\cite{blum2010lexical,dewan2015towards,mamun2016detecting} \end{tabular}& \begin{tabular}[c]{@{}c@{}}~\cite{dewan2017facebook,basnet2015towards,mamun2016detecting}\\~\cite{chu2013protect,nepali2016you,khonji2011novel}\\~\cite{dewan2015towards,su2013suspicious,zhao2013cost} \end{tabular}   & \begin{tabular}[c]{@{}l@{}} ~\cite{darling2015lexical,dewan2017facebook}\\~\cite{lee2014users,dewan2015towards} \end{tabular} &   &
                           \\ \cline{2-9} 
                           & \textbf{100,000s} &   &  &  &~\cite{xiong2015mird,gabriel2016detecting,zhang2016url} & ~\cite{ma2011learning} &  &\\ \cline{2-9} 
                           & \textbf{$\geq$1M }&   &  &  &  &\begin{tabular}[c]{@{}l@{}} 
~\cite{popescu2015study,verma2017s,ma2011learning}\\~\cite{lin2013malicious,xiong2015mird,chhabra2011phi}\\~\cite{lee2013warningbird,lee2014users,bahnsen2017classifying} \end{tabular} &  & ~\cite{yuan2013tfd}\\  \cline{2-9} 
                           & \textbf{Feed} &   &  &  &  & & ~\cite{blum2010lexical},~\cite{jeun2013collecting} &\\ 
                           \cline{2-9} 
                           & \textbf{N/A} &   &  &  &  & &  & \begin{tabular}[c]{@{}l@{}} \cite{rathod2015comparative,egan2011evaluation,zhang2014web}\\ \cite{feroz2014examination,canova2015learn,sananse2015phishing}\\ \cite{feroz2015phishing} \end{tabular}\\
\specialrule{.2em}{.1em}{.1em}                           
\multirow{3}{*}{\textbf{Spam}} & \textbf{1,000s}     &   &   &    &   & ~\cite{sha2015limited} &  &                           \\ \cline{2-9} 
                           & \textbf{10,000s}   &     &  & ~\cite{ma2009beyond}  &~\cite{mamun2016detecting}    &~\cite{verma2017s} &                       &     \\ \cline{2-9} 
                           & \textbf{Feed}  & &  &   &    &     &   ~\cite{blum2010lexical},~\cite{jeun2013collecting}   &                  \\ 
\specialrule{.2em}{.1em}{.1em}
\multirow{2}{*}{\textbf{Malware}} & \textbf{1,000s}     &   & ~\cite{le2011phishdef,vu2016firstfilter}  &    &   & \begin{tabular}[c]{@{}l@{}} ~\cite{darling2015lexical,xiong2015mird}\\~\cite{sha2015limited,lin2013malicious} \end{tabular} &                           &  \\ \cline{2-9} 
                           & \textbf{10,000s}   &     &  &   & ~\cite{mamun2016detecting}   & &     &                       \\ 
 \specialrule{.2em}{.1em}{.1em} 
\end{tabular}}
\end{table*}

\begin{table*}[h]
\centering
\caption{Range of dataset size used by each work for evaluation, separately for legitimate, phishing and malicious websites}
\label{table-webpage-dataset-size}
\resizebox{2\columnwidth}{!}{
\begin{tabular}{|c|c|c|c|c|c|c|c|}
\hline
                           &          & \multicolumn{6}{c|}{\textbf{Legitimate}}                                             \\ \hline
&          & \textbf{100s} & \textbf{1,000s} & \textbf{10,000s} & \textbf{100,000s} & \textbf{$\geq$1 M}     & \textbf{N/A} \\ \specialrule{.2em}{.1em}{.1em}
\multirow{5}{*}{\textbf{Phishing}}  & \textbf{100s}     & \begin{tabular}[c]{@{}c@{}}\cite{tan2014phishing, xu2014gemini, Mohammad2014}\\\cite{he2011efficient, chiew2015utilisation,huang2010mitigate}\end{tabular} & \begin{tabular}[c]{@{}c@{}}\cite{zhang2017phishing, yue2013fine, zhang2017two, dong2010defending}\\\cite{Xiang:2011:CFM:2019599.2019606}\end{tabular}      & & & & \\ \cline{2-8} 
 & \textbf{1,000s}   & \begin{tabular}[c]{@{}c@{}}\cite{Gowtham2014, mohammad2014intelligent, gowtham2014comprehensive}\\\cite{moghimi2016new}\end{tabular}                        & \begin{tabular}[c]{@{}c@{}}\cite{abdelhamid2014phishing, tan2016phishwho, abbasi2015enhancing, mao:alarm}\\\cite{ stringhini2013shady, thabtah2016dynamic, liu2010automatic, xiang2010hierarchical}\\\cite{ ramesh2014efficacious,pao2012malicious,miyamoto2008evaluation,raodetection,el2017detection} \end{tabular}& \cite{geng2015combating, zhang:textual} & \cite{marchal2016know, aggarwal2012phishari} & &  \begin{tabular}[c]{@{}c@{}}\cite{zhang2014domain, varshney2016phish}\\\cite{aburrous2010intelligent, mao2017phishing} \end{tabular}  \\ \cline{2-8} 
& \textbf{10,000s}  & & & \begin{tabular}[c]{@{}c@{}}\cite{marchal2012proactive, zhuang2012intelligent, thakur2014catching}\\\cite{feroz2015examination,wardman2011high,choi2011detecting, lee2016phishing} \end{tabular}& \cite{dong2015beyond}                        & \cite{pao2012malicious}  & \begin{tabular}[c]{@{}c@{}}\cite{mohaisen2015towards,wenyin2012antiphishing}\\\cite{ramesh2017intelligent}\end{tabular}\\ \cline{2-8} 
  & \textbf{100,000s} &                                            & & & & \cite{whittaker2010large} & \cite{britt2012clustering}\\ \cline{2-8} 
& \textbf{N/A} &  & \cite{vargas:enemies} & & & & \begin{tabular}[c]{@{}c@{}}\cite{nguyen2013detecting,maurer2012using,mohammad2014intelligent,rajab2017new} \\\cite{thomas2011design,fang2015proactive,corona2017deltaphish}\\\cite{barraclough2013intelligent,rosiello2007layout,liang2016cracking}\\\cite{shahriar2012trustworthiness,wardman2009identifying,bannur2011judging}\end{tabular}\\ \specialrule{.2em}{.1em}{.1em}
\multirow{3}{*}{\textbf{Malicious}} & \textbf{100s}     &   & & & &    \cite{liang2009malicious} & \\ \cline{2-8} 
& \textbf{1,000s}   & & \cite{yue2013fine} & & & & \cite{mohaisen2015towards}        \\ \cline{2-8} 
& \textbf{10,000s}  & & \cite{stringhini2013shady} & \cite{kosba2014adam} & \cite{canali2011prophiler, xu2014evasion}    &               &              \\ \specialrule{.2em}{.1em}{.1em}
\end{tabular}}
\end{table*}

\begin{table*}[h]
\centering
\caption{Range of dataset size used by each work for evaluation, separately for legitimate and phishing emails}
\label{table-email-dataset-size}
\begin{tabular}{|c|c|c|c|c|c|c|c|}
\hline
                           &          & \multicolumn{6}{c|}{\textbf{Legitimate}}                                             \\ \hline
&          & \textbf{100s} & \textbf{1,000s} & \textbf{10,000s} & \textbf{100,000s} & $\geq$\textbf{1 M}     & \textbf{N/A} \\ \specialrule{.2em}{.1em}{.1em}
\multirow{5}{*}{\textbf{Phishing}}  & \textbf{100s}     & \multicolumn{1}{c|}{\begin{tabular}[c]{@{}c@{}}\cite{chowdhury:multilayer, orman:semantics,Olivo:2013,chandrasekaran2006phishing}
\end{tabular}} & \begin{tabular}[c]{@{}c@{}}\cite{fette2007learning,Chiang:2015:AMC:2778890.2779103,Ammar:online,Andronicus:classification}\\ \cite{Ammar:evolving}
\end{tabular}
& & & &
\\ \cline{2-8} 
 & \textbf{1,000s}   & & \begin{tabular}[c]{@{}c@{}}\cite{Ammar:online,DBLP:journals/corr/YasinA16,Isredza:phishing,verma:phish,verma:detecting,fergus:feature}\\\cite{khonji:lexial,masoumeh:feature,toolan:ensembles}\\
 \end{tabular}&  \cite{vermaH13,adre:new,isredzaRahmi:profiling,bergholz2008improved}&  & &  \begin{tabular}[c]{@{}c@{}}\cite{john:applying, Dazeley2010,yearwood:profiling,Seifollahi2017,verma:detecting} \\\cite{husak:phigaro,kim:semantic}\end{tabular} 
 \\ \cline{2-8} 
& \textbf{10,000s}  & & \cite{ramanathan} & \begin{tabular}[c]{@{}c@{}}\cite{Figueroa2017,DBLP:conf/ecir/GanstererP09}
\end{tabular}&                         & & \\ \cline{2-8} 
  & \textbf{100,000s} &                                            & & & &  & 
  \\ \cline{2-8} 
& \textbf{N/A} &  &  \begin{tabular}[c]{@{}c@{}} \cite{Dazeley2010,Seifollahi2017}\\
\end{tabular}&  \begin{tabular}[c]{@{}c@{}} \cite{huang2017gossip} 
\end{tabular}&\cite{Stringhini2015} & & \cite{Islam:2013:MPD:2405859.2406232}
\\ \specialrule{.2em}{.1em}{.1em}
\multirow{1}{*}{\textbf{Spam}}
& \textbf{100s}     &   &\cite{Chiang:2015:AMC:2778890.2779103} &  & &   &    
\\
\cline{2-8}
& \textbf{1,000s}     &   &\cite{fergus:feature} & & &   & \\

\cline{2-8}
& \textbf{10,000s}     &   & & \cite{DBLP:conf/ecir/GanstererP09,adre:new,Stringhini2015} & &   & \\
\specialrule{.2em}{.1em}{.1em}

\multirow{1}{*}{\textbf{Malicious}} & \textbf{100s}     & 
& \cite{Chiang:2015:AMC:2778890.2779103} & & &   &    \\ 
\specialrule{.2em}{.1em}{.1em}
\multirow{3}{*}{\textbf{Spear}}  & \textbf{100s}     &  &  & & \cite{Stringhini2015}&   &   \\ 
 \cline{2-8}
 & \textbf{1000s}     &  &  & \cite{Han:2016:ASP:2851613.2851801}& &   &   \\ 
 \cline{2-8}
  & \textbf{N/A}  & \cite{Laszka:2015:OPF:2887007.2887140} & & & \cite{Duman:emailprofiler}& \cite{Ho2017DetectingCS}  & \cite{Laszka:2016:MSF:3015812.3015893}   \\ 
\specialrule{.2em}{.1em}{.1em}
\end{tabular}
\end{table*}

\begin{table*}[!htb]
 	\caption{Description of Features Extracted from URLs, Websites (marked with \textbf{uW}) and Emails (marked with \textbf{uE}). N - Number of papers using each feature}
     \label{tab:urlfeatures}
     \centering
     \resizebox{2\columnwidth}{!}{
     \begin{threeparttable}
 	\begin{tabular}{|c|l|l|l|l|}
 	\hline
 \textbf{Type}      & \textbf{Features} & \textbf{N} & \textbf{Used in} & \textbf{Literature}  \\ 
 \hline \hline
 \multirow{7}{*}{\textbf{Lexical}}   & URL length & 43 & U,W,E & \begin{tabular}[c]{@{}l@{}}  ~\cite{darling2015lexical,egan2011evaluation,ma2009beyond,jeun2013collecting,ma2011learning,feroz2014examination,vermaK15, blum2010lexical,mamun2016detecting,jeeva2016intelligent,zhang2016url,le2011phishdef,feroz2015phishing,basnet2015towards,lin2013malicious,sananse2015phishing,gyawali2011evaluating,bahnsen2017classifying}\\~uW:~\cite{el2017detection,manek2014demalfier,rajab2017new,marchal2016know, pao2012malicious, xu2014evasion, mohaisen2015towards, choi2011detecting, kosba2014adam, raodetection, yue2013fine,lee2016phishing,abdelhamid2014phishing,zhang2017two,thomas2011design,Mohammad2014,thabtah2016dynamic,aggarwal2012phishari,mohammad2014intelligent,moghimi2016new,barraclough2013intelligent,aburrous2010intelligent,thakur2014catching} uE:\cite{verma:detecting}\end{tabular}  \\
 \cline{2-5} & Length of URL params\tnote{a} & 15 & U,W & \begin{tabular}[c]{@{}l@{}}  ~\cite{darling2015lexical,egan2011evaluation,ma2009beyond,dewan2015towards,ma2011learning,zhao2013cost,su2013suspicious,vu2016firstfilter,blum2010lexical,le2011phishdef,feroz2015phishing,bahnsen2017classifying}\\ ~uW:~\cite{zhuang2012intelligent,pao2012malicious,choi2011detecting}\end{tabular} \\
 \cline{2-5} & Token/Word Frequency in URL & 22 & U,W,E & \begin{tabular}[c]{@{}l@{}} ~\cite{rathod2015comparative,darling2015lexical,egan2011evaluation,ma2009beyond,gyawali2011evaluating,jeeva2016intelligent,chu2013protect,blum2010lexical,zhao2013cost,vu2016firstfilter, ma2011learning, sananse2015phishing,feroz2015phishing,marchal:phishstorm,marchal:phishscore,khonjiIJ11}\\~uW: \cite{marchal2016know, kosba2014adam, huang2010mitigate,thomas2011design,choi2011detecting}, uE: \cite{Andronicus:classification}
 \end{tabular}\\
 \cline{2-5} &Black-List Word frequency in URL & 14 & U,W,E & \begin{tabular}[c]{@{}l@{}}  ~\cite{darling2015lexical,ma2009beyond,zhang2016url,bahnsen2017classifying},~uW:~\cite{xu2014gemini,marchal2012proactive,tan2016phishwho,Xiang:2011:CFM:2019599.2019606},~uE: \cite{khonji:lexial,Ammar:online,adre:new,isredzaRahmi:profiling,Andronicus:classification,bergholz2008improved,fergus:feature}\end{tabular}
\\
 \cline{2-5} & Digit/Letter Frequencies/Ratio& 7 & U &\begin{tabular}[c]{@{}l@{}}  ~\cite{darling2015lexical,lin2013malicious,egan2011evaluation,guan2009anomaly,zhao2013cost,mamun2016detecting,vermaK15} \end{tabular} \\
 \cline{2-5} & Number of dots (.) & 30 & U,W,E & \begin{tabular}[c]{@{}l@{}}~\cite{rathod2015comparative,ma2009beyond,feroz2014examination,pradeepthi2014performance,le2011phishdef,basnet2015towards,mamun2016detecting,jeeva2016intelligent,ma2011learning,gyawali2011evaluating}, uE:\cite{khonji:lexial,chowdhury:multilayer,adre:new,isredzaRahmi:profiling,Andronicus:classification,DBLP:journals/corr/YasinA16,bergholz2008improved,Olivo:2013,fergus:feature,toolan:ensembles}\\ ~uW:\cite{pao2012malicious,aggarwal2012phishari,mohammad2014intelligent,moghimi2016new,he2011efficient,gowtham2014comprehensive,zhang2014domain,barraclough2013intelligent,Xiang:2011:CFM:2019599.2019606,raodetection,zhang2017phishing} 
 \end{tabular} \\
 \cline{2-5} & Character\tnote{b}~ frequency in URL path & 25 &  U,W,E & \begin{tabular}[c]{@{}l@{}}  ~\cite{darling2015lexical,ma2009beyond,gyawali2011evaluating,lin2013malicious,basnet2015towards,mamun2016detecting,jeeva2016intelligent,sananse2015phishing}, uE: 
\cite{chowdhury:multilayer, Islam:2013:MPD:2405859.2406232} \\ ~uW: \cite{marchal2016know, mohaisen2015towards, choi2011detecting, kosba2014adam, yue2013fine, lee2016phishing, canali2011prophiler, miyamoto2008evaluation, abdelhamid2014phishing,moghimi2016new,zhang2014domain,he2011efficient,Mohammad2014,thabtah2016dynamic}
 \end{tabular}\\
  \cline{2-5} & Kolmogorov Complexity & 2 & U,W & \begin{tabular}[c]{@{}l@{}}~\cite{vermaK15}~uW: \cite{pao2012malicious} \end{tabular} \\
  \cline{2-5} & Character N-grams & 2 & U & \begin{tabular}[c]{@{}l@{}}~\cite{verma2017s,yuan2013tfd} \end{tabular} \\
   \cline{2-5} & URL Entropy & 1 & U & \begin{tabular}[c]{@{}l@{}}~\cite{bahnsen2017classifying} \end{tabular} \\
  \cline{2-5} & \begin{tabular}[c]{@{}l@{}}Edit distance, KL-Divergence,\\ KS-test\end{tabular} &3 & U,W & \begin{tabular}[c]{@{}l@{}}~\cite{vermaK15,bahnsen2017classifying}, uW: \cite{thakur2014catching} \end{tabular} \\
  
 \hline
 \multirow{6}{*}{\textbf{Obfuscation}}            &  Frequency of Special Characters & 32 & U,W,E & \begin{tabular}[c]{@{}l@{}} ~\cite{mamun2016detecting,garera2007framework,gyawali2011evaluating,feroz2014examination,pradeepthi2014performance,basnet2015towards,jeeva2016intelligent,bahnsen2017classifying}, uE: \cite{khonji:lexial,DBLP:conf/ecir/GanstererP09,fette2007learning,Isredza:phishing,isredzaRahmi:profiling,fergus:feature} 
 \\ ~uW: \cite{el2017detection,Mohammad2014, manek2014demalfier, pao2012malicious, xu2014evasion, raodetection, mohaisen2015towards,Xiang:2011:CFM:2019599.2019606,kosba2014adam,zhang2017two,Gowtham2014,rajab2017new,abdelhamid2014phishing,mohammad2014intelligent,he2011efficient,gowtham2014comprehensive,zhang2014domain,aburrous2010intelligent}    \end{tabular} 
 \\
 \cline{2-5} & Obfuscation in IP Address & 8 & U,W,E &\begin{tabular}[c]{@{}l@{}}~\cite{egan2011evaluation, guan2009anomaly, popescu2015study, ma2009beyond,zhao2013cost,gabriel2016detecting}, uW:\cite{rajab2017new}, uE: \cite{khonji:lexial}\end{tabular}     \\
  \cline{2-5} & Encodings in URL & 9 & U,W,E &  \begin{tabular}[c]{@{}l@{}} ~\cite{mamun2016detecting,garera2007framework,gyawali2011evaluating,feroz2014examination,jeeva2016intelligent}, uW:\cite{zhang2014domain,Mohammad2014,mohammad2014intelligent}, uE: \cite{chowdhury:multilayer}                 \end{tabular}     \\
 \cline{2-5} & Shortening in URL & 7 & U,W & \begin{tabular}[c]{@{}l@{}}
 	~\cite{rathod2015comparative, gupta2014bit,chhabra2011phi,nepali2016you,zhang2016url}, uW: \cite{rajab2017new,el2017detection}

 	                   \end{tabular}      \\
 \cline{2-5} & Spoofing in URL path &3 & U,W & \begin{tabular}[c]{@{}l@{}}
 	~\cite{mamun2016detecting,garera2007framework}, uW:~\cite{Mohammad2014}\end{tabular}       \\
 \cline{2-5} & Source/destination URLs mismatch & 5 & U,W,E & \begin{tabular}[c]{@{}l@{}}
 	~\cite{lee2013warningbird, gupta2014bit,chhabra2011phi}, uW: \cite{thomas2011design}, uE: \cite{Dazeley2010} 
 	            \end{tabular}      \\
 \cline{2-5} & IP address instead of domain name & 55 & U,W,E & \begin{tabular}[c]{@{}l@{}}
 ~\cite{rathod2015comparative,mamun2016detecting,gyawali2011evaluating,darling2015lexical,ma2009beyond,zhao2013cost,blum2010lexical,pradeepthi2014performance,le2011phishdef,basnet2015towards,jeeva2016intelligent,egan2011evaluation,guan2009anomaly,lin2013malicious,vermaK15,bahnsen2017classifying}\\~uW: \cite{raodetection,zhang2017two,el2017detection,zhang2017phishing, manek2014demalfier, Xiang:2011:CFM:2019599.2019606,kosba2014adam, miyamoto2008evaluation,he2011efficient}\\~uW:~\cite{canali2011prophiler,mohaisen2015towards,rajab2017new,yue2013fine,stringhini2013shady, abdelhamid2014phishing,zhang2014domain, whittaker2010large, Gowtham2014, bannur2011judging,Mohammad2014,thabtah2016dynamic,liang2016cracking,mohammad2014intelligent,moghimi2016new,barraclough2013intelligent,aburrous2010intelligent} \\ ~uE: ~\cite{fette2007learning,Isredza:phishing,isredzaRahmi:profiling, Ammar:online,chowdhury:multilayer,masoumeh:feature, orman:semantics,chandrasekaran2006phishing,DBLP:journals/corr/YasinA16, bergholz2008improved,Olivo:2013,fergus:feature,toolan:ensembles} 
 \end{tabular}     
      \\ 
      \cline{2-5} & Hostname obfuscation~\tnote{f} &15 & U,W,E & \begin{tabular}[c]{@{}l@{}}~\cite{rathod2015comparative,darling2015lexical,egan2011evaluation,guan2009anomaly,zhao2013cost,ma2011learning,sha2015limited,astorino2016malicious,basnet2015towards,mamun2016detecting,garera2007framework}, uW:~\cite{barraclough2013intelligent,aburrous2010intelligent}, uE: \cite{chandrasekaran2006phishing,DBLP:journals/corr/YasinA16}
      \end{tabular} \\   
 \hline
 \multirow{3}{*}{\textbf{DNS based}}   &  Misspelled/Bad domain name &8 & U & \begin{tabular}[c]{@{}l@{}}
 	~\cite{egan2011evaluation,nguyen2014novel,popescu2015study,ma2009beyond,jeun2013collecting,zhao2013cost, khonji2011novel,zhang2016url} \end{tabular} \\
 \cline{2-5} & Top Level Domain features & 22 & U,W,E & \begin{tabular}[c]{@{}l@{}}~\cite{egan2011evaluation,popescu2015study,ma2009beyond,gyawali2011evaluating,blum2010lexical,astorino2016malicious,popescu2016practical, lin2013malicious,lee2014users},\\~uW: \cite{kosba2014adam,Xiang:2011:CFM:2019599.2019606, stringhini2013shady,barraclough2013intelligent,marchal2012proactive,gowtham2014comprehensive,thabtah2016dynamic,liang2016cracking,thakur2014catching}, uE: \cite{Stringhini2015}, \cite{chowdhury:multilayer,fergus:feature,john:applying} 
 \end{tabular} \\
 \cline{2-5} & TTL\tnote{c}~ value of DNS & 9 & U,W,E & \begin{tabular}[c]{@{}l@{}}~\cite{ma2009beyond,basnet2012mining,lee2014users,ma2011learning}, uW: \cite{canali2011prophiler, lee2016phishing,Gowtham2014,zhang2017phishing},~uE:~\cite{huang2017gossip} \end{tabular} \\
 \cline{2-5} & Age of Domain & 13 & U,W,E& ~\cite{pradeepthi2014performance}, uW:~\cite{el2017detection,Gowtham2014,thabtah2016dynamic,abdelhamid2014phishing,zhang2017two,mohammad2014intelligent,gowtham2014comprehensive,zhang2014domain,Xiang:2011:CFM:2019599.2019606,raodetection,Mohammad2014}, uE: \cite{chowdhury:multilayer} \\
 \cline{2-5} & Ranking\tnote{d}~ based features & 30 & U,W,E& \begin{tabular}[c]{@{}l@{}} ~\cite{nguyen2013detecting, nguyen2014novel,marchal:phishstorm,marchal:phishscore,pradeepthi2014performance, chu2013protect, sananse2015phishing, garera2007framework, khonji2011novel, basnet2015towards, ma2011learning, guan2009anomaly,bahnsen2017classifying}, uE:~\cite{huang2017gossip,chowdhury:multilayer}\\~uW:~\cite{el2017detection,lee2016phishing,yue2013fine,whittaker2010large,thabtah2016dynamic,tan2014phishing, choi2011detecting, Gowtham2014, rajab2017new, he2011efficient,gowtham2014comprehensive,Xiang:2011:CFM:2019599.2019606,raodetection,tan2016phishwho,thakur2014catching} \end{tabular}\\
 \hline
 \multirow{8}{*}{\begin{tabular}[c]{@{}c@{}}\textbf{Hostname}\\ \textbf{based}\end{tabular}} & Frequency of Tokens & 12 & U,W & \begin{tabular}[c]{@{}l@{}} ~\cite{darling2015lexical,egan2011evaluation,zhao2013cost,ma2011learning,sha2015limited,astorino2016malicious,pradeepthi2014performance,chu2013protect,blum2010lexical,gyawali2011evaluating,garera2007framework}, uW: \cite{whittaker2010large}\end{tabular}    \\
\cline{2-5} &Longest Token& 6 & U & \begin{tabular}[c]{@{}l@{}} ~\cite{darling2015lexical,egan2011evaluation,feroz2014examination,sha2015limited,chu2013protect,lin2013malicious} \end{tabular}     \\
\cline{2-5} & Frequency of Digits & 3 & U  & \begin{tabular}[c]{@{}l@{}} ~\cite{sha2015limited,astorino2016malicious,lin2013malicious} \end{tabular}   \\
\cline{2-5} & Frequency of Special characters  &8 & U,W &  \begin{tabular}[c]{@{}l@{}} ~\cite{darling2015lexical,guan2009anomaly,zhao2013cost,rathod2015comparative,ma2011learning,lin2013malicious}, uW:~\cite{Gowtham2014,Mohammad2014}
\end{tabular}  \\
\cline{2-5} & Port presence & 18 & U,W,E &  \begin{tabular}[c]{@{}l@{}} ~\cite{le2011phishdef,basnet2015towards,egan2011evaluation,su2013suspicious,sananse2015phishing,lee2014users,gyawali2011evaluating} \\~uW: \cite{mohaisen2015towards, kosba2014adam,rajab2017new,bannur2011judging,zhang2017phishing,thabtah2016dynamic}, uE: \cite{khonji:lexial,Ammar:online,Ammar:evolving,isredzaRahmi:profiling,fergus:feature} 
\end{tabular} \\
\cline{2-5} & HTTPS or HTTP  & 15 & U,W & \begin{tabular}[c]{@{}l@{}} ~\cite{jeeva2016intelligent,blum2010lexical,gyawali2011evaluating}, uW: \cite{marchal2016know,el2017detection,bannur2011judging,rajab2017new, raodetection, lee2016phishing,thabtah2016dynamic,mohammad2014intelligent,moghimi2016new,he2011efficient,aburrous2010intelligent,thakur2014catching}
\end{tabular} \\
\hline
\multirow{4}{*}{\begin{tabular}[c]{@{}c@{}}\textbf{IP address}\\ \textbf{based}\end{tabular}} & Characters and digits in IP address  & 3 & U &\begin{tabular}[c]{@{}l@{}} ~\cite{egan2011evaluation, basnet2012mining,burnap2015real} \end{tabular}  \\
\cline{2-5} & IP address in octal/hex form & 5 & U,E &\begin{tabular}[c]{@{}l@{}} \cite{rathod2015comparative}, \cite{guan2009anomaly,garera2007framework,blum2010lexical},~uE:~\cite{Ammar:online}
\end{tabular}     \\
\cline{2-5} & Blacklisted IP address & 5 & U,W & \begin{tabular}[c]{@{}l@{}} ~\cite{ma2009beyond,ma2011learning}, uW: \cite{Gowtham2014,lee2016phishing,bannur2011judging}   \end{tabular}    \\
\cline{2-5} & Whitelisted IP address & 3 & U,W & \begin{tabular}[c]{@{}l@{}}~\cite{popescu2016practical, nguyen2013detecting}, uW: \cite{Gowtham2014}   \end{tabular}    \\
\cline{2-5} &IP records and prefix checking  & 4 & U,W,E & \begin{tabular}[c]{@{}l@{}} ~\cite{ma2009beyond,zhao2013cost},~uW:~\cite{el2017detection}, uE: \cite{chowdhury:multilayer}  
\end{tabular}   \\
\hline
\multirow{4}{*}{\begin{tabular}[c]{@{}c@{}}\textbf{WHOIS}\\ \textbf{properties}\end{tabular}} &Registration details& 14 & U,W & \begin{tabular}[c]{@{}l@{}}~\cite{egan2011evaluation,ma2009beyond,zhao2013cost,basnet2012mining,le2011phishdef,sananse2015phishing}, uW: \cite{kosba2014adam, fang2015proactive, Mohammad2014,mohaisen2015towards,bannur2011judging,dong2010defending,tan2014phishing,thabtah2016dynamic} \end{tabular}   \\
\cline{2-5} & Creation/Update/Expiration time  & 27 & U,W,E & \begin{tabular}[c]{@{}l@{}}~\cite{egan2011evaluation,ma2009beyond,zhao2013cost,basnet2012mining,le2011phishdef,sananse2015phishing,ma2011learning,chu2013protect,gupta2014bit,guan2009anomaly}, ~uE: \cite{fette2007learning,yearwood:profiling,chowdhury:multilayer} \\~uW:~\cite{kosba2014adam, mohaisen2015towards, lee2016phishing, canali2011prophiler, manek2014demalfier,zhang2017two,abdelhamid2014phishing,Gowtham2014,yue2013fine, miyamoto2008evaluation,rajab2017new,zhang2014domain, Mohammad2014,aggarwal2012phishari}\end{tabular}  \\ 
\cline{2-5}
& AS number          &  8  & U,W,E                                                                            & ~\cite{feroz2014examination,le2011phishdef,egan2011evaluation}, uW:\cite{kosba2014adam, mohaisen2015towards,whittaker2010large,geng2015combating}, uE: \cite{chowdhury:multilayer} \\  \cline{2-4} 
\cline{2-5} & Status of WHOIS entry & 7 & U & \begin{tabular}[c]{@{}l@{}}~\cite{egan2011evaluation,ma2009beyond,zhao2013cost,basnet2012mining,le2011phishdef,sananse2015phishing,ma2011learning}\end{tabular} \\
\hline
\multirow{2}{*}{\begin{tabular}[c]{@{}c@{}}\textbf{Geographic}\\ \textbf{properties}\end{tabular}} & Location\tnote{e}~ of IP address origin & 20 & U,W,E &\begin{tabular}[c]{@{}l@{}}~\cite{ma2009beyond,zhao2013cost,vu2016firstfilter,astorino2016malicious,basnet2015towards,le2011phishdef,ma2011learning,eshete2014webwinnow} \\~uW: \cite{kosba2014adam, mohaisen2015towards, canali2011prophiler, pao2012malicious,whittaker2010large, manek2014demalfier,bannur2011judging,stringhini2013shady,gowtham2014comprehensive,thomas2011design}, uE: \cite{DBLP:conf/ecir/GanstererP09,Ho2017DetectingCS} 
\end{tabular}  \\
\cline{2-5} & Nature/speed of connection & 5 & U & \begin{tabular}[c]{@{}l@{}}~\cite{ma2009beyond,astorino2016malicious,basnet2015towards,egan2011evaluation,ma2011learning}    \end{tabular}  \\
\hline
\multirow{6}{*}{\begin{tabular}[c]{@{}c@{}}\textbf{Shortened}\\\textbf{URL}\\ \textbf{features}\end{tabular}} & Initial and Landing URL & 6 & U,W &\begin{tabular}[c]{@{}l@{}}~\cite{popescu2015study,sandracoordinator}, uW: \cite{stringhini2013shady,barraclough2013intelligent,aburrous2010intelligent,thomas2011design}\end{tabular} \\
\cline{2-5} & Redirects on each page & 5 & U,W & \begin{tabular}[c]{@{}l@{}}~\cite{cao2016detection,burnap2015real,chhabra2011phi}, uW: \cite{thomas2011design,el2017detection}
\end{tabular} \\
\cline{2-5} & URL Redirect Chain length & 16 & U,W & \begin{tabular}[c]{@{}l@{}}~\cite{sandracoordinator,gupta2014bit,burnap2015real,lee2013warningbird,eshete2014webwinnow}, uW: \cite{manek2014demalfier,thomas2011design, xu2014evasion, choi2011detecting, stringhini2013shady, Mohammad2014,thabtah2016dynamic,abdelhamid2014phishing,aggarwal2012phishari,mohammad2014intelligent, thakur2014catching} \end{tabular} \\
\cline{2-5} & Frequency of entry point URL & 4 & U & \begin{tabular}[c]{@{}l@{}}~\cite{lee2013warningbird,sandracoordinator,gupta2014bit,chhabra2011phi}\end{tabular} \\
\cline{2-5} & Position of entry point URLs & 3 & U & \begin{tabular}[c]{@{}l@{}}~\cite{sandracoordinator,cao2016detection,lee2013warningbird}  \end{tabular}  \\
\cline{2-5} & \# of URLs/Landing URLs & 2 & U & \begin{tabular}[c]{@{}l@{}}~\cite{lee2013warningbird,sandracoordinator} \end{tabular} \\
\cline{2-5} & \# of domain names and IPs & 6 & U,E & \begin{tabular}[c]{@{}l@{}}~\cite{burnap2015real,lee2013warningbird,chhabra2011phi,nepali2016you}, uE:
\cite{chowdhury:multilayer,isredzaRahmi:profiling}

\end{tabular} \\
\hline
\end{tabular}
\begin{tablenotes}
\item[a] \textit{A URL usually has the following parts: scheme, host, path, query string.}\\ \textit{Source:}https://www.ibm.com/support/knowledgecenter/en/SSGMCP\_5.1.0/com.ibm.cics.ts.internet.doc/topics/dfhtl\_uricomp.html
\item[b] \textit{Characters like period, slash, etc.} 
\item[c] \textit{Time to live}
\item[d] \textit{Alexa Ranking, PageRank scores, search engines lookup }
\item[e] \textit{Continent/country/city}
\item[f] \textit{Hexadecimal encoding}
\end{tablenotes}
\end{threeparttable}}
\end{table*}

\begin{table*}[!htb]
\centering
\caption{Description of the features used in phishing website detection. N - Number of papers using each feature}
\label{table-webpage-features}
\begin{tabular}{|c|l|c|l|}
\hline
\textbf{Feature Source}               & \textbf{Features}      & \textbf{N}                                                                                                      & \textbf{Literature}                                                                                                                                                                                                            \\ \hline \hline
\multirow{17}{*}{\textbf{Network}}   & Domain registration info.   & 8                                                                                        & \cite{bannur2011judging, tan2014phishing, yue2013fine,zhang2017two,vargas:enemies, manek2014demalfier,zhang2014domain,barraclough2013intelligent}                                          
\\ \cline{2-4}
& Registrar ID       & 3                                                                                     & \cite{kosba2014adam, mohaisen2015towards,barraclough2013intelligent} \\ \cline{2-4} 

                             & \# of Nameservers     & 3                                                                                              & \cite{manek2014demalfier, choi2011detecting,aggarwal2012phishari}                                                                                                                                                                                           \\ \cline{2-4} 
                             & DNS record       & 12                                                                                                 & \cite{canali2011prophiler, lee2016phishing,xiang2010hierarchical,zhang2017two,abdelhamid2014phishing,zhang2014domain,Mohammad2014,mohammad2014intelligent,geng2015combating,ramesh2014efficacious,barraclough2013intelligent,el2017detection}                                                                                                                     \\ \cline{2-4} 
                             & \# of DNS queries       & 1                                                                                            & \cite{xu2014evasion}                                                                                                                                                                                                \\ \cline{2-4} 
                             & \begin{tabular}[c]{@{}l@{}}HTTP header (content-type, content-length, \\ Server, X-Powered-By, ...)\end{tabular} & 4 &\cite{bannur2011judging, xu2014evasion, mohaisen2015towards, kosba2014adam}                                                                                                                                         \\ \cline{2-4}                              
                             & Alexa rank      & 5                                                                                                   & \cite{marchal2016know, dong2010defending, lee2016phishing,Mohammad2014,zhang2017two}      
                              \\ \cline{2-4} 
                             & Gmail Reputation  & 1                                                                                                        & \cite{whittaker2010large}  
                             
                             \\ \cline{2-4} 
                             & \begin{tabular}[c]{@{}l@{}}\# of bytes/packets transferred, duration              \end{tabular}          & 2                                                             & \cite{xu2014evasion, choi2011detecting}                                                                                                                                                                             \\ \cline{2-4} 
                             & Fake HTTP protocol      & 1                                                                                           & \cite{abdelhamid2014phishing}                                                                                                                                                                            \\ \cline{2-4} 
                             & \# of IP/port upon complete  download                            & 1                                                   & \cite{xu2014evasion}                                                                                                                                                                                                \\\cline{2-4} & SSL Certification attributes & 9& \cite{dong2015beyond,he2011efficient,Gowtham2014,Mohammad2014,bannur2011judging,thabtah2016dynamic,abdelhamid2014phishing,gowtham2014comprehensive,aburrous2010intelligent}\\ 
                             \specialrule{.2em}{.1em}{.1em}
\multirow{16}{*}{\textbf{HTML}}       & \# of various tag types                                          & 10                                                   & \cite{manek2014demalfier, bannur2011judging, marchal2016know, choi2011detecting, yue2013fine, lee2016phishing,zhuang2012intelligent,vargas:enemies,canali2011prophiler,liang2016cracking}                                                                                                    \\ \cline{2-4} 
                             & HTML tag attributes    & 10                                                                                             & \cite{bannur2011judging,zhuang2012intelligent,wenyin2012antiphishing,varshney2016phish,zhang2017two,he2011efficient,vargas:enemies,thabtah2016dynamic,corona2017deltaphish,Xiang:2011:CFM:2019599.2019606}                                                                                                                                                                                           \\ \cline{2-4} 
                             & Term Frequency        & 13                                                                                              & \cite{manek2014demalfier, bannur2011judging, tan2014phishing, huang2010mitigate, miyamoto2008evaluation,whittaker2010large,zhang2017two,liu2010automatic,he2011efficient,tan2016phishwho,ramesh2014efficacious,Xiang:2011:CFM:2019599.2019606,zhang2017phishing}                                                                                                           \\ \cline{2-4} 
                             & \# of element out of place                                              & 2                                            & \cite{manek2014demalfier,Xiang:2011:CFM:2019599.2019606}                                                                                                                                                                                           \\ \cline{2-4} 
                             & \# of small/hidden elements    & 2                                                                                            & \cite{canali2011prophiler,yue2013fine}                                                                                                                                                                                          \\ \cline{2-4} 
                             & \# of suspicious elements   & 3                                                                                        & \cite{canali2011prophiler, lee2016phishing,shahriar2012trustworthiness}                                                                                                                                                                         \\ \cline{2-4} 
                             & \# of suspicious objects                                                    & 2                                           & \cite{canali2011prophiler,lee2016phishing}                                                                                                                                                                                          \\ \cline{2-4} 
                             & \# of internal/external links                                              & 22                                           & \begin{tabular}[c]{@{}l@{}}\cite{bannur2011judging, xu2014evasion, choi2011detecting, lee2016phishing,whittaker2010large,Gowtham2014, Mohammad2014, yue2013fine,canali2011prophiler,abdelhamid2014phishing,zhang2017two,el2017detection,thabtah2016dynamic} \\\cite{liang2016cracking,liu2010automatic,mohammad2014intelligent,he2011efficient,corona2017deltaphish,ramesh2014efficacious,zhang2014domain,Xiang:2011:CFM:2019599.2019606,zhang2017phishing}   \end{tabular}
                           \\ \cline{2-4} & NULL links on site and footer & 2 & \cite{raodetection,zhang2017phishing}
                           \\ \cline{2-4} 
                             & More than one head tag/document                                          & 2                                                    & \cite{yue2013fine,canali2011prophiler}                                                                                                                                                                             \\ \cline{2-4} 
                             & invisible frames                                                        & 5                                            & \cite{liang2009malicious, choi2011detecting, yue2013fine,canali2011prophiler,abdelhamid2014phishing}                                                                                                                                                           \\ \cline{2-4} 
                             & \begin{tabular}[c]{@{}l@{}}\# of specific file type\\ (image, video, binary, system file, ...)\end{tabular}   & 4      & \cite{bannur2011judging, marchal2016know, pao2012malicious, mohaisen2015towards}                                                                                                                                                      \\ \cline{2-4} 
                             & Visual      & 9                                                                                                     & \cite{bannur2011judging, huang2010mitigate,shahriar2012trustworthiness,zhuang2012intelligent,chiew2015utilisation,geng2015combating,liu2010automatic,corona2017deltaphish,mao2017phishing}                                                                                                                                                                         \\ \cline{2-4} 
                             & \# of iframes                                                            & 9                                          & \cite{marchal2016know, xu2014evasion, choi2011detecting, yue2013fine, lee2016phishing,rajab2017new, canali2011prophiler,thabtah2016dynamic,el2017detection}                                                                                                                              \\ \cline{2-4} 
                             & DOM-tree                                                                & 3                                            & \cite{rosiello2007layout,mao:alarm,Xiang:2011:CFM:2019599.2019606}                                                                                                                                                                                           \\ \cline{2-4} 
                             & ActiveX function                                                        & 1                                            & \cite{lee2016phishing}                                                                                                                                                                                              \\ \cline{2-4} 
                             & Right click disabled                                                    & 8                                            & \cite{lee2016phishing,Mohammad2014,thabtah2016dynamic,mohammad2014intelligent,barraclough2013intelligent,aburrous2010intelligent,rajab2017new,el2017detection}
                              
                   \\ \cline{2-4} 
                             &      Server Form Handler                                  & 2                                               & \cite{Mohammad2014,el2017detection}         
                              \\ \cline{2-4} 
                             &      Login form detection                                  & 4                                               & \cite{Gowtham2014,Xiang:2011:CFM:2019599.2019606,xu2014gemini,zhang2017two}                             \\ \cline{2-4} 
                             &      External term frequency                                  & 1                                               & \cite{marchal2016know}     
                              \\
                             \specialrule{.2em}{.1em}{.1em}
\multirow{9}{*}{\textbf{Javascript}} & \begin{tabular}[c]{@{}l@{}}keywords to words ratio\\ \# of suspicious string\end{tabular}    & 2                                                                                         & \cite{canali2011prophiler, yue2013fine}                                                                                                                                                                             \\ \cline{2-4} 
                             & \# of long strings (\textgreater 40, \textgreater51)            & 3                                                       & \cite{canali2011prophiler, xu2014evasion, yue2013fine}                                                                                                                                                              \\ \cline{2-4} 
                             & \begin{tabular}[c]{@{}l@{}}decoding routines, shellcode detection, \\ \# of iframe strings   \end{tabular}        & 1                                                                                       & \cite{canali2011prophiler}                                                                                                                                                                                          \\ \cline{2-4} 
                             & \begin{tabular}[c]{@{}l@{}}\# of DOM modifying functions,\\\# of event attachment        \end{tabular}                                 & 3                                              & \cite{canali2011prophiler, yue2013fine, lee2016phishing}                                                                                                                                                            \\ \cline{2-4} 
                             & \# of suspicious objects                                              & 3                                            & \cite{canali2011prophiler, yue2013fine,choi2011detecting}                                                                                                            \\ \cline{2-4} 
                             & \# of scripts                                                            & 2                                           & \cite{xu2014evasion,liang2016cracking}                                                                                                                                                                                                \\ \cline{2-4} 
                             & \# of func. (eval, setInterval, OnMouseOver...)              & 9                                                   & \cite{xu2014evasion, choi2011detecting, yue2013fine, lee2016phishing,Mohammad2014,rajab2017new,thabtah2016dynamic,mohammad2014intelligent,aburrous2010intelligent}                                                                                                                                               \\ 
                        \specialrule{.2em}{.1em}{.1em}
  \multirow{2}{*}{\textbf{Others}}   & \begin{tabular}[c]{@{}l@{}}Ray Scan Method and \\ Webpage Layout Similarity \end{tabular}   & 1                                                                                        & \cite{fang2015proactive}                                          
\\ \cline{2-4}
& Markov Chain and DISCO         &   1  & \cite{marchal2012proactive}\\ \hline
\end{tabular}
\end{table*}

\begin{table*}[tbh!]
\centering
\caption{Description of the features used in phishing email detection}
\label{table-email-features}
\begin{threeparttable}
\begin{tabular}{|c|l|l|l|}
\hline
\textbf{Feature Source} &\textbf{Features}&\textbf{N} & \textbf{Literature}                                        \\ \hline\hline
\multirow{18}{*}{\textbf{Header}}& \textit{``Message-ID''} & 3 & \cite{Isredza:phishing,satheesaan:email,verma:phish} \\ \cline{2-4} 
              & \textit{``Received''} fields                                                                                                 & 1& \cite{verma:detecting}
              \\ \cline{2-4} 
                             & \textit{``From'', ``Mail from'', ``Sender'', ``Mail-to'', ``Delivered To''}                                                         & 6& \cite{masoumeh:feature,Ho2017DetectingCS,Duman:emailprofiler,fergus:feature}
                             \cite{kim:semantic,Han:2016:ASP:2851613.2851801}                                                                                                                                           \\ \cline{2-4} 
                             & Authentication-Results (SPF, DKIM, etc.)       &1     & \cite{verma:detecting}
                             \\
                                                    \cline{2-4} 
                             &\begin{tabular}[c]{@{}l@{}} \textit{``Subject''} features (length of the subject, \# of words,\\\# of characters, vocabulary richness, etc.) \end{tabular}& 8& 
             \begin{tabular}[c]{@{}l@{}}\cite{Duman:emailprofiler,chandrasekaran2006phishing,khonji:lexial,Han:2016:ASP:2851613.2851801,masoumeh:feature}\\
\cite{isredzaRahmi:profiling, Islam:2013:MPD:2405859.2406232,fergus:feature} \end{tabular}
                              \\\cline{2-4} 
                             & Blacklisted words in \textit{``Subject''} & 4& \cite{khonji:lexial,Isredza:phishing,Ammar:online,fergus:feature} 
                              \\
                               \cline{2-4} 
                             & \# of words and/or characters in the \textit{``Send''} field&1& \cite{isredzaRahmi:profiling} 
                             \\
                               \cline{2-4} 
                             &\textit{``Sender''} domain $\neq$ \textit{``Reply-to''} domain
                             & 4&\cite{khonji:lexial,Ammar:online,isredzaRahmi:profiling,fergus:feature}
                             \\
                             \cline{2-4} 
                             & \textit{``Sender''} domain $\neq$ \textit{``Message-Id''} domain& 1&\cite{Isredza:phishing}
                             \\
                               \cline{2-4} 
                             & \textit{``Sender'' / ``From''} $\neq$  email's modal domains& 5&\cite{Andronicus:classification,khonji:lexial,Ammar:evolving,isredzaRahmi:profiling,fergus:feature}
                              \\
                               \cline{2-4}
                            
                             & Timestamp, ``Sent date'' & 3&
                             \cite{Han:2016:ASP:2851613.2851801,Stringhini2015,Duman:emailprofiler}   
                               
                               
                             \\
                             \cline{2-4}
                            
                             & Source IP, Autonomous System Number, Origin Country & 1  & \cite{Han:2016:ASP:2851613.2851801}     
                             \\
                        \cline{2-4}
                            
                             &\textit{``Subject''}: Fwd, Reply & 3 & \cite{Ammar:online,isredzaRahmi:profiling,fergus:feature} 
                             \\      
                             \cline{2-4}
                            
                             &Interaction habits & 1 & \cite{Stringhini2015}      \\ 
                              \cline{2-4}
                            
                             &\textit{``Cc''}, \textit{``BCc''} fields& 1 & \cite{Duman:emailprofiler}       
                             \\
                              \cline{2-4}
                            
                             & \begin{tabular}[c]{@{}l@{}}
                             \textit{``X-Mailer''}, \textit{``X-Originating-IP''}, \textit{``X-Originating-hostname''},\\\textit{``X-spam-flag'},\textit{``X-virus-scanned'}
                             \end{tabular}& 1 & \cite{Duman:emailprofiler}       
                             \\
                             
                                    \specialrule{.2em}{.1em}{.1em} 
\multirow{45}{*}{\textbf{Body}}    & \begin{tabular}[c]{@{}l@{}}  Lexical Features (\# of words, \# of unique words,\\
\# of characters, Tokens\tnote{a}, regular expressions, etc.) \end{tabular}                                                                                &12  & \begin{tabular}[c]{@{}l@{}} \cite{chandrasekaran2006phishing,Duman:emailprofiler,Laszka:2015:OPF:2887007.2887140,isredzaRahmi:profiling,chowdhury:multilayer}
\\
\cite{Han:2016:ASP:2851613.2851801,Stringhini2015,Islam:2013:MPD:2405859.2406232,fergus:feature} 
\\
\cite{DBLP:journals/corr/YasinA16,verma:detecting,vermaH13}
\end{tabular}
\\

 \cline{2-4} 
                             & Function words (count, frequency distribution, etc.)  &4                                                                                                & \cite{chandrasekaran2006phishing,Stringhini2015,isredzaRahmi:profiling,Han:2016:ASP:2851613.2851801}               
                                                                                                      \\ 

\cline{2-4}

          & Style metrics (number of paragraphs in the email, Yule metric, etc.) &1& \cite{Stringhini2015}

\\

\cline{2-4}
                               
                             & \begin{tabular}[c]{@{}l@{}}Topic in the body, Latent Semantic Indexing,\\Readability Indexes\end{tabular}& 3&\cite{Han:2016:ASP:2851613.2851801,bergholz2008improved,ramanathan}

\\ 

\cline{2-4} 
                             &\begin{tabular}[c]{@{}l@{}} NLP (Part-of-Speech tags, Named entities,\\ Wordnet properties, etc.)   \end{tabular}                                                                 & 5& \cite{Duman:emailprofiler,verma:detecting,vermaH13,DBLP:journals/corr/YasinA16,ramanathan}       

                          \\   \cline{2-4} & \begin{tabular}[c]{@{}l@{}} Semantic Network Analysis  \end{tabular} &1 & \cite{kim:semantic}                                                                                 
\\ \cline{2-4} 
                             & Vocabulary richness                                                  & 3 &\cite{chandrasekaran2006phishing,isredzaRahmi:profiling,fergus:feature}
                             \\ \cline{2-4} & Urgency, reward, threat language in the body                                                                           &2 & \cite{kim:semantic,chandrasekaran2006phishing}
                             \\ 
                             
                            \cline{2-4} 
                             & Greeting in the message                                                                                &4& \cite{chandrasekaran2006phishing,chowdhury:multilayer,Duman:emailprofiler,john:applying}    
                              \\ \cline{2-4} 
                            & Signature in the message &3 & \cite{chowdhury:multilayer,Duman:emailprofiler,john:applying}   \\
                             \cline{2-4} 
                              & Farewell in the message  &1 & \cite{Duman:emailprofiler}   
        \\ \cline{2-4}    
             & Presence of both ``From:'' and ``To:'' in email body & 1 & \cite{Islam:2013:MPD:2405859.2406232}
                         
                           \\
              \cline{2-4}
             & \# of linked to domains &1 & \cite{Andronicus:classification} 
                         
                           \\ \cline{2-4}

                             &URL features& 18& \begin{tabular}[c]{@{}l@{}} \cite{chandrasekaran2006phishing,verma:detecting,Ammar:online,khonji:lexial,yearwood:profiling,Olivo:2013}
                \\             
\cite{Dazeley2010,Stringhini2015,chowdhury:multilayer,masoumeh:feature,john:applying} 
                             \\\cite{Ammar:evolving,adre:new,Ho2017DetectingCS,orman:semantics, DBLP:journals/corr/YasinA16,bergholz2008improved,toolan:ensembles,fergus:feature}                             
                             \end{tabular}
                               \\ \cline{2-4} 
                             & HTML features &13
                             & \begin{tabular}[c]{@{}l@{}} \cite{Andronicus:classification,yearwood:profiling,Ammar:online,chowdhury:multilayer,adre:new, Islam:2013:MPD:2405859.2406232,toolan:ensembles} 
                             \\ 
                   \cite{masoumeh:feature,Dazeley2010, DBLP:journals/corr/YasinA16,fergus:feature,Olivo:2013,john:applying}
                             \end{tabular} 
                               \\ \cline{2-4} 
                             & Presence and/or \# of forms in email body& 6 &\begin{tabular}[c]{@{}l@{}}  \cite{yearwood:profiling,chowdhury:multilayer,adre:new,isredzaRahmi:profiling,fergus:feature,john:applying}
                             \end{tabular}
                             
                                                            \\ \cline{2-4} 
                             & Blacklisted words in the message&10& \begin{tabular}[c]{@{}l@{}} \cite{Andronicus:classification,khonji:lexial,isredzaRahmi:profiling,Ammar:online,adre:new}\\
                             \cite{Ammar:evolving,masoumeh:feature, DBLP:journals/corr/YasinA16, bergholz2008improved,fergus:feature}  

                           \end{tabular}
                             \\
                       \cline{2-4} 
                             & Scripts/JavaScripts features in email body&16& \begin{tabular}[c]{@{}l@{}} \cite{DBLP:conf/ecir/GanstererP09,khonji:lexial,yearwood:profiling,Ammar:online, Islam:2013:MPD:2405859.2406232,fergus:feature}
                  
                             \\                           \cite{Dazeley2010,chowdhury:multilayer,Andronicus:classification,fette2007learning,isredzaRahmi:profiling,toolan:ensembles}
                             \\                     \cite{masoumeh:feature,adre:new,bergholz2008improved,john:applying} \end{tabular}
                             \\   
                             \cline{2-4} 
                             & \# of onClick events in email body& 4&\cite{DBLP:conf/ecir/GanstererP09,Ammar:online,isredzaRahmi:profiling,fergus:feature} 
                             \\    
                             \cline{2-4} &   Features from images/logos in the message & 6&\begin{tabular}[c]{@{}l@{}} \cite{yearwood:profiling,Dazeley2010,kim:semantic,chowdhury:multilayer,adre:new,Islam:2013:MPD:2405859.2406232}\\
             \end{tabular}
                             
                             \\    
                             \cline{2-4} & \begin{tabular}[c]{@{}l@{}} Mention of the sender \end{tabular} &1 & \cite{kim:semantic}
                         \\ 
                         \cline{2-4} & \# of links in email body
                         &  10&\begin{tabular}[c]{@{}l@{}}\cite{yearwood:profiling,chowdhury:multilayer,isredzaRahmi:profiling,Ammar:evolving,Andronicus:classification}\\           \cite{adre:new,Dazeley2010,fergus:feature,toolan:ensembles,john:applying}\end{tabular}\\
                            \cline{2-4}
             &\# of images (links)   &9 & \begin{tabular}[c]{@{}l@{}}\cite{masoumeh:feature,adre:new,Ammar:evolving,isredzaRahmi:profiling, Islam:2013:MPD:2405859.2406232} \\ \cite{DBLP:journals/corr/YasinA16, bergholz2008improved,fergus:feature,john:applying} \end{tabular} 
             \\

                         \cline{2-4} & \# of tables in email body& 4&\cite{yearwood:profiling,chowdhury:multilayer,Dazeley2010,john:applying}\\
                         
                         \cline{2-4} &Recipient's email address in email body&1& \cite{Han:2016:ASP:2851613.2851801}\\
                         \cline{2-4} & Phishing terms weight& 2 &\cite{Dazeley2010,DBLP:journals/corr/YasinA16}
                         \\
                          \cline{2-4} & TFIDF& 4 &\cite{Dazeley2010,chowdhury:multilayer,john:applying,Seifollahi2017}\\
                         
                                \cline{2-4} & Email size & 3& \cite{Chiang:2015:AMC:2778890.2779103,chowdhury:multilayer,john:applying}
                                \\
                                \cline{2-4} &\#  of email body parts& 2&\cite{adre:new,bergholz2008improved}\\
                               
                                 \cline{2-4} & \begin{tabular}[c]{@{}l@{}}MIME Version or Content-Type: (Multipart/Alternative,\\ text/plain, text/html)\end{tabular}& 10&\begin{tabular}[c]{@{}l@{}} \cite{Ammar:online,adre:new,Ammar:evolving,fette2007learning,isredzaRahmi:profiling}\\     
                                 \cite{masoumeh:feature,Chiang:2015:AMC:2778890.2779103,yearwood:profiling,Duman:emailprofiler,Islam:2013:MPD:2405859.2406232} \end{tabular}\\
                                  \cline{2-4} &JavaScript PopUp Windows& 2&\cite{Ammar:online,masoumeh:feature}\\
                     \cline{2-4} 
                             
                             &  Link displayed $\neq$ Link in destination                                                    & 10& \begin{tabular}[c]{@{}l@{}}\cite{Dazeley2010,Ammar:evolving,Andronicus:classification,fette2007learning} \\ \cite{orman:semantics,chandrasekaran2006phishing, DBLP:journals/corr/YasinA16,bergholz2008improved,Olivo:2013,john:applying}
                             \end{tabular}
                              \\
                             \cline{2-4} 
                             & Img links $\neq$ spoofed target address &3 & \cite{Ammar:online,adre:new, Olivo:2013}   \\ 
                             \cline{2-4} 
                             
                             & Hidden text in the email, Salting techniques &3 & \cite{chowdhury:multilayer,adre:new, Islam:2013:MPD:2405859.2406232}   \\
                              \cline{2-4} 
                              
                              
               \cline{2-4}
               &\# of internal/external links   &6 & \cite{masoumeh:feature,adre:new,Ammar:evolving,isredzaRahmi:profiling,bergholz2008improved,fergus:feature}
                       

                                 \\\specialrule{.2em}{.1em}{.1em}
\multirow{1}{*}{\textbf{External}}        & SpamAssassin           &5 & \cite{khonji:lexial,Ammar:online,adre:new,fette2007learning, bergholz2008improved} 
\\  
        \specialrule{.2em}{.1em}{.1em} 
\multirow{2}{*}{\textbf{Attachment}}        & Size in bytes of attachment                                               & 1& \cite{Han:2016:ASP:2851613.2851801}
\\ 
                         \cline{2-4} &Number of attachments & 2& \cite{Dazeley2010, chandrasekaran2006phishing}\\ 
\hline
                       
 \end{tabular}
 
 \begin{tablenotes}
\item[a] \textit{Semantic relations can also be used to add tokens}
\end{tablenotes}
 \end{threeparttable}  
 \end{table*}

\end{document}